%% file: main.tex
\documentclass[twocolumn]{aastex63}

\usepackage{amsmath}
\input{objects}
\DefineObj[2MASS~J0045$+$16]{2m0045}{2MASS~J00452143$+$1634446}
\DefineObj[2MASS~J0501$-$00]{2m0501}{2MASS~J05012406$-$0010452}
\DefineObj[2MASS~J1425$-$36]{2m1425}{2MASS~J14252798$-$3650229}

\DefineObj[2MASS~J0001$+$1535]{2m0001}{2MASS~J00011217$+$1535355}
\DefineObj[2MASS~J0030$-$1450]{2m0030}{2MASS~J0030300$-$145033}

\DefineObj[2MASS~J0031$+$5749]{2m0031}{2MASS~J00310928$+$5749364}
\DefineObj[2MASS~J0153$-$6744]{2m0153}{2MASS~J01531463$-$6744181}
\DefineObj[2MASS~J0326$-$2102]{2m0326}{2MASS~J03264225$-$2102057}
\DefineObj[2MASS~J0342$-$6817]{2m0342}{2MASS~J0342162$-$681732}
\DefineObj[2MASS~J0349$+$0635]{2m0349}{2MASS~J03492367$+$0635078}
\DefineObj[2MASS~J0355$+$1133]{2m0355}{2MASS~J03552337$+$1133437}
\DefineObj[2MASS~J0447$-$1216]{2m0447}{2MASS~J04473039$-$1216155}

\DefineObj[2MASS~J0459$-$2853]{2m0459}{2MASS~J04590034$-$2853396}
\DefineObj[2MASS~J0506$+$5236]{2m0506}{2MASS~J05065012$+$5236338}
\DefineObj[2MASS~J0642$+$4101]{2m0642}{2MASS~J06420559$+$4101599}
\DefineObj[2MASS~J0718$-$6415]{2m0718}{2MASS~J07180871$-$6415310}

\DefineObj[2MASS~J0809$+$4434]{2m0809}{2MASS~J08095903$+$4434216}
\DefineObj[2MASS~J0951$-$8023]{2m0951}{2MASS~J09512690 $-$8023553}
\DefineObj[2MASS~J1551$+$0941]{2m1551}{2MASS~J15515237$+$0941148}
\DefineObj[2MASS~J1647$+$5632]{2m1647}{2MASS~J16471580$+$5632057}

\DefineObj[2MASS~J1741$-$4642]{2m1741}{2MASS~J17410280$-$4642218}
\DefineObj[2MASS~J2002$-$0521]{2m2002}{2MASS~J20025073$-$0521524}
\DefineObj[2MASS~J2117$-$2940]{2m2117}{2MASS~J21171431$-$2940034}
\DefineObj[2MASS~J2154$-$1055]{2m2154}{2MASS~J21543454$-$1055308}

\DefineObj[2MASS~J2206$+$3301]{2m2206+33}{2MASS~J22062520$+$3301144}
\DefineObj[2MASS~J2206$-$4217]{2m2206-42}{2MASS~J2206450$-$421721}
\DefineObj[WISE~J2216$+$1952]{2m2216}{WISE~J221628.62$+$195248.1}
\DefineObj[2MASS~J2322$-$6151]{2m2322}{2MASS~J23225299$-$6151275}
\DefineObj[2MASS~J2343$-$3646]{2m2343}{2MASS~J23433470$-$3646021}

\DefineObj[2MASS~J2139$+$02]{2m2139}{
2MASS~J21392676$+$0220226}

\DefineObj[SIMP~J0136$+$09]{s0136}{
SIMP J01365662+0933473}

\DefineObj[WISE~J0047$+$6803]{w0047}{WISE~J004701.06$+$680352.1}

\DefineObj[2MASS~J2244$+$2043]{2m2244}{2MASS~J2244316$+$204343}

\DefineObj[SDSS~1110$+$0116]{s1110}{SDSS~111010$+$011613}

\usepackage{gensymb}
\usepackage{soul}
\submitjournal{AAS Journals}

\shorttitle{Let the Great World Spin}
\shortauthors{Vos et al.}
\graphicspath{{./}{figures/}}

\begin{document}

\title{Let the Great World Spin: Revealing the Stormy, Turbulent Nature of Young Giant Exoplanet Analogs with the Spitzer Space Telescope}

\correspondingauthor{Johanna Vos}
\email{jvos@amnh.org}

\author[0000-0003-0489-1528]{Johanna M. Vos}
\affil{Department of Astrophysics, American Museum of Natural History, Central Park West at 79th Street, NY 10024, USA}

\author[0000-0001-6251-0573]{Jacqueline K. Faherty}
\affil{Department of Astrophysics, American Museum of Natural History, Central Park West at 79th Street, NY 10024, USA}

\author[0000-0002-2592-9612]{Jonathan Gagn\'e}
\affil{Plan\'etarium Rio Tinto Alcan, Espace pour la Vie, 4801 av. Pierre-de Coubertin, Montr\'eal, Québec, Canada}
\affil{Institute  for  Research  on  Exoplanets, Universit\'e  de  Montr\'eal, 2900  Boulevard  \'Edouard-Montpetit Montr\'eal,  QC  Canada  H3T  1J4}

\author[0000-0002-5251-2943]{Mark Marley}
\affiliation{University of Arizona Department of Planetary Sciences and Lunar and Planetary Laboratory}

\author[0000-0003-3050-8203]{Stanimir Metchev}
\affil{Department of Physics \& Astronomy and Centre for Planetary Science and Exploration, The University of Western Ontario, London, Ontario N6A 3K7, Canada}

\author[0000-0002-8916-1972]{John Gizis}
\affil{Department of Physics and Astronomy, University of Delaware, Newark, DE 19716, USA}

\author[0000-0002-3252-5886]{Emily L. Rice}
\affil{Department of Astrophysics, American Museum of Natural History, Central Park West at 79th Street, NY 10024, USA}
\affil{Macaulay Honors College, 35 W. 67th Street, New York, NY 10023, USA}

\author[0000-0002-1821-0650]{Kelle Cruz}
\affil{Department of Astrophysics, American Museum of Natural History, Central Park West at 79th Street, NY 10024, USA}
\affil{Department of Physics and Astronomy, Hunter College, City University of New York, 695 Park Avenue, New York, NY 10065, USA}
\affil{Flatiron Institute, 162 Fifth Avenue, New York, NY 10010, USA}



\begin{abstract}
We present a survey for photometric variability in young, low-mass brown dwarfs with the Spitzer Space Telescope. The 23 objects in our sample show robust signatures of youth and share properties with directly-imaged exoplanets. We present three new young objects: \obj*{2m0349}, \obj*{2m0951} and \obj*{2m0718}. We detect variability in 13 young objects, and find that young brown dwarfs are highly likely to display variability across the L2--T4 spectral type range. In contrast, the field dwarf variability occurrence rate drops for spectral types $>$L9.  We examine the variability amplitudes of young objects and find an enhancement in maximum amplitudes compared to field dwarfs. We speculate that the observed range of amplitudes within a spectral type may be influenced by secondary effects such as viewing inclination and/or rotation period.  We combine our new rotation periods with the literature to investigate the effects of mass on angular momentum evolution. While high mass brown dwarfs ($>30 M_{\mathrm{Jup}}$) spin up over time, the same trend is not apparent for lower mass objects ($<30 M_{\mathrm{Jup}}$), likely due to the small number of measured periods for old, low-mass objects. The rotation periods of companion brown dwarfs and planetary-mass objects are consistent with those of isolated objects with similar ages and masses, suggesting similar angular momentum histories. Within the AB Doradus group, we find a high variability occurrence rate and evidence for common angular momentum evolution. The results are encouraging for future variability searches in directly-imaged exoplanets with facilities such as the James Webb Space Telescope and 30-meter telescopes.
\end{abstract}



\section{Introduction} \label{sec:intro}
The atmospheres of brown dwarfs and exoplanets with temperatures $<2300 $K are characterised by the presence of condensate clouds, which shape their emergent colors and spectra. Based on the handful of directly-imaged exoplanets discovered and investigated to date \citep[e.g. $\beta$~Pictoris~b, HR8799~bcde, 2M1207~b, 51~Eri~b;][]{Lagrange2010, Marois2008, Marois2010, Chauvin2004,Macintosh2015}, it is clear that in-depth investigations of these planets will hinge on a thorough understanding of condensate cloud chemistry and its relationship with surface gravity, temperature and metallicity.

Time-resolved atmospheric monitoring of brown dwarfs and exoplanets directly probes atmospheric features such as condensate clouds. In the last decade, our understanding of clouds in brown dwarf atmospheres has been revolutionized by a number of infrared studies of rotationally modulated variability. Large surveys \citep{Buenzli2014, Radigan2014, Wilson2014, Metchev2015, Eriksson2019} have revealed that variability is  common across L and T spectral types \citep{Metchev2015}, with a possible enhancement in variability amplitude and occurrence rate across the L/T transition \citep{Radigan2014}. {First results from dwarf monitoring of Y dwarfs with the Spitzer Space Telescope suggest that variability is also common at the lowest temperatures  \citep{Cushing2016,Esplin2016,  Leggett2016}.} Multiwavelength characterization studies have provided insight into the vertical atmospheric structure of brown dwarfs and planetary-mass objects, since different wavelengths probe different atmospheric pressures. For example, comparing variability amplitudes at different wavelengths provides information on the types of clouds present \citep{Artigau2009, Radigan2012, Apai2013,Kellogg2017, Biller2018} and phase shifts between light curves of different wavelengths have revealed multiple cloud layers \citep{Buenzli2012, Biller2013}.

{Physical mechanisms that can explain observed spectroscopic and photometric variations without clouds have also been proposed. Models presented by \citet{Morley2014} for mid-late T dwarfs and Y dwarfs show that hot spots can lead to variability. \citet{ Tremblin2020} suggest that variability may be driven by temperature fluctuations that arise as a result of thermochemical instabilities. Moreover, brown dwarfs are known to exhibit magnetically-driven aurorae, which may also drive infrared variations \citep[e.g.][]{Hallinan2015,Kao2016, Pineda2017}. Spectroscopic variability monitoring will be necessary to confirm the spectral signatures predicted by each model and to disentangle the effects of each atmospheric process. }

Photometric and spectroscopic monitoring is likely to reveal important insights into the atmospheres of directly-imaged exoplanets. Indeed, there have already been a number of variability detections in wide-orbit companions with masses in the range $\sim10-30~M_{\mathrm{Jup}}$ \citep{Zhou2016, Zhou2018, Zhou2020, Naud2017}. For the closer in directly-imaged exoplanets, there have been two attempts to detect photometric variability in the HR8799~bc planets \citep{Apai2016,Biller2021}. However, neither study detected variability in the planets, and placed upper limits of $5-10\%$ on the variability amplitudes. The James Webb Space Telescope (JWST) will enable unprecedented sensitivity and stability for future variability observations of directly-imaged exoplanets such as HR8799~bcde and $\beta$~Pictoris~b.

Young, isolated brown dwarfs are emerging as a bridge population to giant exoplanets. This population is vastly easier to study without the glare of a bright host star, and exist in greater numbers than the current population of known directly-imaged exoplanets \citep{Allers2013, Cruz2009, Faherty2016, Gagne2015c, Liu2016}. The colors and spectra of this new population are remarkably similar to the directly-imaged exoplanets; they exhibit redder near-infrared colors than higher mass field brown dwarfs with the same spectral type and appear fainter  than their equivalent temperature, more massive counterparts \citep{ Faherty2016, Liu2016}. These differences are generally explained by enhanced atmospheric clouds and/or vertical atmospheric mixing \citep[e.g.][]{Barman2011, Marley2012}, arising from lower surface gravity of the younger objects. Since surface gravity appears to have a large effect on the atmospheric and cloud properties of brown dwarfs and directly-imaged exoplanets, it follows that variability may be distinct for the low-gravity population.

A number of variability studies have focused on young, low-gravity free-floating and companion objects, and initial studies of young free-floating objects suggest that variability may be enhanced compared to field brown dwarfs. The first detection of cloud-driven variability in a planetary-mass object for the 23 Myr old $\beta$~Pictoris moving group member PSO~J318.5$-$22 showed the largest variability amplitude to date in an L dwarf \citep{Biller2015}. This initial discovery was followed by a notable sample of low-gravity objects showing high-amplitude variability \citep{Lew2016, Zhou2016, Vos2018, Zhou2020, Bowler2020}.
\citet{Metchev2015} note a potential enhancement in amplitude in a sample of five low-gravity L3-L5.5 brown dwarfs at $3.6~\mu$m and $4.5~\mu$m. \citet{Vos2020} also note an apparent amplitude enhancement for young late-L dwarfs at $4.5~\mu$m based on a compilation of all of the objects monitored with Spitzer to date. Recently, the high-amplitude variables \obj{s0136} and \obj{2m2139} were identified as members of the $\sim200$~Myr old Carina-Near moving group \citep{Gagne2017,ZJZhang2021}. 
There is also evidence for a higher variability occurrence rate in young brown dwarfs. \citet{Vos2019} carried out a large, ground-based survey for $J$-band variability in isolated. low-gravity dwarfs, finding that the young population has a variability occurrence rate of $30^{+16}_{-8}\%$, compared to $11^{+11}_{-4}\%$ of field dwarfs. It is clear that a large, unbiased survey for variability in young, low-gravity objects is necessary to test the tentative amplitude and occurrence rate enhancements reported to date. 

In this paper, we present the first large survey for variability in young, low-gravity brown dwarfs in the mid-infrared, and the most sensitive survey to date at any wavelength. We describe the sample in Section \ref{sec:sample}, including the discovery of three new brown dwarfs with evidence of youth in Section \ref{sec:newbds}. We describe the Spitzer observations, data reduction and light curve analysis in Sections \ref{sec:observations} and \ref{sec:variabilityanalysis}. We discuss the youth and variability properties of each target in detail in Section \ref{sec:youth+var}. We discuss the variability occurrence rate of young brown dwarfs in comparison with field brown dwarfs in Section \ref{sec:variability_occurrence}, and the variability amplitudes in Section \ref{sec:variability_properties}. We put the rotation rates of our young sample in context with literature data in Section \ref{sec:rotation_rates}, and specifically discuss the variability properties of AB Doradus members in Section \ref{sec:ABDor}. Finally, we lay out some lessons for variability searches in directly-imaged exoplanets in Section \ref{sec:lessons}. We summarize our results in Section \ref{sec:conclusions}.

\begin{figure}[tb]
   \centering
   \includegraphics[scale=0.55]{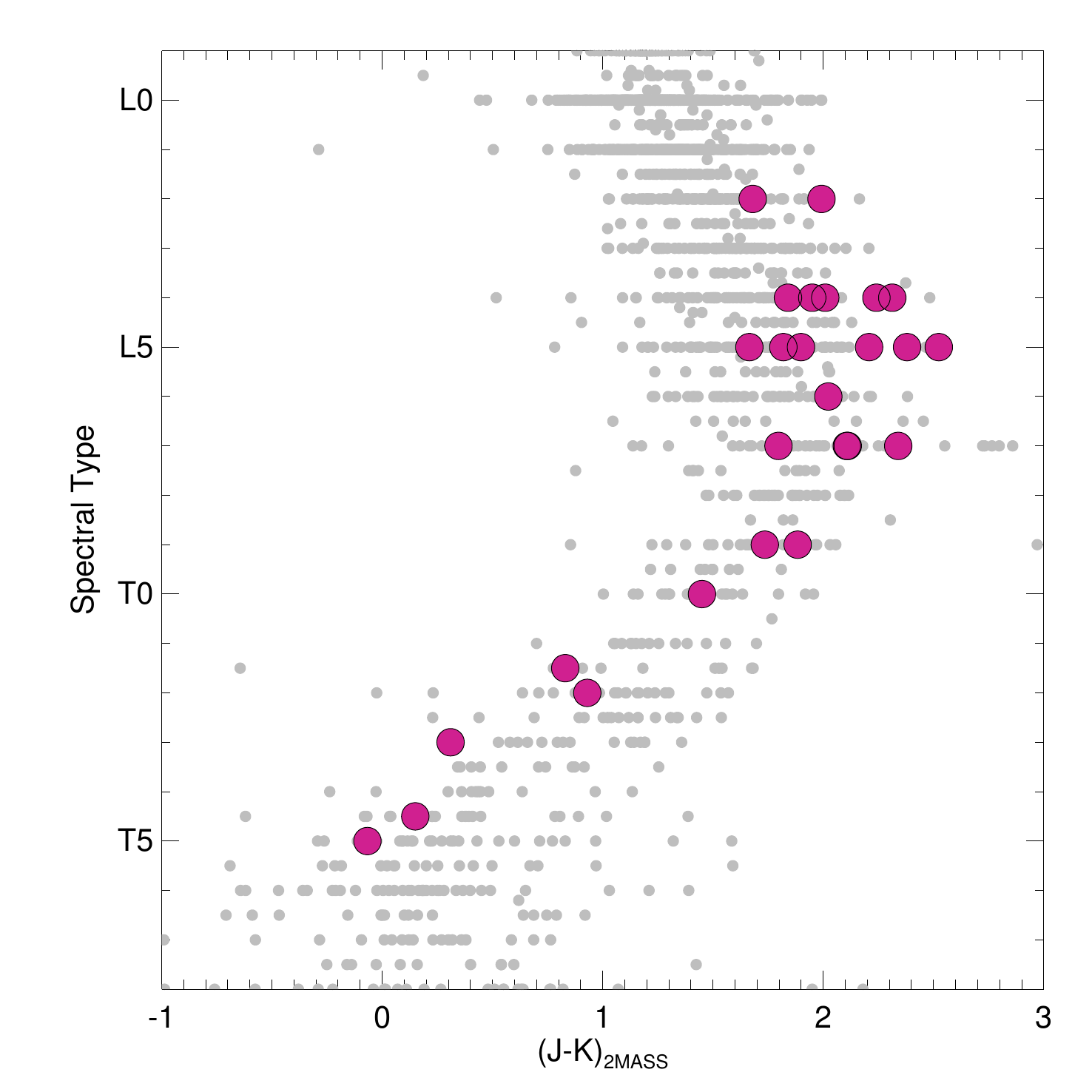}
   \caption{Spectral type--color diagram of our sample of young, low-gravity brown dwarfs (large pink circles) compared to the field population (small grey circles). Our Spitzer sample consists of 26 L3--T5 brown dwarfs with evidence of youth and/or low-gravity. The L dwarfs appear redder than the field L dwarf sequence, which is a signature of their low surface gravity. The field brown dwarf population is drawn from \citet{Best2020}.}
   \label{fig:spt_color_sample}
\end{figure}

\section{An Age-Calibrated Sample of Exoplanet Analogs} \label{sec:sample}
We designed our survey to test the variability properties of young, low-gravity brown dwarfs by comparing our results to the field brown dwarfs observed by \citet{Metchev2015}. The similar sample-eps-converted-to.pdf size, observing strategy and magnitudes of our sample and the \citet{Metchev2015} sample allow us to robustly compare variability properties between them. We present our sample, spectral types and magnitudes in Table \ref{tab:sample} and show our sample in a color-magnitude diagram in Figure \ref{fig:spt_color_sample}. Our sample spans spectral types L2 -- T5, and the L dwarfs in the sample appear redder than the field L dwarf sequence, often a signature of their low gravity.

\subsection{Three new brown dwarfs with evidence of youth from the BASS Ultracool Survey} 
\label{sec:newbds}

We present the discovery of three brown dwarfs which display evidence of youth: {\obj*{2m0349}, \obj*{2m0718} and \obj*{2m0951}.} All three objects were discovered as candidate young objects through the BASS-Ultracool survey (Gagn\'e, J. et al., in preparation) based on their kinematics, which are presented in Table \ref{tab:kinematics} and discussed in detail in Section \ref{sec:youth+var}. Here we present their near-IR  FIRE/Prism spectra and investigate their spectral signatures of youth.

\begin{figure*}[tb]
   \centering
   \includegraphics[scale=0.35]{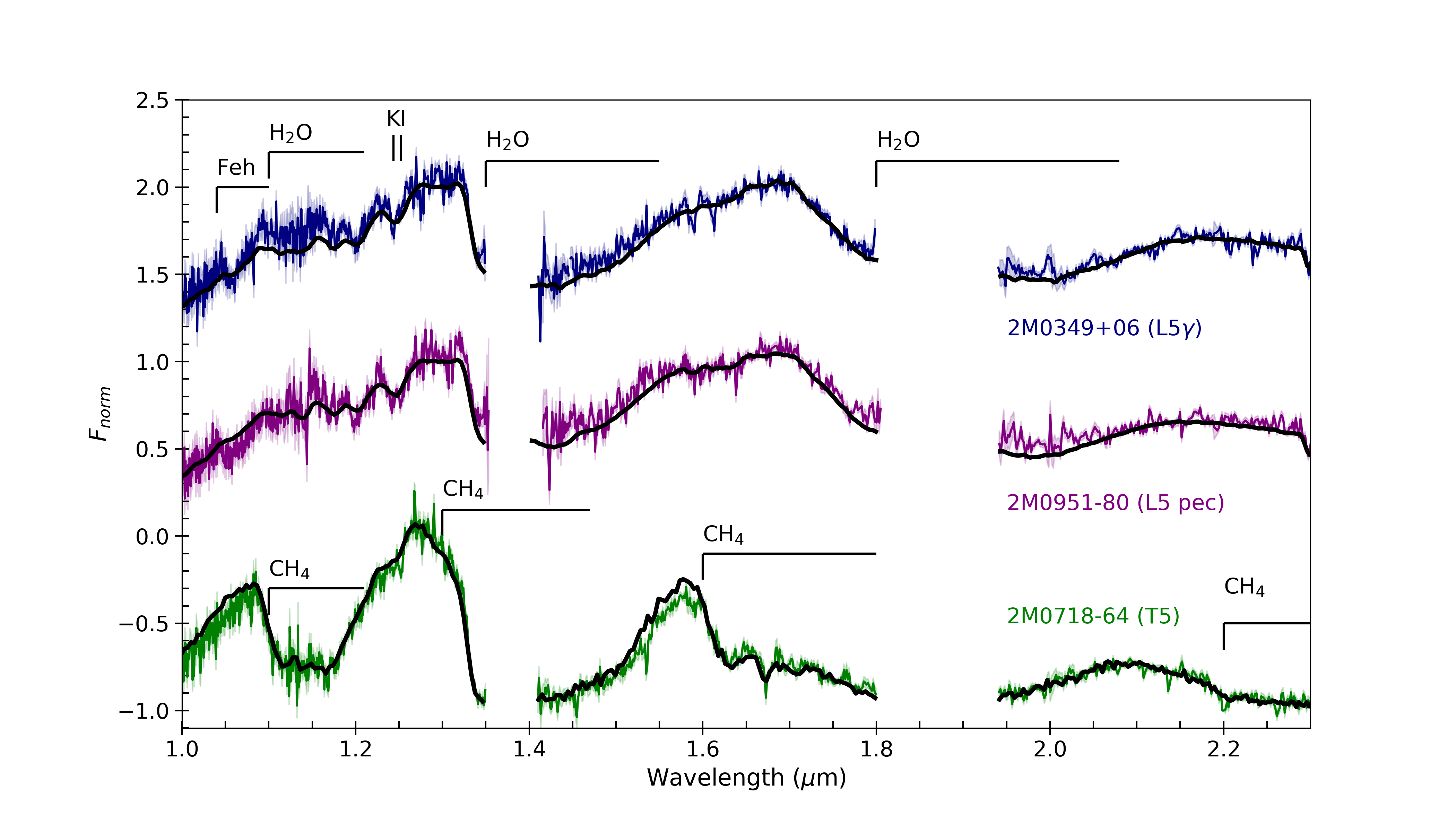}
   \caption{FIRE/Prism spectra of three new young brown dwarfs presented in this paper --  \obj{2m0349} (L5$\gamma$), \obj{2m0718} (T5) and \obj{2m0951} (L5 (pec)). {Spectral templates for an L5$\gamma$, and L5 and a T5 are overplotted in black.} The spectral resolution ranges from 500 at $J$ to 300 at $K$. These objects were discovered and classified as young as part of the BASS-Ultracool survey (Gagn\'e, J. et al. in preparation). }
   \label{fig:prism_spectra}
\end{figure*}

\subsubsection{FIRE/Prism Spectra: Observations and Analysis}

We used the FIRE \citep{Simcoe2013} spectrograph on the 6.5 m Baade Magellan telescope to obtain near-infrared spectra of the three targets to investigate their spectral properties. Observations were taken on 2018 Feb 01, 2015 Nov 24 and 2016 Jan 22 for \obj{2m0349}, \obj{2m0718} and \obj{2m0951} respectively using prism longslit mode, using ABBA nods in each case. Quartz lamps, ThAr lamps and telluric standards were taken either before our after each target. Airmass ranged from 1.2-1.6 for the three observing sequences. All FIRE/Prism data were reduced and combined using a modified version of the FireHose package as described in \citet{Gagne2015a}. 
We show the $R\sim400$ FIRE/prism spectra of these three targets in Figure \ref{fig:prism_spectra}.

\subsubsection{FIRE/Prism Spectra: Signatures of Youth}

We  perform a visual spectral classification on each of our new discoveries and an index-based classification on \obj{2m0349} and \obj{2m0951} to assess the evidence of youth and low-gravity {by comparing our L-type spectra to spectral standard templates reported by \citet{Cruz2018} and T-type spectrum to known T-type spectral standards.}

Based on a visual classification of our FIRE/Prism spectrum of \obj{2m0349},  we find that it is best matched by an L5$\gamma$ spectrum.  Using the \citet{Allers2013} indices, we find that the spectrum is classified as an L4 {\sc{vl-g}}. Thus, the spectrum provides strong evidence of youth. 

We assign \obj{2m0951} a visual classification of L5 (pec). The spectrum is not well matched by the spectral type templates, particularly in the $H$-band, where there is a `bump' at $\sim1.6~\mu$m that is not shown in the L5 templates. The \citet{Allers2013} index-based classification assigns an L1.6 {\sc{int-g}} spectral type. The disparity between the two spectral types suggests that the peculiarity of the spectrum prohibits a robust spectral classification. If we force the \citet{Allers2013} to use an L5 spectral type the gravity classification is {\sc{int-g}}.  

Visually, \obj{2m0718} is best matched by a T5 spectral type. It is very red, emitting a higher flux in $K$-band compared to the T5 template. Interestingly, the spectrum of the planetary-mass AB Doradus member SDSS J111010.01+011613.1 \citep{Gagne2015b} provides an excellent fit to our spectrum. With a spectral type of T5, we cannot perform the index-based classification of \citet{Allers2013} on this target. 



\begin{table*}
\hskip-1.cm \begin{tabular}{llllllll}
\hline \hline 
Target Name    &Spt Opt&     SpT IR & Refs & $J_{\mathrm{2MASS}}$ & $K_{\mathrm{2MASS}}$ & $(J-K)$ & [3.6] mag$^{\mathrm{a}}$             \\ [0.5ex]  \hline 
\obj*{2m0001} &\nodata &L4$\beta$            &1         & 15.46        & 13.62        & 1.84             & $12.613\pm0.004$       \\
\obj*{2m0030} & L7 &L4-L6$\beta$            & 2, 3   & 16.28       & 14.48       & 1.80            & $13.365\pm0.008$   \\
\obj*{2m0031} & \nodata &L9            &4          & 14.95        & 13.22       & 1.74            & $11.956\pm0.003$     \\
\obj*{2m0153} & L2 &L3$\beta$     & 3,6        & 16.41       & 14.42        & 1.99            & $13.349\pm0.005$      \\
\obj*{2m0326} & L5$\beta$ &L5$\beta/\gamma$  & 1,3        & 16.13        & 13.92       & 2.208         & $12.559\pm0.003$   \\
\obj*{2m0342} &L4$\gamma$& \nodata            & 1        & 16.854       & 14.541       & 2.313            & $13.579\pm0.006$   \\
\obj*{2m0349} & \nodata &L5$\gamma$     &   TW        & 16.742       & 15.077       & 1.665            & $14.181\pm0.007$                     \\
\obj*{2m0355} &L5$\gamma$ &L3-L6$\gamma$    & 3, 5          & 14.05        & 11.526       & 2.524            & $10.339\pm0.002$ \\
\obj*{2m0447} & \nodata&T2            &7           & 16.48        & 15.55        & 0.93             & $11.868\pm0.002$   \\
\obj*{2m0459} & \nodata&L7            &8      & 17.429       & 15.318       & 2.111            & $13.924\pm0.006$   \\
\obj*{2m0506} & \nodata&T4.5          &7           & 15.75        & 15.6         & 0.15             & $14.251\pm0.008$    \\
\obj*{2m0642} & \nodata&L9            &7         & 16.164       & 14.28        & 1.884            & $12.888\pm0.013$ \\
\obj*{2m0718} & \nodata&T5            & TW          & 16.62       & 16.69       & -0.07           & $15.440\pm0.017$ \\
\obj*{2m0809} & \nodata&L6 pec        &1       & 16.44        & 14.42       & 2.02            & $13.015\pm0.006$ \\
\obj*{2m0951} & \nodata&L5 pec     &   TW        & 17.098       & 15.28       & 1.82            & $14.554\pm0.008$ \\
\obj*{2m1551} &L4$\gamma$&$>$L5$\gamma$   &3       & 16.32       & 14.31        & 2.009            & $13.243\pm0.004$ \\
\obj*{2m1647} & L7&L9 pec(red)        &9   & 16.59        & 14.48       & 2.107                    & $13.180\pm0.005$  \\
\obj*{2m1741} &\nodata &L6-L8$\gamma$   &3, 10       & 15.78        & 13.44        & 2.34             & $11.936\pm0.003$ \\
\obj*{2m2002} &L5$\beta$ &L5-7$\gamma$          &3     & 15.32       & 13.42       & 1.90            & $12.161\pm0.002$ \\
\obj*{2m2117} &\nodata &T0            &7          & 15.6         & 14.15        & 1.45             & $13.014\pm0.004$ \\
\obj*{2m2154} &L4$\beta$ &L5$\gamma$            &1,3        & 16.44        & 14.12       & 2.24            & $12.979 \pm0.004$ \\
\obj*{2m2206+33}&\nodata & T1.5          &7          & 16.58        & 15.75        & 0.83             & $14.630\pm0.011$  \\
\obj*{2m2206-42}&L4$\gamma$ &L4$\gamma$   &3   & 15.56        & 13.61        & 1.95             & $12.539\pm0.003$  \\
\obj*{2m2216} & \nodata & T3            &7          & 16.59        & 16.28        & 0.31             & $14.182\pm0.007$   \\
\obj*{2m2322} &L2$\gamma$  & L3$\gamma$   &1,5          & 15.54        & 13.86        & 1.68             & $12.893\pm0.003$    \\
\obj*{2m2343} &\nodata & L3-L6$\gamma$ & 1        & 16.57        & 14.19        & 2.38             & $12.835\pm0.004$       \\  [0.5ex]  \hline              
\end{tabular}
\caption{The survey consists of 26 brown dwarfs with spectral and/or kinematic evidence of youth. }
\tablecomments{{a: 3.6~$\mu$m magnitudes were calculated  using a 2-pixel aperture and applying aperture corrections as detailed in the Spitzer calibration documentation. The quoted magnitude is the mean magnitude, and the error is the standard deviation of the corrected light curve, and thus includes both noise and astrophysical variability.}}
\label{tab:sample}
\tablerefs{(1) \citet{Gagne2015c}, (2) \citet{Kirkpatrick2000}, (3) \citet{Faherty2016},
(4) \citet{Best2013}, (5) \citet{Cruz2009}, (6) \citet{Reid2008}, (7) \citet{Best2015}, (8) \citet{Schneider2017},
(9) \citet{Kirkpatrick2011}, (10) \citet{Schneider2014}, (TW) This work. }
\end{table*}

\subsection{YMG Membership Probabilities of the Sample}
\begin{figure}[tb]
   \centering
   \includegraphics[scale=0.5]{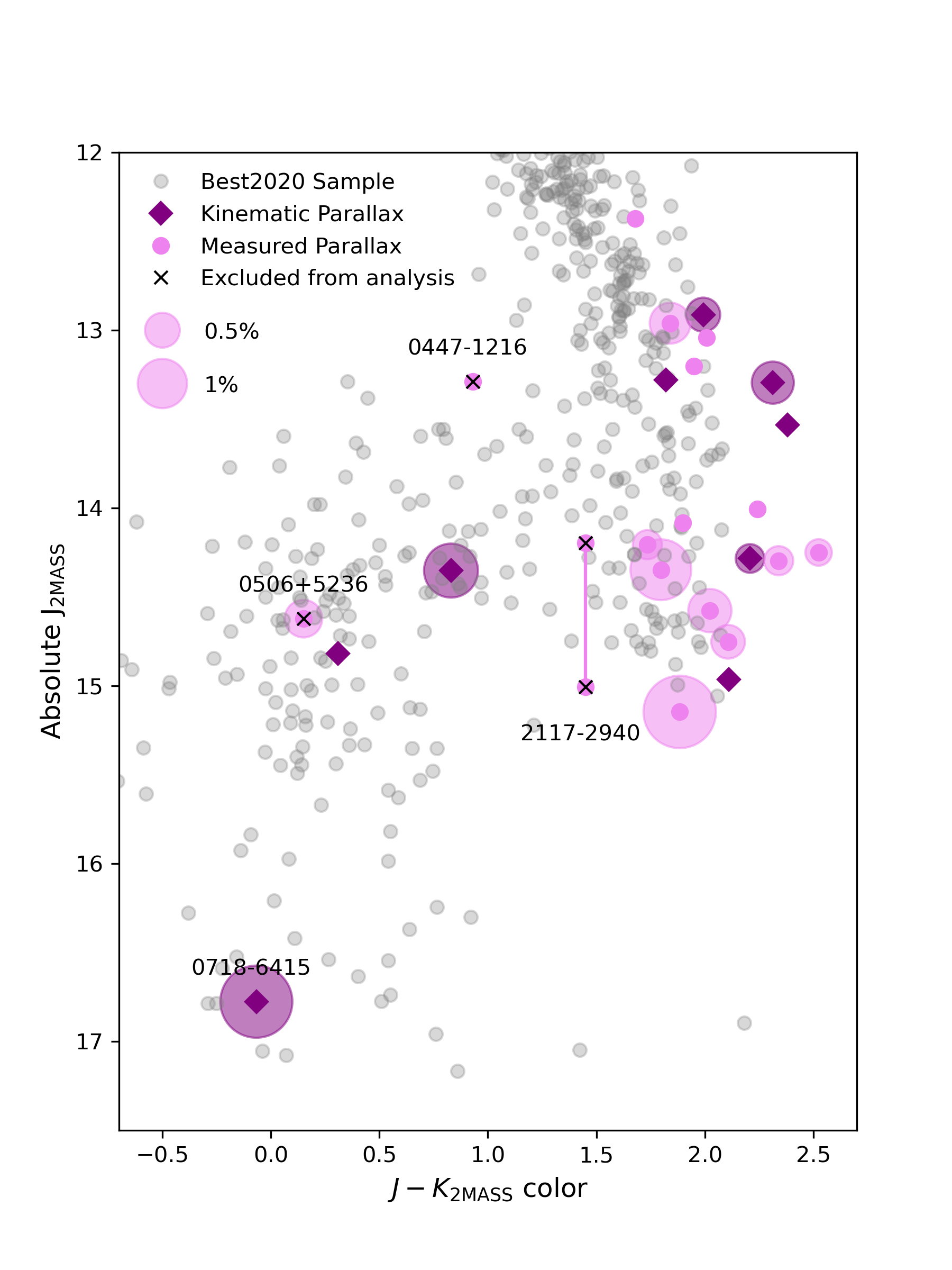}
   \caption{Spitzer variability sample shown in a color-magnitude diagram for objects with measured parallaxes (pink circles) and kinematic parallaxes (purple diamonds), compared to the larger population of brown dwarfs. {For sources identified as significantly variable in Section \ref{sec:variabilityanalysis}, their amplitudes are proportional to the partially transparent symbol size. $0.5\%$ and $1\%$ variability amplitudes are shown in the top left for reference. } We exclude three sources from our variability analysis which are denoted by black crosses. \obj{2m0506} and \obj{2m0447} have kinematics that are consistent with field membership and \obj{2m2117} has highly discrepant parallax measurements. Most of the sample falls in their expected positions relative to the field brown dwarf sample. \obj{2m0447} is over-luminous compared to other field brown dwarfs, possibly due to unresolved binarity.  The T5 dwarf \obj{2m0718} is the faintest and coldest object in the sample. }
   \label{fig:CMD_sample}
\end{figure}

For each target in our sample we assess the likelihood of kinematic membership in nearby young moving groups. Proper motions, radial velocities and parallaxes used in our kinematic analysis are shown in Table \ref{tab:kinematics}. We use these as inputs for the BANYAN~$\Sigma$ tool \citep{Gagne2018b}, which assesses the probability that a source is a member of a young moving group based on its observed kinematics. We additionally use the LACEwING tool \citep{Riedel2017} to check for consistency, but for clarity we only present the BANYAN $\Sigma$ results. We consider the following moving groups: 
TW Hydra \citep[TWA, $10\pm3~$Myr;][]{Bell2015}, 
$\beta$~Pictoris \citep[$\beta$ Pic, $22\pm6~$Myr;][]{Shkolnik2017}, 
Columba \citep[COL, $42^{+6}_{-4}~$Myr;][]{Bell2015}, 
Tucana-Horologium \citep[THA, $45\pm4~$Myr;][]{Bell2015}, 
Carina \citep[CAR, $45\pm4~$Myr;][]{Bell2015}, 
Argus \citep[ARG, 30--50~Myr;][]{Torres2008}, 
AB Doradus \citep[ABDMG, 110--150~Myr;][]{Luhman2007, Barenfeld2013}
and Carina-Near \citep[CARN, $200\pm50~$Myr;][]{Zuckerman2006}. Field objects are estimated to have ages $>1$~Gyr.
We show the kinematic information and results from  membership tools in Table \ref{tab:kinematics}. We use the BANYAN~$\Sigma$ results to make a final assessment on each target using the following categories:
\begin{itemize}
    \item Bona fide member ({BF}): an object with full kinematic information (parallax, proper motion, radial velocity), membership probability $>90\%$ and spectroscopic signs of youth.
    \item High-likelihood member ({HLM}): An object with membership probability $>90\%$ but does not have full kinematic information.
    \item Candidate member ({CAN}): An object that shows greater probability of moving group membership than being a field object, with missing kinematic information.
    \item Field object ({FLD}): An object that is kinematically eliminated from falling into a known nearby moving group based on current kinematic information, or if the spectrum looks unambiguously field age.
    \item Young field object ({YNG-FLD}): An object that is kinematically eliminated from falling into a known nearby moving group based on current kinematic information but other observational features point to its youth.
    \item Ambiguous object ({AM}): An object that requires updated astrometric precision because it could either belong to more than one group or it cannot be differentiated from the field.
\end{itemize}
As shown in Table \ref{tab:kinematics}, our sample is composed of three bona fide members (\obj{2m0355}, \obj{2m1741} and \obj{2m2322}), seven high-likelihood members, eight candidate members, two field objects, four young field objects and two ambiguous members. We discuss the kinematic and spectral indicators of youth for each object in detail in Section \ref{sec:youth+var}. In Figure \ref{fig:CMD_sample} we plot the full sample on a color-magnitude diagram.

\input{kinematics_v2.tex}

\section{Spitzer Observations and Data Reduction} \label{sec:observations}
We used the Infrared Array Camera \citep[IRAC;][]{Fazio2004} to observe our targets
in the Channel 1 ($3.6~\mu$m) band as part of the Cycle 14 Program: ``The Young and the Restless: Revealing the Turbulent, Cloudy Nature of Young Brown Dwarfs and Exoplanets'' (PID: 14128, PI: J Faherty). 
The observations were designed following the recommendations for obtaining high precision photometry from the Spitzer Science Center, and following the many examples of brown dwarf variability monitoring in the literature \citep{Metchev2015, Apai2017, Biller2018, Vos2018, Vos2020, Allers2020, Zhou2020}. Science images were obtained in staring mode, using 12 s exposures for a total monitoring duration of 20 hr.  Science observations were preceded by a 30-minute dithered sequence to remove the initial slew settling that occurs when acquiring a new target, and followed by a 10-minute dithered sequence to capture the most accurate dark image for each observation. 

We follow the same data reduction outlined in \citet{Vos2020}; we measured photometry from the Basic Calibrated Data (BCD) images produced by the Spitzer Science Center using pipeline version S19.2. The {\sc{box\_centroider.pro}} routine was used to find the centroids of the target and reference stars of similar brightness in the field of view. Aperture photometry was performed on the target and reference stars using apertures with radii of $2.0-5.0$ pixels, in steps of $0.25$. We choose the final aperture size that produces the lowest rms target light curve. Outliers were identified and rejected from the raw light curves using a $6\sigma$ clip.

\begin{figure}[tb]
   \centering
   \includegraphics[scale=0.47]{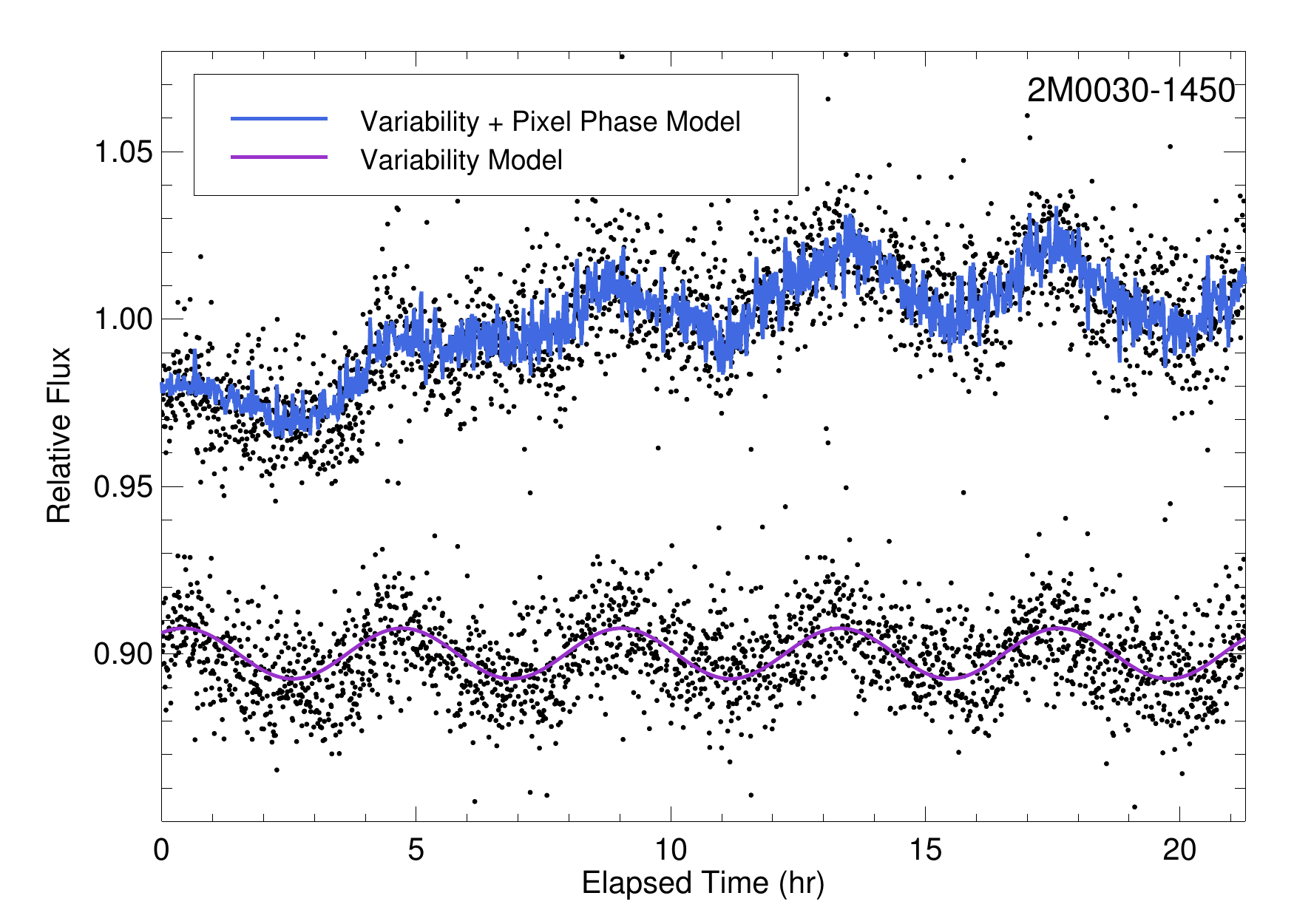}
   \caption{Raw light curve (top) and corrected light curve for \obj{2m0030}. The combined astrophysical variability and pixel phase model is shown in blue and the astrophysical variability model is shown in purple. }
   \label{fig:correction_process}
\end{figure}

\textit{Spitzer}/IRAC photometry is known to exhibit a systematic effect due to intra-pixel sensitivity variations, also known as the pixel phase effect. 
Intra-pixel sensitivity variations combined with telescope pointing fluctuations result in raw photometry that is highly correlated with the $x$ and $y$ sub-pixel coordinates. 
We model the pixel phase effect using a cubic function of the $x$ and $y$ coordinates \citep{Knutson2008, Heinze2013, Metchev2015, Vos2020}:
\begin{equation}\label{eq:pixelphase}
\begin{split}
    f(x,y) = P_0 + P_1x + P_2y + P_3 xy + P_4 x^2 + P_5 y^2 \\+ P_6x^3 + P_7 y^3 + P_8x^2y + P_9 xy^2
\end{split}
\end{equation}
where $f(x,y)$ represents the measured flux, $P_i$ are the fitted coefficients, and $x$ and $y$ are the sub-pixel coordinates. 
{We correct the light curves of the target and reference stars using Equation \ref{eq:pixelphase}, and find that the fitted coefficients $P_i$ are similar for the target and reference stars.} We find that the correction significantly decreases the correlation between the flux and pixel position for each observation. {In Figure \ref{fig:correction_process} we show an example target light curve before and after correction using the cubic function listed in Equation \ref{eq:pixelphase}.} This is the final step in producing corrected light curves.

\subsection{Determining variability parameters using MCMC}\label{sec:MCMC}
For objects showing variability {(the process for identifying variability is presented in Section \ref{sec:variabilityanalysis})}, we use the Markov Chain Monte Carlo (MCMC) {\texttt{emcee}} package \citep{fm2013} to constrain the variability parameters such as rotation period and variability amplitude.  For the MCMC analysis we use 1000 walkers with 8000 steps, and discard the initial 1000 steps as the burn-in sample. We check for convergence by visually inspecting the resulting chains for each parameter to ensure that they are consistent with Gaussian noise. Additionally, we check that there is no significant difference between the parameter constraints obtained from the first and second halves of each chain. Based on these two checks we find that the  model converges well for each variable light curve. We use a sinusoidal model for all of our targets, except for \obj{2m0030} and \obj{2m0642}, whose light curves favor a two-term Fourier function.
We show our measured variability parameters along with their 1$\sigma$ uncertainties in Table \ref{tab:varparameters}. Best-fit variability models are overplotted in pink in Figure \ref{fig:var_lcs}. 

We plot all of our variability detections on a color - magnitude diagram in Figure \ref{fig:CMD_sample}, where the symbol area is proportional to the amplitude of variability. This figure does not show any obvious trends between spectral type and amplitudes and we observe a range of amplitudes throughout  the L--T spectral sequence. We present a more in-depth discussion of variability amplitudes in Section \ref{sec:variability_properties}. 


\section{Variability Analysis}\label{sec:variabilityanalysis}
Following the approach outlined in \citet{Vos2020}, we use two independent methods to identify variable signals in each light curve - a periodogram significance test and the Bayesian Information Criterion (BIC). We briefly outline both methods below.

\begin{figure}[tb]
   \centering
   \includegraphics[scale=0.7]{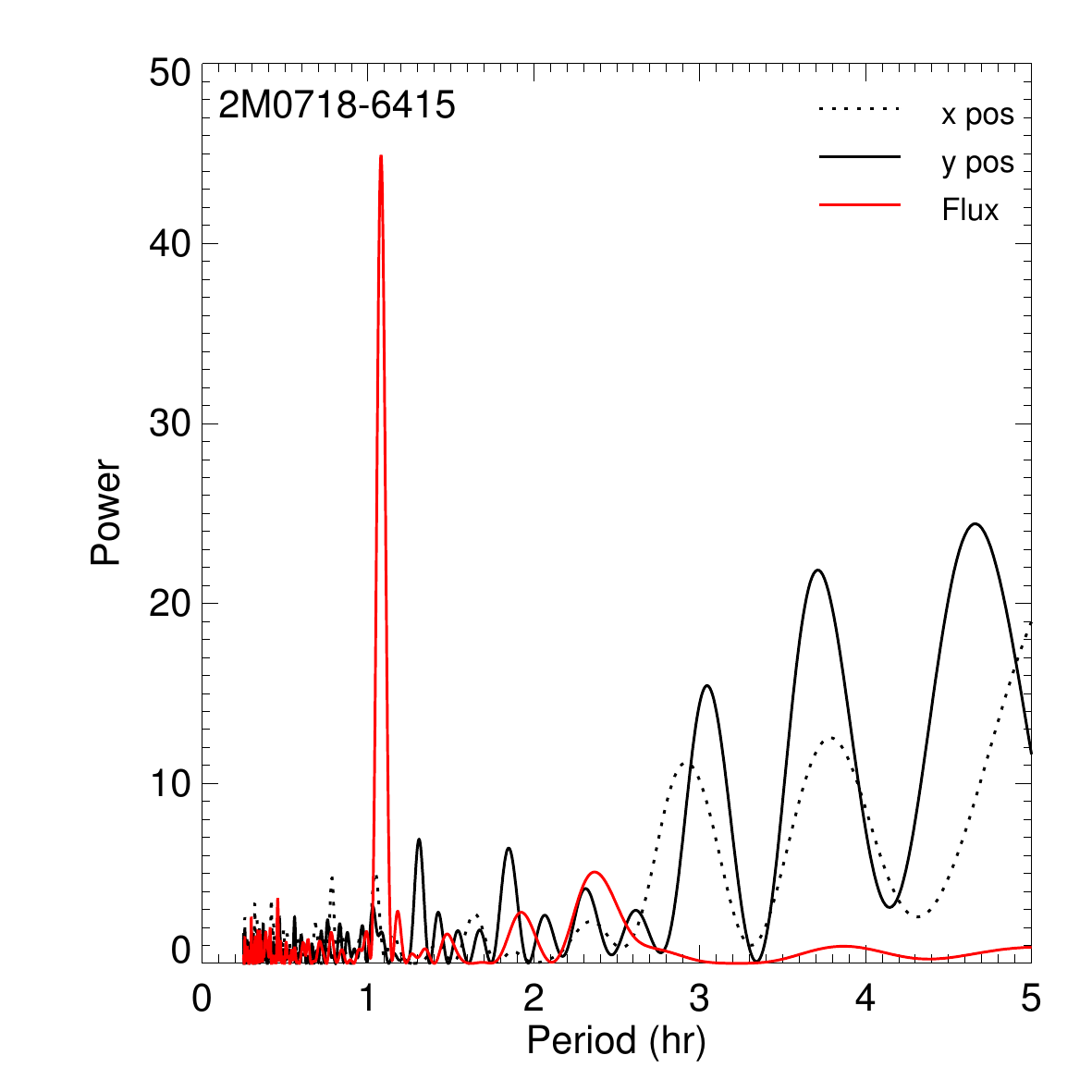}
   \caption{Periodogram of the light curve (red) as well as $x$ and $y$ pixel positions. For variable targets, we confirm that the periodicity is not correlated with the periodicity of the $x$ and $y$ pixel coordinates. }
   \label{fig:0718_periodoram}
\end{figure}

\subsection{Variability Detection with Periodogram Analysis}
We calculate the Lomb-Scargle periodograms of our target and reference star light curves \citep{Scargle1982} to assess the significance of their brightness fluctuations. For each observation we calculate the $95\%$ and $99\%$ significance thresholds by simulating non-variable light curves from our observed reference stars. 
The simulated, non-variable light curves are created by by randomly rearranging the indices of the reference star light curves, which produces simulated light curves with Gaussian noise similar to our observed light curves. We consider a target to be variable only if the periodogram peak falls above the $99\%$ significance threshold. {For variable targets, we also perform a visual check to ensure that the periodicity of the object is not correlated with the periodicity of the $x$ and $y$ pixel coordinates. In Figure \ref{fig:0718_periodoram}, we this comparison for the object \obj{2m0718}}.

{\subsection{Variability Detection Using the Bayesian Information Criterion} \label{sec:BIC}}

{We additionally use the BIC to identify significant variability in each observation \citep{Schwarz1978}.  The BIC is defined as}
\begin{equation}
    \mathrm{BIC}=-2~\mathrm{ln}~\mathcal{L}_\mathrm{max} + k ~\mathrm{ln}~ N
\end{equation}
{where $\mathcal{L}_\mathrm{max}$ is the maximum likelihood achievable by the model, $k$ is the number of parameters in the model and $N$ is the number of datapoints used in the fit \citep{Schwarz1978}. }

For each observation, we calculate $\Delta\mathrm{BIC} = \mathrm{BIC}_{flat} - \mathrm{BIC}_{\mathrm{sin}}$ to assess whether the variable sinusoidal or non-variable flat model is favored by the data. The BIC penalizes the sinusoidal model for having additional parameters compared with the flat model. $\Delta\mathrm{BIC}>0$ indicates that the sinusoidal model is favored and $\Delta\mathrm{BIC}<0$  indicates that the non-variable, flat model is favored. A $|\Delta\mathrm{BIC}|$ value between 0 and 6 indicates that one model is positively favored over the other, a value between 6 and 10 indicates that one model is strongly favored over the other and values above 10 indicate that one model is very strongly favored over the other \citep{Schwarz1978}. We conservatively consider light curves with $\Delta\mathrm{BIC}>10$ to be a significant detection of variability for this survey.

We additionally use the BIC to determine whether any of our light curves favor a more complex variability model. For each light curve we use the BIC to test both a sinusoidal model and a two-term Fourier model. Such models have been used to describe more complex brown dwarf light curves in the literature \citep{Heinze2013,Yang2016,Vos2018}, and may be driven by atmospheres with multiple atmospheric features and/or evolving cloud structures. We find that a two-term Fourier series is favored for two of our variable light curves - the L7 dwarf \obj{2m0030} and the L9 dwarf \obj{2m0642}.

\subsection{Comparison of $\Delta$BIC and Periodogram Methods}\label{sec:BICvsP}

\begin{figure*}[tb]
   \centering
   \includegraphics[scale=0.7]{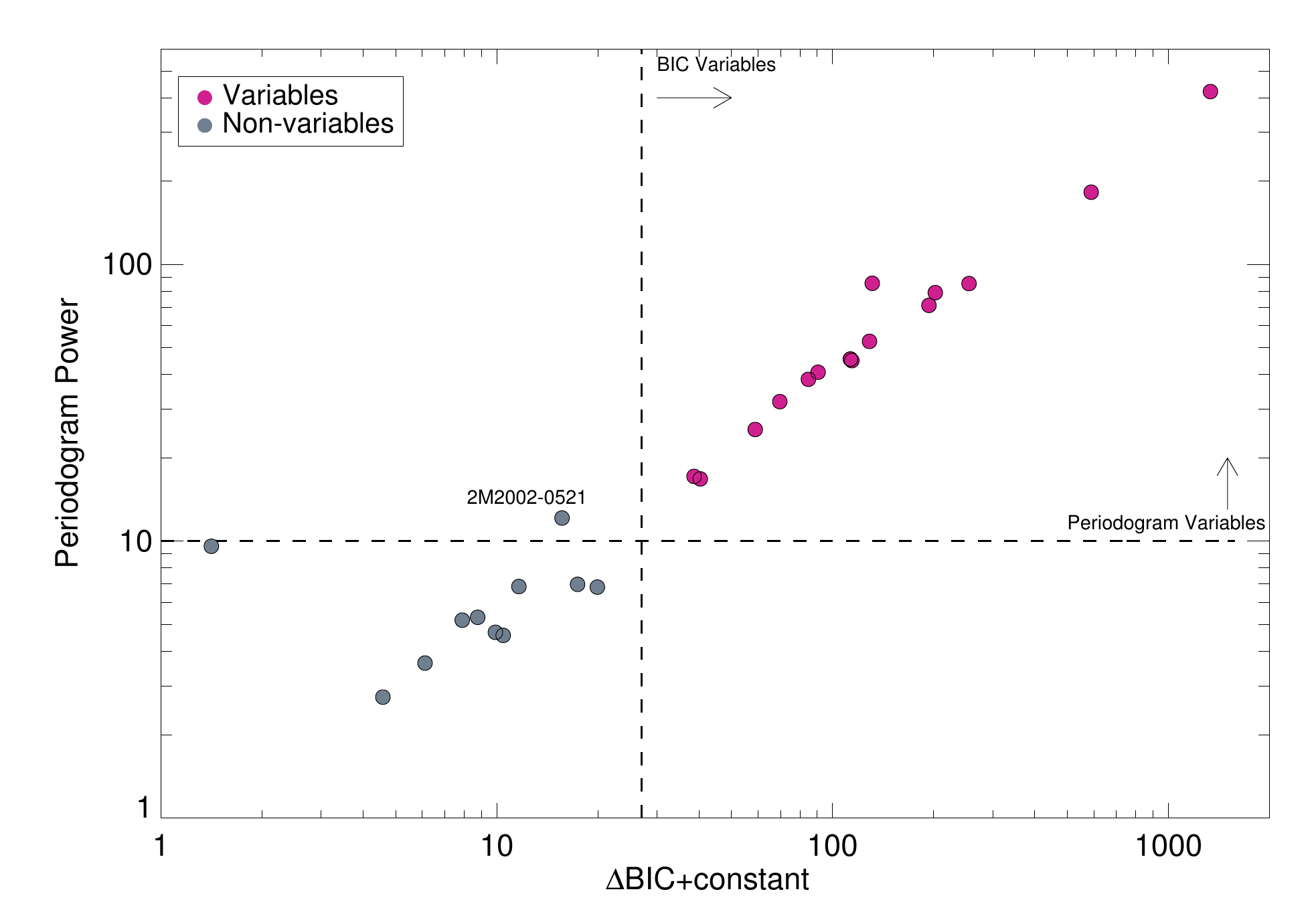}
   \caption{Periodogram power plotted against $\Delta$BIC for all of our survey targets.  Variable objects are shown in pink. Non-variable objects are shown in grey. A constant has been added to the $\Delta$BIC values. The dashed lines show the significance thresholds for each test. Objects placed in the upper left of the figure are considered variable by both methods, while objects in the lower left are classified as non-variable using both methods. We find that both methods generally identify the same variables, except for the object \obj{2m2002}, which is significantly variable according the the periodogram power method but not the $\Delta$ BIC test. }
   \label{fig:BICvsP}
\end{figure*}

\begin{figure*}[tb]
   \centering
   \includegraphics[scale=0.7]{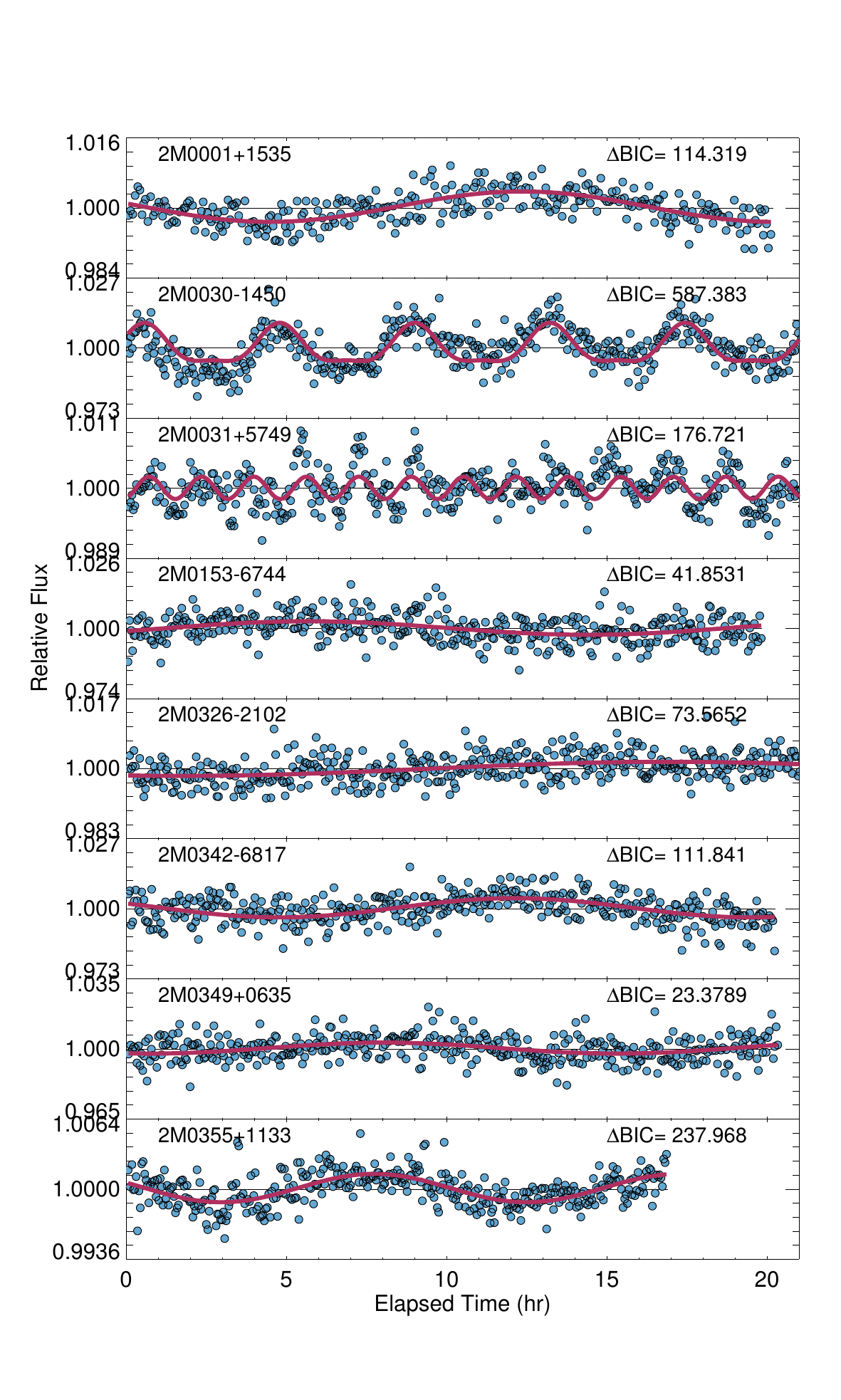}
    \includegraphics[scale=0.7]{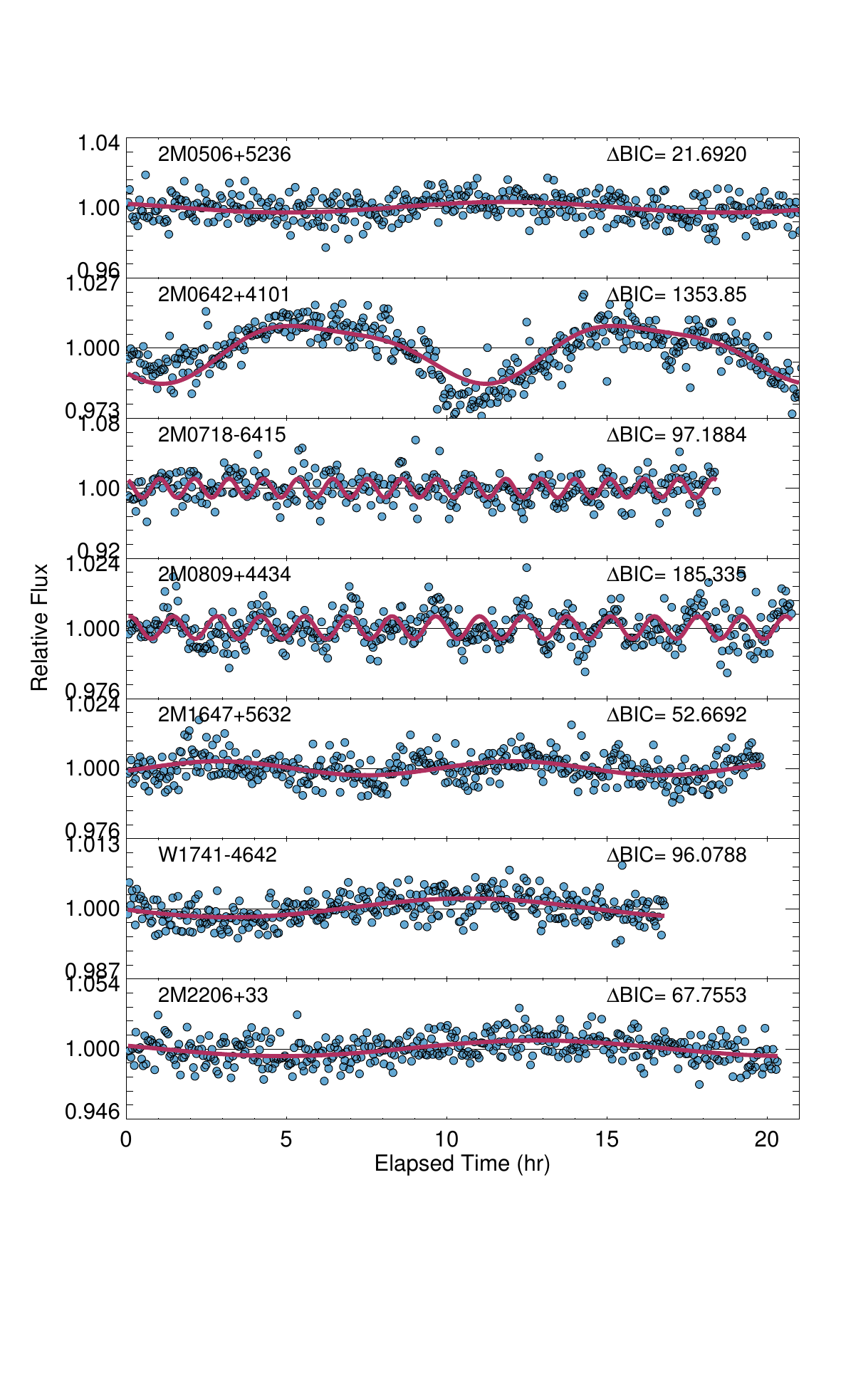}\\
   \caption{Positive variability detections from our survey. The blue points show 2.5 min cadence. $\Delta$BIC values show that a variable model is favored in each case. A sinusoidal model is preferred for the majority of our variables, and a two-term Fourier series is favored for \obj{2m0030} and \obj{2m0642}. }
   \label{fig:var_lcs}
\end{figure*}

\begin{figure*}[tb]
   \centering
   \includegraphics[scale=0.7]{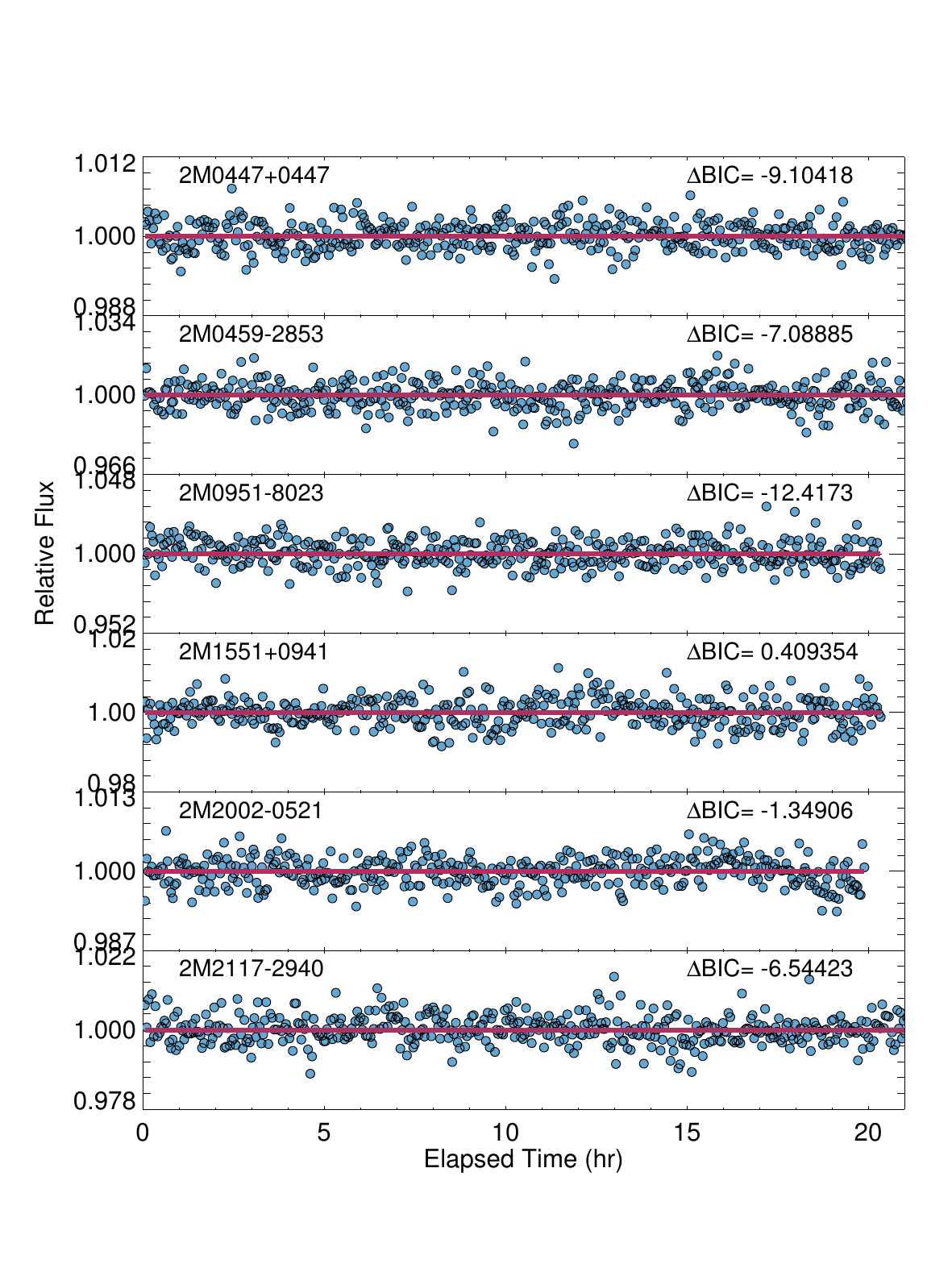}
    \includegraphics[scale=0.7]{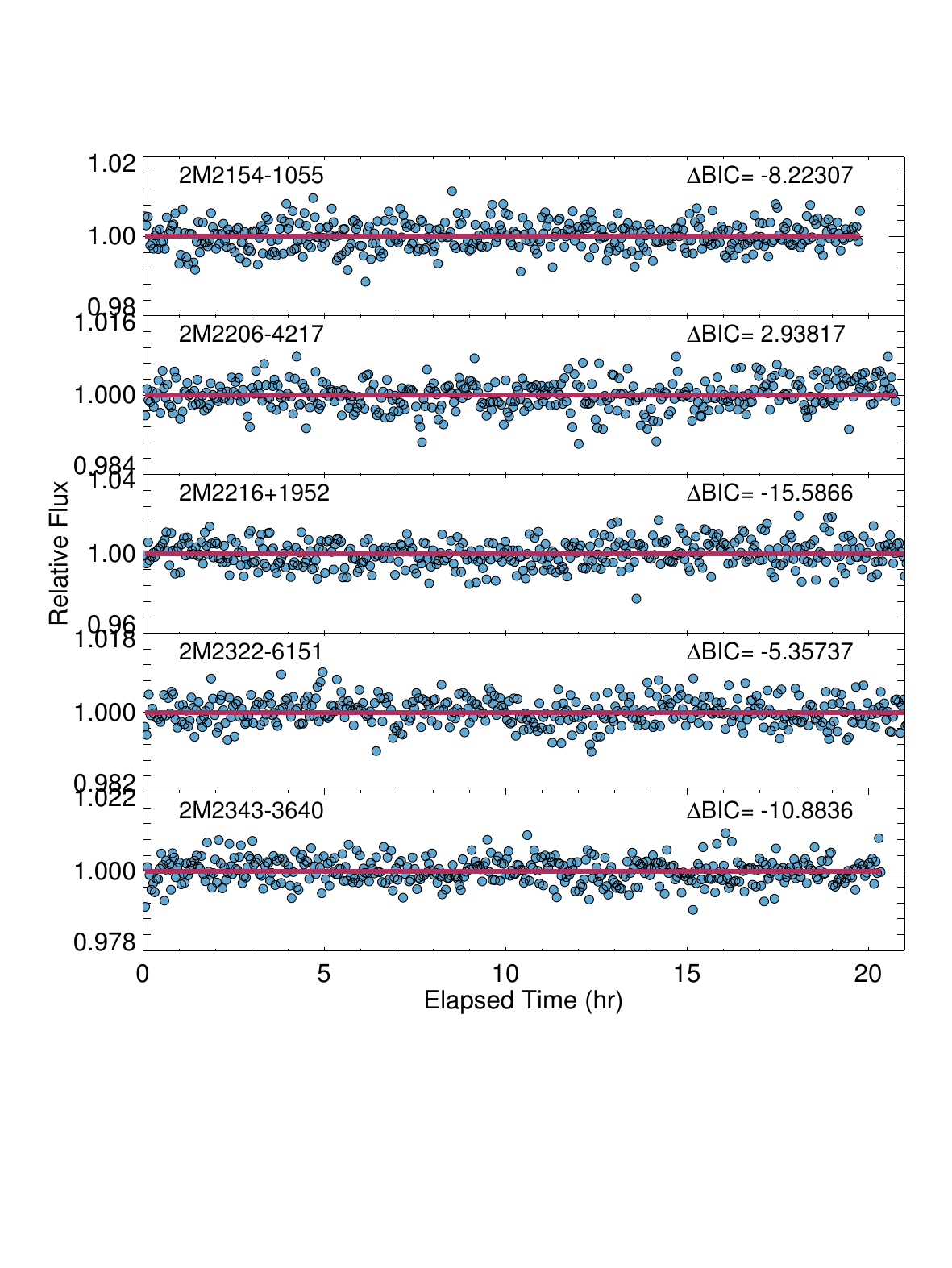}\\
   \caption{Non-detections of variability from our survey. The blue points show the light curve binned to 2.5 min cadence. }
   \label{fig:nonvar_lcs}
\end{figure*}

In Figure \ref{fig:BICvsP} we compare the results of both variability identification methods. Objects placed in the upper right of the figure are considered variable by both methods, while objects in the lower left are classified as non-variable using both methods. We find that in general both methods agree on which light curves show significant variability as well as the relative significance of their variability detections. One object, \obj{2m2002}, is found to be variable using the periodogram method but not the $\Delta$BIC test. We conservatively count it as non-variable for the remainder of the paper. We show our positive variability detections in Figure \ref{fig:var_lcs}, and non-detections in Figure \ref{fig:nonvar_lcs}. In total, we detect variability in 15 objects, and do not detect variability in 11 objects.

{\subsection{Modeling Red Noise Contributions Using Gaussian Processes} \label{sec:GP}}

{While our two variability identification methods are in agreement, we must ensure that the detected variability is astrophysical in nature, and not due to correlated, or ``red'' noise. Red noise can mimic the signatures of astrophysical variability.}

{\citet{Littlefair2017} presents a framework to determine between astrophysical variability and red noise using Gaussian Processes (GPs). GPs have recently been used in the exoplanet and stellar subfields of astronomy to model both light curve systematics and rotational modulations \citep[e.g.][]{Gibson2012b,  Aigrain2016, Littlefair2017, Angus2017}. }

{For this work, we compare two GP models computed using the \texttt{celerite2}  package \citep{Foreman-Mackey2017, Foreman-Mackey2018}. The first exhibits variability driven by red noise, as modeled by the Matern-3/2 kernel: }
\begin{equation}
    k(\tau)=\sigma^2 \left( 1 + \frac{\sqrt 3\tau}{\rho} \right) exp \left( -\frac{\sqrt 3\tau}{\rho}\right)
\end{equation}
{where $\sigma$ and $\rho$ are the GP hyperparameters.} {The second model exhibits rotational variability as described by a kernel that is a mixture of two simple harmonic oscillator kernels given by:}

\begin{equation}
\begin{aligned}
    Q_1      = \frac{1}{2} + Q_0 +\delta Q  \\
    \omega_1 = \frac{4\pi Q_1}{P\sqrt{4Q_1^2-1}} \\
    S_1      = \frac{\sigma^2}{(1+f)\omega_1Q_1} 
\end{aligned}
\end{equation}
for the primary term, and 
\begin{equation}
\begin{aligned}
    Q_2      = \frac{1}{2} + Q_0  \\
    \omega_2 = \frac{8\pi Q_1}{P\sqrt{4Q_1^2-1}} \\
    S_2      = \frac{f \sigma^2}{(1+f)\omega_2Q_2} 
\end{aligned}
\end{equation}

{for the secondary term. Its hyperparameters are easier to interpret that for the previous model. $\sigma$ represents the standard deviation of the process, $P$ represents the primary period of variability, $Q_0$ represents the quality factor for the secondary oscillation, $\delta Q$ is the difference between quality factors for the first and second modes, and $f$ is the fractional amplitude of the secondary mode compared to the primary. }

{For each of our variable light curves we use \texttt{emcee} \citep{fm2013} to fit the GP model and compare models using the $\Delta BIC$ framework described previously. For each model we initialize 32 walkers and run for 5000 steps after discarding a 2000 burn-in sample. Most light curves are fit moderately well by both models, but the rotation model in particular has a lot of parameters that are not constrained by the data, particularly for the longer period rotators. We present the $\Delta BIC_{\mathrm{Mat3/2}-\mathrm{Rot}}$ for objects identified as variable in Section 
\ref{sec:BICvsP} in Table \ref{tab:varparameters}. We  find that all of the previously identified variables are preferred by the rotation model over the red noise model, apart from \obj{2m0326}, the target identified with to have a long period. The upward trend is fit by the both models equally well, but since the Matern-3/2 kernel has fewer paramters it is favored. We conclude that this target requires longer-term monitoring to identify the cause of its upward flux trend, and leave it out of the statistical analysis.}

\subsection{Simultaneous Fit for Intra-Pixel Sensitivity and Variability Parameters}
The pixel-phase effect can, in principle, be covariant with astrophysical variability \citep[e.g.][]{Heinze2013}. As described in \citet{Vos2020}, we also perform a simultaneous fit of the intra-pixel phase effect (Equation \ref{eq:pixelphase}) and astrophysical variability. We again use the {\texttt{emcee}} package \citep{fm2013} to explore the parameter space. We find that fitting these effects simultaneously identifies the same variables and non-variables, and results in variability parameters that are broadly consistent with those presented in Table \ref{tab:varparameters}. However, for the simultaneous fit we consistently find a higher correlation between the final light curve and the $x,y$ pixel positions compared to the two-step fitting method; the Kendall $\tau$ correlation coefficient for the simultaneous fit can be up to an order of magnitude higher than the two-step method, and occasionally the correlation with pixel position increases compared to the raw flux. For this reason we favor light curves corrected using the two-step method in this paper, as the corrected photometry is less likely to be affected by intra-pixel sensitivity.

\begin{table*}[]
\begin{tabular}{lcccccc}
\hline \hline 

Name                     & SpT$^{\mathrm{a}}$    & {[}3.6{]} Amp  & Period    & Periodogram  & $\Delta$BIC$_{\mathrm{flat-sin}}$$^{\mathrm{b}}$  & $\Delta BIC_{\mathrm{Mat3/2}-\mathrm{Rot}}$$^{\mathrm{c}}$ \\ 
                        &         &    (\%)         &    (hr)          & Power         &    & \\ [0.5ex]\hline 

\obj{2m0001}  & L4$\beta$     & $0.69\pm0.04$        & $15.75 \pm 0.37$                              & 85.4           & 114.3    &41.3\\
\obj{2m0030}    & L7     & $1.52\pm0.06$        & $4.22\pm0.02$                                 & 182.3           & 572.2        &2.4\\
\obj{2m0031}  & L9     & $0.35\pm0.03$        & $1.64\pm0.01$                                 & 40.8            & 73.6           &23.7\\
\obj{2m0153}  & L2     & $0.48\pm0.07$        & $17.63^{+1.13}_{-0.94}$                     & 25.3          & 41.9               &10.5\\
\obj{2m0326}  & L5$\beta$     & $0.34\pm0.04$        & $32.84^{+4.60}_{-4.62}$                    & 40.8            & 73.6       &-4.4\\
\obj{2m0342}   & L4$\gamma$     & $0.73\pm0.07$        & $14.73^{+0.51}_{-0.49}$                     & 52.7           & 111.8    &43.5\\
\obj{2m0349}  & L5$\gamma$    & $0.53\pm0.09$        & $14.62^{+1.08}_{-1.00}$                     & 16.8           & 23.4       &30.1\\
\obj{2m0355}  & L5$\gamma$   & $0.26\pm0.02$        & $9.53\pm0.19$                                 & 85.2            & 85.2     &20.7\\
\obj{2m0447}  & T2     & $<0.93$                & \nodata                                    & 5.2           & -9.1              &\nodata\\
\obj{2m0459}  & L7:     & $<0.47$                & \nodata                                    & 4.7           & -7.1             &\nodata\\
\obj{2m0506}  & T4.5   &$ 0.58\pm0.10$        & $13.77^{+0.82}_{-0.77}$                     & 17.1           & 21.7              &23.1\\
\obj{2m0642}  & L9(red)     & $2.16\pm0.16$        & $10.11\pm0.06$                                & 421.1           & 1317.8    &49.9\\
\obj{2m0718}  & T5     & $2.14\pm0.21$        & $1.080^{+0.004}_{-0.003}$                                 & 44.9           .2    &47.5\\
\obj{2m0809}  & L6p     & $0.77\pm0.06$        & $1.365\pm0.004$                               & 79.1           & 185.3          &61.2\\
\obj{2m0951} & L5(pec)     & $<0.92$                 &\nodata                                     & 2.7           & -12.4        &\nodata\\
\obj{2m1551}  & L4$\gamma$     & $<0.39$                & \nodata                                     & 7.0           & 0.4      &\nodata\\
\obj{2m1647}  & L7 & $0.47\pm0.06$       & $9.234^{+0.25}_{-0.23}$                      & 32.0           & 52.7                  &2.8\\
\obj{2m1741}  & L6-L8$\gamma$     & $0.35\pm0.03$        & $15.00^{+0.71}_{-0.57}$                      & 45.5       & 96.1      &32.9\\
\obj{2m2002}  & L5$\beta$   & $<0.28$                &  \nodata                                    & 12.1           & -1.3       &\nodata\\
\obj{2m2117}  & T0     & $<0.38$                &  \nodata                                    & 4.6           & -6.5             &\nodata\\
\obj{2m2154}  & L4$\beta$     & $<0.39$                &  \nodata                                   & 5.3           & -8.2       &\nodata\\
\obj{2m2206+33}  & T1.5   & $1.20\pm0.13$        & $15.91^{+0.62}_{-0.61}$                   & 38.4           & 67.8             &38.5\\
\obj{2m2206-42}   & L4$\beta$     & $<0.33$                  &   \nodata                                     & 6.8      & 2.9    &\nodata\\
\obj{2m2216}    & T3     & $<0.97$              &  \nodata                                     & 9.6           & -15.6           &\nodata\\
\obj{2m2322}  & L2$\gamma$   & $<0.41$                &   \nodata                                   & 6.9           & -5.4       &\nodata\\
\obj{2m2343}  & L3-L6$\gamma$  & $<0.37$                &  \nodata                                    & 3.6           & -10.9    &\nodata\\ [0.5ex]  \hline         
\end{tabular}
\caption{Variability parameters for objects in the survey.}
\label{tab:varparameters}
\tablecomments{a: Spectral types and gravity classifications are from optical data unless only infrared was available. \\{b: Positive $\Delta BIC_{\mathrm{flat-sin}}$ values correspond to light curves preferred by the variable model over the flat model. \\ c: Positive $\Delta BIC_{\mathrm{Mat3/2}-\mathrm{Rot}}$ values correspond to light curves preferred by the rotation model over the red noise model.}}
\end{table*}

\subsection{Survey Sensitivity}

{We estimate the sensitivity of each observation to simulated sinusoidal variability following the methods outlined in \citet{Vos2019,Vos2020}. 
We inject sinusoidal signals with a ranges of amplitudes and periods into light curves with Guassian distributed noise similar to that of each target. The simulated light curves have amplitudes of $0.05-0.5\%$, rotation periods of $0.5-40~$hr and random phase shifts. We then analyze the simulated light curves using the periodogram analysis described in Section \ref{sec:variabilityanalysis} and calculate the detection probability as the percentage of light curves with a given variability amplitude and period that produces a periodogram power above the significance threshold.} 
Due to the range of magnitudes in our sample, the sensitivity of each observation can change significantly. 

\begin{figure*}[tb]
   \centering
   \includegraphics[scale=0.55]{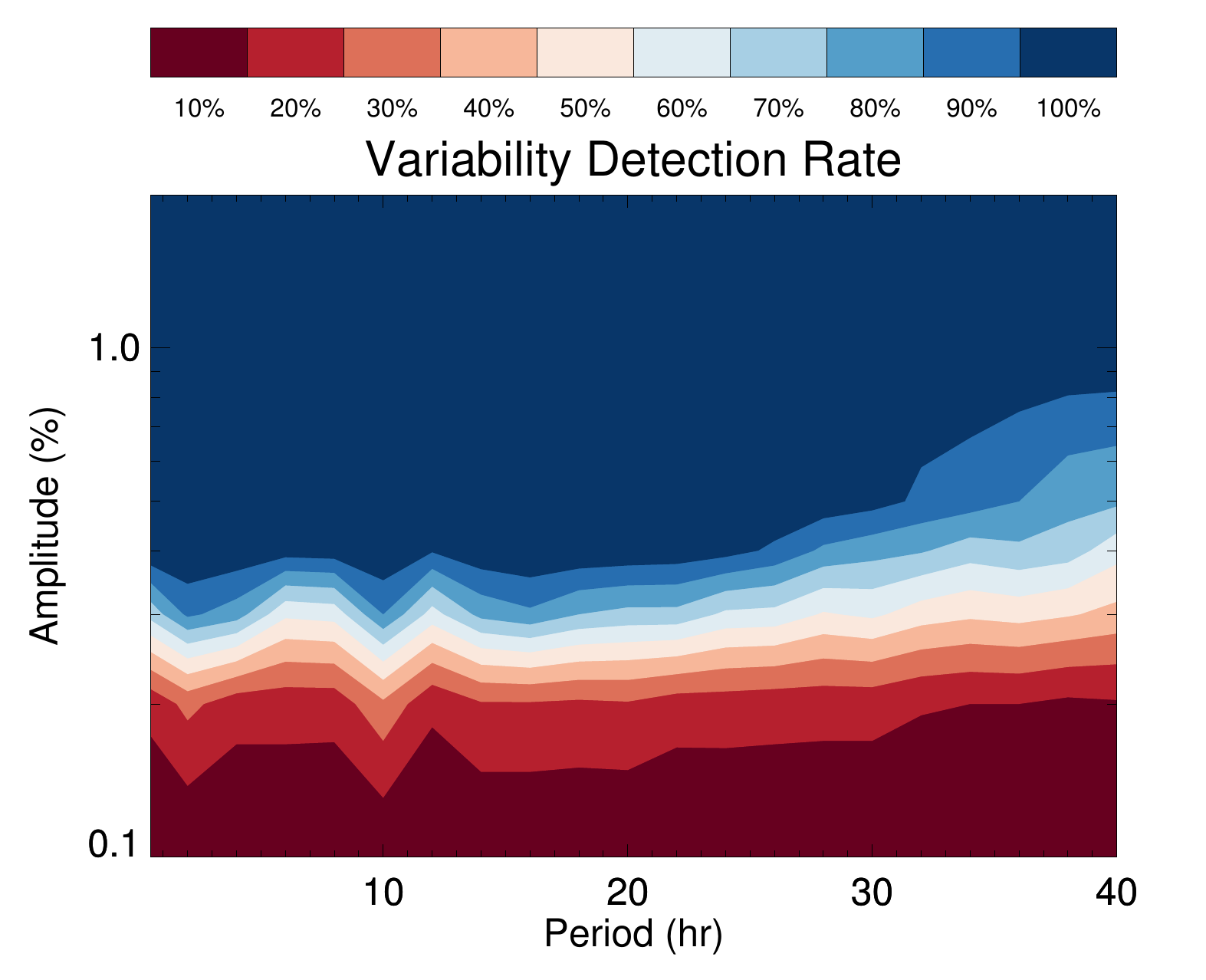}
    \includegraphics[scale=0.55]{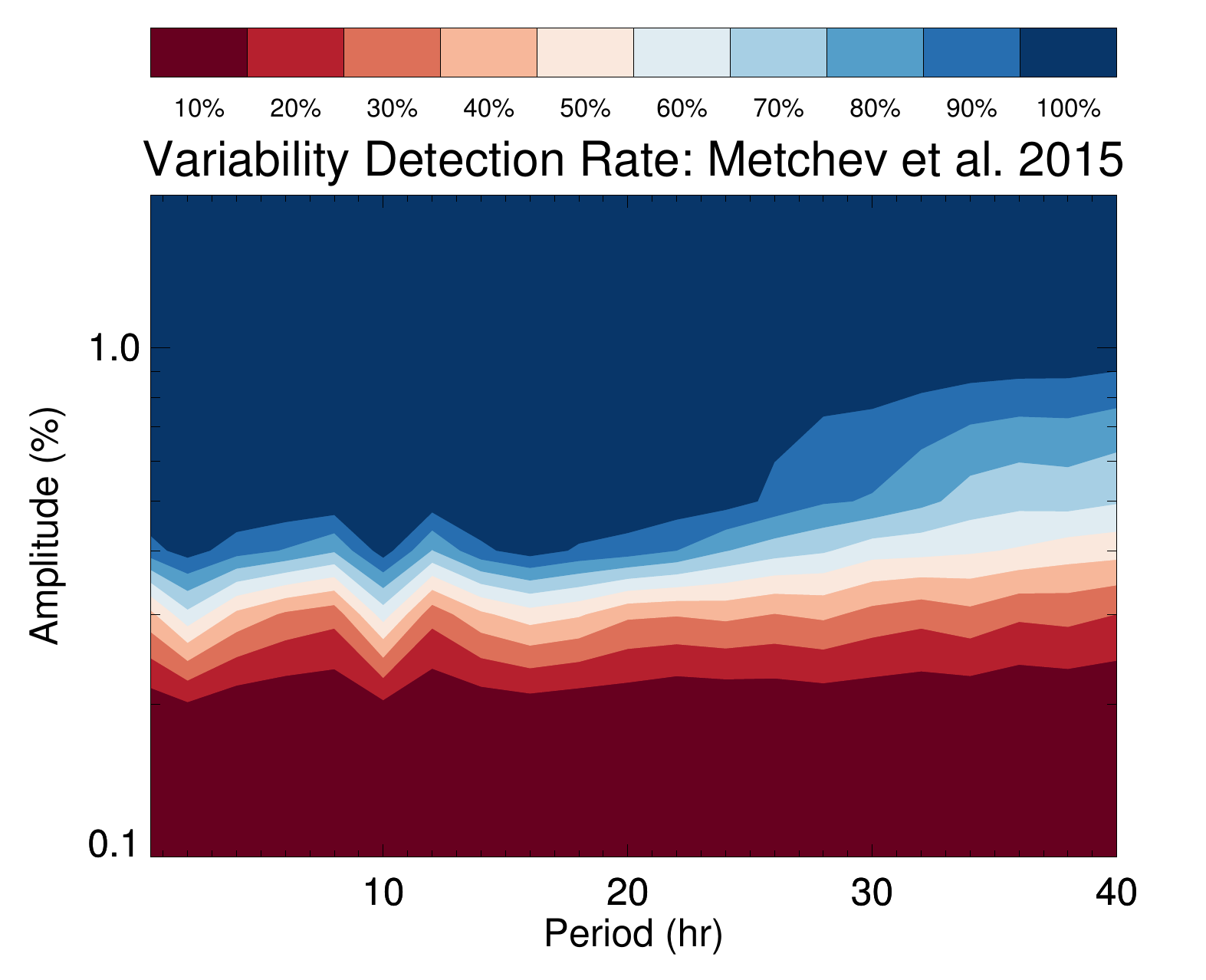}
   \caption{Median survey sensitivity as a function of variability amplitude and rotation rate for this survey (left) and the \citet{Metchev2015} survey. Contours show the percentage of recovered signals for each amplitude and period. The detection rates achieved by both variability surveys are very similar, and thus can be compared robustly. For non-variable objects, we assign upper limits on the variability amplitude as the 90\% contour at 10 hr.}
   \label{fig:survey_sensitivity}
\end{figure*}

For non-variable targets, we determine an upper limit on the variability amplitude as the 90\% contour boundary at 10 hours. \citet{Metchev2015} also set their upper limits assuming 10 hr periods, so our upper limits can be directly compared with their values. The left panel of Figure \ref{fig:survey_sensitivity} shows the median sensitivity for the survey. Contours represent the rate of detection for signals of a given amplitude and period. 
The right panel of Figure \ref{fig:survey_sensitivity} shows our calculated median sensitivity of the \citet{Metchev2015} survey for comparison, calculated in the same way. We note a slight improvement in the sensitivity of our survey compared to the \citet{Metchev2015}. Both surveys have similar observation lengths ($\sim20$~hr) and the median $3.6~\mu$m magnitude for both samples is 13.1--13.2.
The slightly higher sensitivity achieved by our survey is likely due to the fact that the \citet{Metchev2015} survey swapped between the [$3.6~\mu$m] and [$4.5~\mu$m] during each monitoring observation, or the differing data reduction and intra-pixel sensitivity correction methods. It is also possible that the amplitude of telescope motion for our observations was less than that of the \citet{Metchev2015} survey; the telescope drift during our ~$\sim20~$hr observations was consistently below $0.5$ pixels for the observations presented in this survey, resulting in low-level intra-pixel effects. The sensitivity plot for each observation is used in Section \ref{sec:variability_occurrence} to estimate the variability occurrence rates of young and field brown dwarfs. 

\section{Youth and Variability Properties of Inidividual Objects}\label{sec:youth+var}
In this section we discuss the youth and variability properties of each object in the sample individually.

{\obj{2m0001} } was first identified  as a candidate member of the AB Doradus moving group by \citet{Gagne2015c}. They classify \obj{2m0001} as a L4$\beta$ spectral type, with an {\sc {int-g}} gravity score \citep{Allers2013}. \citet{Faherty2016} classified \obj{2m0001} as an ambiguous member as it lacked a parallax and radial velocity. 
With updated proper motion and parallax measurements from Gaia DR2 \citep{Gaia2018}, we find that \obj{2m0001} is a high-likelihood member of the AB Doradus moving group (Table \ref{tab:kinematics}). We note that the proper motions and a parallax presented in \citet{Best2020}, which are within $2-3\sigma$ of the Gaia DR2 values, favor field over AB Doradus membership. 
\citet{Vos2019} monitored \obj{2m0001} in their $J$-band survey, but did not detect variability {above $\sim2\%$} during their $3.75$~hr observation. In this work we detect significant $3.6~\mu$m variability with an amplitude of $0.69\pm0.04\%$ and a period of $15.75\pm0.37$~hr. Its long period likely prohibits detection in shorter observations -- {the sensitivity map for the light curve reported by \citet{Vos2019} shows that sinusoidal signals with amplitudes of $>8-10\%$ would have been detected with a period $\sim15$~hr.} Its long period is also consistent with other known variables from the AB Doradus moving group such as WISE~J0047+68 \citep[$16.4\pm0.2$ hr;][]{Vos2018}.

{\obj{2m0030}} was given an optical classification of L7 by \citet{Kirkpatrick2000}, an infared spectral type of L4-L6$\beta$ by \citet{Gagne2015c}, and was assigned a {\sc{fld-g}} gravity class by \citet{Liu2016}. Its kinematic measurements in Table \ref{tab:kinematics} give it a high probability of membership in Argus according to BANYAN~$\Sigma$. We classify it as a high-likelihood member of Argus. 
\obj{2m0030} has been searched extensively for optical and infrared variability from ground-based telescopes in the literature \citep{Enoch2003, Clarke2008,Radigan2014}. Only \citet{Enoch2003} find a marginal detection of variability in the $i$ band with a period of $~1.5$ hr. \citet{Clarke2008} and \citet{Radigan2014} {report light curves with $1-4\%$ precision and} report no evidence for variability. We detect variability in its light curve, with an amplitude of $1.46\pm0.06\%$ and a rotation period of $4.29\pm0.02$ hr. The light curve appears stable over 5 rotations, however the non-detections reported in the literature to date suggest the possibility of an evolving-light curve. \obj{2m0030} is one of two variables whose light curve favors a two-term Fourier fit over a sinusoidal model.

{\obj{2m0031}/PSO~J282.7576+59.5858} was given an L9 spectral type by \citet{Best2013}, who note that it is not matched well by any of their spectral templates. They suggest that the spectrum may be a blended spectrum of multiple objects, however the spectral indices do not classify it as a binary. \citet{Best2020} also report a parallax of $71\pm5$ mas. Using the kinematic measurements in Table \ref{tab:kinematics}, BANYAN~$\Sigma$ reports that \obj{2m0031} is a high-likelihood member (98\% probability) of the Carina-Near moving group. 
\obj{2m0031} has not been monitored for variability in the past. Our new Spitzer light curve reveals variability with an amplitude of $0.35\pm0.03\%$ with a rotation period of $1.64\pm0.01$ hr. The light curve shows evidence of light curve evolution during the 20 hr observations, with the amplitude ranging from $\sim0.3-1\%$. The periodogram shows two narrow peaks at $\sim1.6$ hr, which is suggestive of differential rotation \citep{Apai2017}. This target adds to the small but growing number of L/T transition brown dwarfs that show light curve evolution on rotational timescales \citep[e.g.][]{Apai2017, Apai2021}.

{\obj{2m0153}} was classified as a L3$\beta$ with a {\sc{vl-g}} gravity class by \citet{Gagne2015c}, who also identified it as a likely member of the  Tucana-Horologium moving group. Using the kinematics presented in Table \ref{tab:kinematics}, both BANYAN~$\Sigma$ and LACEwING identify it as a likely Tucana-Horologium member, however parallax and radial velocity measurement are needed to confirm this object as a member. We classify \obj{2m0153} as a candidate member of Tucana-Horologium. 
Our Spitzer light curve shows significant variability with a $0.48\pm0.07\%$ amplitude and a period of $17.63^{+1.13}_{-0.94}$ hr. 

{\obj{2m0326}} has been identified as a highly probable member of AB Doradus by \citet{Gagne2014} who also assign a L5 $\beta/ \gamma$ spectral type, and note that the spectrum is very red for its spectral type. \citet{Cruz2007} suggest that \obj{2m0326} is likely younger than $500$ Myr based on the strength of its Li absorption. Our updated kinematic analysis which uses new proper motions from \citep{Marocco2021} agrees with these findings. We classify \obj{2m0326} as a high-likelihood member of AB Doradus but a radial velocity is needed to confirm membership. 
Our Spitzer monitoring shows a slowly increasing brightness during the observations. The variability is highly significant, with a periodogram power of $\sim40$ and a $\Delta$BIC of $\sim74$. Our MCMC variability fitting finds a period of $33\pm5~$hr, however this is extremely uncertain since we did not cover a full period. {The GP analysis presented in Section \ref{sec:GP} finds that the signal may be explained by correlated red noise in the data. Whether the variability is rotationally modulated, related to a longer-term astrophysical process or caused by systematics is an interesting question, which additional long-term monitoring will answer.}

{\obj{2m0342}} is an L4$\gamma$ with strong membership probability in the Tucana-Horologium association \citep{Gagne2015c} and \citet{Faherty2016} classify it as a high-likelihood member. Using proper motions from \citet{Gagne2015a}, we recover similar results, finding that  BANYAN~$\Sigma$ favors membership in Tucana-Horologium. We classify \obj{2m0342} as a high-likelihood member of Tucana-Horologium.  The light curve of \obj{2m0342}  displays significant variability over the observation. With an estimated period of $14.7\pm0.5$ hr, we cover just over one rotation period during our Spitzer observation.

{\obj{2m0349}} is one of three likely young brown dwarfs presented in this work for the first time. As discussed in Section \ref{sec:newbds}, \obj{2m0349} is an L5$\gamma$, with prominent alkali absorption lines that are suggestive of low-gravity. 
We use BANYAN~$\Sigma$ to assess membership using measured kinematic information. Based on sky positions and proper motions from \citep{Marocco2021} (shown in Table \ref{tab:kinematics}), we find a combined probability of $75\%$ that \obj{2m0349} is a member of either AB Doradus or $\beta$~Pictoris. Including the radial velocity measured from a FIRE/Echelle spectrum (Gagn\'e, J. et al. in preparation) reduces the probabilities of AB Doradus and $\beta$~Pictoris membership to $25\%$ and $19\%$ respectively.  While the kinematic analysis doesn't conclusively identify moving group membership, the low-gravity signatures in its spectrum lend support to its likely youth. We consider \obj{2m0349} as an ambiguous member for the remaining analysis. We detect significant variability in our Spitzer light curve, with an amplitude of $0.53\pm0.09\%$ and a period of $14.62^{+1.1}_{-1.0}$ hr. 

{\obj{2m0355}} is one of the prototypical isolated exoplanet analogs. It was identified as a high-likelihood member of the AB Doradus moving group by \citet{Faherty2013}, and confirmed by follow-up measurements and analysis by \citet{Liu2016, Faherty2016}.  \citet{Cruz2009} assign an L5$\gamma$ spectral type. \citet{Faherty2013} noted its extremely red near-infrared colors and underluminosity compared to field brown dwarfs. These peculiarities were first noted by \citet{Metchev2006} for the young companion HD203030~B \citep{Miles-Paez2017}, and are shared by directly-imaged exoplanets such as 2M1207~b and HR8799~bcd. Our kinematic analysis, which makes use of Gaia proper motions and parallax confirms that it is a member of AB Doradus using BANYAN~$\Sigma$. We thus confirm \obj{2m0355} as a bona fide member of AB Doradus. 
We measure significant variability in its $3.6~\mu$m light curve. We measure a variability amplitude of $0.26\pm0.02\%$ and a period of $9.53\pm0.19$ hr. \obj{2m0355} joins W0047+68 and 2M2244+20 \citep{Vos2018} as variable bona fide members of the AB Doradus moving group. We specifically focus on the variability properties of AB Doradus members in Section \ref{sec:ABDor}.

{\obj{2m0447}/PSO~J071.8769--12.2713} was discovered by \citet{Best2015} as a T2 candidate of the $\beta$~Pictoris moving group. {\citet{Best2020} measure new proper motions and parallax for this target which determine that it is likely a field object using BANYAN~$\Sigma$.  Proper motions from CatWISE2020 \citep{Marocco2021} also agree that \obj{2m0447} is likely a field object. We conclude that \obj{2m0447} is likely a field object and leave it out of the statistical analysis of variability in young objects in Section \ref{sec:variability_occurrence}.} In Figure \ref{fig:CMD_sample} we plot our sample on a color-magnitude diagram and find that \obj{2m0447} is over-luminous compared to the field population; we speculate that this may be due to an unresolved binary companion, as suggested by \citet{Best2020}.  
\citet{Vos2019} monitored this object for variability in ground-based $J$-band observations, detecting significant (4.5\%) variability during a three hour observation. Our Spitzer light curve does not display significant variability and we place an upper limit of $0.9\%$. The variability detection in \citet{Vos2019} followed by a non-detection in this work suggests that \obj{2m0447} may be a variable object whose light curve evolves over time.  Such behavior has been observed in a number of variable L/T transition brown dwarfs \citep[e.g.][]{Apai2017}.  {Another possibility is that \obj{2m0447} exhibits variability that is only apparent near-IR wavelengths, since different wavelengths probe different atmospheric layers. }.

{\obj{2m0459}} was discovered and classified as an L7 object with very red colors by \citet{Schneider2017}. They note spectral features that are consistent with low surface gravity, but its kinematics did not match well with a young moving group. Our updated kinematics, which include proper motions from CatWISE2020 \citep{Marocco2021} and a radial velocity from Gagn\'e, J. et al. (in preparation) indicates that \obj{2m0459} is a high-likelihood member of Argus. Its red near-IR colors, spectral signatures of youth and spectral type are remarkably similar to other young, high-amplitude variables discovered to date such as PSO~J318.5$-$22 \citep{Biller2015, Biller2018} VHS~1256~B \citep{Bowler2020, Zhou2020} and WISE~0047+68 \citep{Vos2018, Lew2016}. 
We do not detect variability in the Spitzer light curve of \obj{2m0459}, and place an upper limit of $<0.47\%$ on the peak-to-peak variability amplitude.

{\obj{2m0506}/PSO~J076.7092+52.6087} was discovered and classified as a T4.5 spectral type by \citet{Best2015}. Based on their measured proper motions, they determined that \obj{2m0506} was a likely member of Argus. We check for moving group membership using new proper motions and parallax from \citet{Best2020}, which favor field membership with a $69\%$ probability. Since this object has a T4.5 spectral type, we cannot examine the spectrum to look for signatures of youth since such signatures are not yet known. We classify this object as a field object and do not include it in the following analysis sections. 
We detect  variability in \obj{2m0506} during our Spitzer observation and measure an amplitude of $0.6\pm0.1\%$ and a rotation period of $13.8\pm0.8$ hr. Although its current kinematics favor field membership, its long rotation period is indicative of youth.

{\obj{2m0642}} was discovered by \citet{Mace2013} as an ``extremely red'' object at the L/T transition. \citet{Best2015} later assign a L9 spectral type based on its $J$-band profile and the depth of its $2.4~\mu$m water absorption feature. Its extremely red colors strongly suggest the presence of large amounts of dusty photospheric condensates \citep{Best2015}. \citet{Best2015} and \citet{Gagne2014} both reported that \obj{2m0642} is a likely AB Doradus member based on its proper motions and positions. 
Our analysis, which uses proper motions  and a new parallax from \citet{Kirkpatrick2020}, reports a $86\%$ probability of AB Doradus membership. We thus consider it a candidate member of AB Doradus. 
With a measured variability amplitude of $2.16\pm0.16\%$, \obj{2m0642} shows the highest [$3.6~\mu$m] amplitude of any L dwarf measured to date, and is comparable to large amplitude [$4.5~\mu$m] variables such as PSO~J318.5$-$22 \citep{Biller2018}. With a $\sim10$ hr period, we observe two full rotations during our Spitzer monitoring. During those two rotations we see significant evolution in the light curve, including the emergence of a $\sim1\%$ dip with a $1$ hr duration in the light curve at 15 hr. This dip may be due to the emergence of a Great Red Spot analog \citep{Apai2017}, or even a transiting satellite \citep{Tamburo2019, Limbach2021}. Long-term follow-up observations could differentiate the cause of such dips in the light curve of \obj{2m0642}.

{\obj{2m0718}}
Using the positions, proper motions and radial velocity shown in Table \ref{tab:kinematics}, BANYAN~$\Sigma$ reports a membership probability of $85\%$ for the $\beta$~Pictoris moving group. 
Assuming it is a member of the $\beta$ Pic moving group, we estimate a mass of $2.6\pm0.3~M_{\mathrm{Jup}}$ using the SED analysis described in Section \ref{sec:SED}.  Coincidentally, with a period of $\sim1.08$ hr, \obj{2m0718} joins a T7 field brown dwarf, 2MASS J03480772$-$6022270 (hereafter 2MASS J0348$-$60), as the fastest rotating brown dwarfs detected to date \citep{Tannock2021}. 2MASS~J0348$-$60 is estimated to be spinning at $\sim45\%$ of its break-up velocity. However,  \obj{2m0718} is likely young, and we have classified it as a candidate $\beta$~Pictoris member. Using our estimated mass and radius from Table \ref{tab:SEDs}, we find that the break-up velocity for \obj{2m0718} is  $\sim2.22$ hr. Thus, if \obj{2m0718} is indeed a $\beta$~Pictoris member it would be spinning faster than its theoretical break-up velocity. It is further possible that he observed variability pattern is produced by more than one dominant spatial feature, as for example seen with the regular positioning of near-equatorial hot spots on Jupiter \citep{dePater2016}. Follow-up high resolution spectroscopy will allow us to determine whether this object truly rotates so rapidly. 
Securing a parallax for \obj{2m0718} and confirming its membership will reveal whether it is a very low-mass object that is potentially in the process of breaking up, or whether it is an older object rotating at a more moderate pace like 2MASS~J0348$-$60.

{\obj{2m0809}} was identified by \citet{Gagne2015c} as an L6 pec (red) brown dwarf, and a candidate member of the Argus moving group. \obj{2m0809} is classified with an {\sc {int-g}} gravity index following the \citet{Allers2013} scheme. \citet{Best2020} provide a new parallax measurement of $42.4\pm3.6$ mas which differs substantially from the predicted parallax of $\sim65$ mas assuming membership in Argus \citep{Gagne2015c}. Using this new parallax,  BANYAN~$\Sigma$ reports \obj{2m0809} as a field object. Due to its low-gravity spectral signatures, we consider \obj{2m0809} to be a young field object. 
The Spitzer light curve of \obj{2m0809} shows significant variability with an amplitude of $0.77\pm0.06\%$ and a period of $1.365\pm0.004$ hr. 

{\obj{2m0951}} is one of the new brown dwarfs presented in this paper. As discussed in Section \ref{sec:newbds}, we assign \obj{2m0951} an L5(pec) spectral type.
Using proper motions from CatWISE2020 \citep{Marocco2021}, and a radial velocity from Gagn\'e, J. et al. (in preparation), our kinematics analysis classifies it as a candidate member of Argus, with a $87\%$ BANYAN~$\Sigma$ probability.  A parallax  measurement is still necessary to confirm membership. We do not detect variability in the Spitzer light curve of \obj{2m0951}. We place an upper limit of $0.9\%$ on its amplitude.

{\obj{2m1551}} shows clear signatures of youth in its spectrum. It has been classified as L4$\gamma$ by \citet{Faherty2013} and L4 {\sc {vl-g}} by \citet{Allers2013}. However, despite its clear spectral signatures of youth, \obj{2m1551} is not an obvious member of any known moving group. \citet{Faherty2016} classified it as having ambiguous membership in young moving groups. Using an updated parallax from \citet{Gaia2018} and an updated radial velocity from Gagn\'e, J. et al. (in preparation), the BANYAN~$\Sigma$ tool reports that \obj{2m1551} is likely a field object. Due to its strong youth indicators, we consider that \obj{2m1551} is a young field object. 
The Spitzer light curve of \obj{2m1551} does not show statistically significant variability during our 20 hour observation. We place an upper limit of $0.39\%$ on the variability amplitude for periods of $\sim10$ hr.

{\obj{2m1647}} was discovered and classified as an L9 (pec) brown dwarf with a very red spectrum by \citet{Kirkpatrick2011}. They report a parallax of $116\pm29$ mas. More recently, \citet{Best2020} report a parallax of $42.9\pm2.1$ mas, which is not in agreement with the original parallax measurement from \citet{Kirkpatrick2011}. \citet{Kirkpatrick2011} identifies \obj{2m1647} as a candidate member of $\beta$~Pictoris or Argus, with a low probability of field membership. However, the updated parallax from \citet{Best2020} results in a 66\% probability that \obj{2m1647} is a field object, and $34\%$ that it is a Carina-Near member. We classify \obj{2m1647} as a young field object. We detect significant variability in our Spitzer light curve, and estimate a variability amplitude of $0.5\pm0.1\%$ with a period of $9.2\pm0.3$ hr.

{\obj{2m1741}} was discovered and classified as L7 (pec) by \citet{Schneider2014}. They also identified \obj{2m1741} as a candidate member of either the AB Doradus or the $\beta$~Pictoris moving group. \citet{Faherty2016} classify it as a L6--L8 $\gamma$ based on its near-infrared spectrum. Recently, \citet{Kirkpatrick2020} provide a new parallax measurement, and report a high probability that \obj{2m1741} is a member of AB Doradus. Our updated analysis, which uses the parallax from \citet{Kirkpatrick2020} and a new radial velocity from Gagn\'e, J. et al. (in preparation) confirms \obj{2m1741} as a member of the AB Doradus moving group, with probability of 99\% using BANYAN $\Sigma$. This agrees well with the signatures of youth in its spectrum \citep{Schneider2014, Faherty2016}. We detect significant variability in the Spitzer light curve of \obj{2m1741}, with an amplitude of $0.35\pm0.03\%$ and a period of $15.0^{+0.7}_{-0.6}$ hr. \citet{Vos2019} previously monitored \obj{2m1741} in $J$-band and did not detect significant variability during a $\sim2.5$ hr observation. However, the sensitivity plot from this observation shows a very small probability ($\sim5\%$) of detecting variability with a period of $15$ hr. 
Its rotation period matches well with the other late-L AB Doradus bona fide members with periods of $9-16$ hr (W0047+68, 2M2244+20; \citet{Vos2018} and \obj{2m0355}; this work), suggesting a common angular momentum history for these objects. We discuss \obj{2m1741} in context with other AB Doradus members in Section \ref{sec:ABDor}. 

{\obj{2m2002}} was classified as an L5-L7$\gamma$ by \citet{Gagne2015c}, but multiple studies have been unable to place it in a young moving group \citep{Gagne2015c, Faherty2016, Vos2019}. Our updated kinematics, which includes a new parallax from Gaia DR2 \citep{Gaia2018} similarly do not place it within a known moving group.
 However, its spectrum displays clear signature of youth, so we classify it as a young field object.
\citet{Vos2019} detect $J$-band variability in this object using UKIRT/WFCAM, but the 4 hour observation did not yield periodicity. Its Spitzer light curve does not show significant variability, and we place an upper limit of $0.3\%$ on $3.6~\mu$m. Follow-up observations may reveal whether \obj{2m2002} shows periods of variability followed by `quiescent' phases, such as those objects reported by \citet{Apai2017}.

{\obj{2m2117}}{/PSO J319.3102-29.6682} was discovered and classified as a T0 dwarf by \citet{Best2015}. More recently, \citet{Best2020} measured a parallax of $52.4\pm6.8$ mas for this object, and report a BANYAN~$\Sigma$ probability of 99\% that is a member of the $\beta$~Pictoris moving group. However, \citet{Kirkpatrick2020} report a parallax of $76.1\pm3.5$ mas, which gives a 99\% probability that \obj{2m2117} is a field object. 
We do not include \obj{2m2117} as a young object in the statistical analysis of variability Section \ref{sec:variability_occurrence} since its membership is unclear. However, in Section \ref{sec:SED} we estimate fundamental parameters for this object for both cases for completeness. 
 We do not detect significant variability in the Spitzer light curve of \obj{2m2117}, and place an upper limit of $0.38\%$ on its amplitude.

{\obj{2m2154}} was identified as an L5$\beta/\gamma$ candidate member of the Argus moving group by \citet{Gagne2014c,Gagne2015c}. \citet{Vos2019} later identified it as a likely member of the Carina-Near moving group. Our updated kinematic analysis finds that BANYAN~$\Sigma$ favors Carina-Near membership, with a membership probability of $80\%$. We thus classify \obj{2m2154} as a candidate Carina-Near member. \obj{2m2154} was monitored for $J$-band variability by \citet{Vos2019}, who find no evidence for variability during a $\sim4$ hr observation. Similarly, our Spitzer light curve does not display any variability during our 20 hour observation, and we place an upper limit of $0.4\%$ on the variability amplitude of this object.

{\obj{2m2206+33}/PSO~J331.6058+33.0207} was discovered and identified as a T1.5 candidate Argus member by \citet{Best2015} based on its kinematics. Our analysis, which uses the proper motions from \citet{Best2015} finds that BANYAN~$\Sigma$ favors Argus ($68\%$). We note that CatWISE2020 \citep{Marocco2021} report discrepant proper motions for this target which result in a higher probability of field membership. For this work we classify \obj{2m2206+33} as a candidate Argus member. 
A future radial velocity and parallax measurement will help to clarify membership. We detect significant variability in \obj{2m2206+33} with an amplitude of {$1.2\pm0.1\%$} and a period of $15.9\pm0.6$ hr. This is one of the longest rotation periods measured for any T dwarf, and adds further support that \obj{2m2206+33} may be young.

{\obj{2m2206-42}} was discovered and classified as an L2 dwarf by \citet{Kirkpatrick2000}. \citet{Kirkpatrick2008} detected lithium in its spectrum, and \citet{Gagne2014} report it as a candidate AB Doradus member. Our updated kinematics analysis, which include new proper motions and a parallax from Gaia DR2 \citep{Gaia2018}, confirms \obj{2m2206-42} as a high-likelihood member of AB Doradus. A radial velocity is needed to confirm its membership. We do not detect variability in our Spitzer light curve of \obj{2m2206-42}, and place an upper limit of $0.33\%$ on the variability amplitude.

{\obj{2m2216}/PSO~J334.1+19} was discovered and identified as a T3 dwarf by \citet{Best2015}, who identify it as a possible comoving companion to the M4 star LSPM J2216+1952. They find that both components are candidate members of $\beta$~Pictoris. Our independent kinematics analysis, which uses proper motions from CatWISE2020 \citep{Marocco2021}, finds a slightly higher BANYAN~$\Sigma$ $\beta$~Pictoris membership probability (85\%) and we classify it as a candidate $\beta$~Pictoris member. We also repeat the BANYAN~$\Sigma$ analysis for the potential comoving primary LSPM J2216+1952 using new Gaia DR2 \citep{Gaia2018} proper motions and parallaxes, which reports a $99\%$ probability that it is a field star. Thus, we conclude that \obj{2m2216} is a candidate $\beta$~Pictoris member, but is likely not a companion to the star LSPM J2216+1952. We do not detect variability in our Spitzer light curve of \obj{2m2216}, and place an upper limit of $1\%$ on the variability amplitude.

{\obj{2m2322}} is an L2$\gamma$ \citep{Gagne2014} that was identified as a possible comoving companion the the M5 star 2MASS J23225240-6151114. \citet{Faherty2016} classify this object as a high-likelihood member of Tucana Horologium. Our updated kinematic analysis combines new proper motions and a parallax from Gaia DR2 \citep{Gaia2018} with a radial velocity measurement from \citet{Faherty2016} to confirm it as a member of the Tucana Horologium moving group. The primary, which lacks a radial velocity, shows a $99\%$ membership in Tucana Horologium also. Thus we confirm that \obj{2m2322} is a member of the Tucana Horologium moving group and consider it highly likely that it is comoving with the M5 star 2MASS~J23225240--6151114. This system will be the subject of an upcoming paper (Faherty, J. K. et al., in preparation). 
We do not detect variability in our Spitzer monitoring of \obj{2m2322} and place a limit of $0.4\%$ on the maximum variability amplitude for this object.

{\obj{2m2343}} was discovered and classified as an L3-L6$\gamma$ dwarf, with a spectrum very similar to \obj{2m0355}. \citet{Faherty2016} identify it as an ambiguous moving group member, which shows probability of membership in the AB Doradus and Tucana Horologium moving groups. We use proper motions from CatWISE2020 \citep{Marocco2021} in our updated kinematic analysis. BANYAN~$\Sigma$ shows moderately high probability of membership in AB Doradus. We classify it as a candidate AB Doradus member. We do not detect significant variability in our Spitzer light curve of \obj{2m2343}, and place an upper limit of $0.4\%$ on its variability amplitude.

\section{The Variability Occurrence Rate of Young Brown Dwarfs in the Mid-infrared} \label{sec:variability_occurrence}
Our variability detections and detection limits can be used to estimate the fraction of young brown dwarfs that exhibit photometric variability \citep{Vos2019}. The occurrence rate analysis presented in this section is described by \citet{Lafreniere2007, Bonavita2013}, and summarized below.

\subsection{Statistical formalism}
We consider variability monitoring observations of $N$ sources enumerated by $j=1....N$. $f$ is the fraction of objects that exhibit variability with amplitude and rotation period in the interval $[a_{\mathrm{min}},a_{\mathrm{max}}]\cap [r_{\mathrm{min}},r_{\mathrm{max}}]$. For this work we consider amplitude ranges of $0.05-3\%$ and rotation periods of $0.5-40$ hrs. $p_j$ is the probability that such variability would be detected from each observation. To calculate $p_j$ we use the variability sensitivity maps, $g(a,r)$, shown in Figure \ref{fig:survey_sensitivity}. We integrate over $g(a,r)$ the considered range in amplitude and period to obtain the probability of detecting a variable signal with $a_{\mathrm{min}}<a<a_{\mathrm{max}}$ and $r_{\mathrm{min}}<r<r_{\mathrm{max}}$ for target $j$:
\begin{equation}
    p_j = \int_{a_{\mathrm{min}}}^{a_{\mathrm{max}}} \int_{r_{\mathrm{min}}}^{r_{\mathrm{max}}} g(a,r) ~ da~ dr
\end{equation}

Using this notation, the probability of detecting variability in target $j$ is the fraction of objects displaying variability multiplied by the probability that such variability would be detected, or $fp_j$. The probability of not detecting variability is $(1-fp_j)$. We denote $d_j$ the detections made by the observations such that $d_j=1$ for a positive variability detection for target $j$ and $d_j=0$ for a non-detection. Then the probability of observing a detection or non-detection {for} a given $f$ is:
\begin{equation}\label{eq:likelihood}
L \big({d_j}|f \big) = \prod\limits_{j=1}^N(1-fp_j)^{1-d_j}(fp_j)^{d_j}
\end{equation}
According to {Bayes' }theorem, the a priori probability density, or prior distribution $p(f)$,  and the likelihood function $L$,  can be used to calculate the posterior distribution $p(f|{d_j})$, the probability density updated in light of the data:
\begin{equation}\label{eq:posterior}
p\big( f|{d_j} \big) = \frac{L \big({d_j}|f \big) p(f)}{\int_{0}^{1}L \big({d_j}|f \big) p(f)\mathrm{d}f}
\end{equation}

Since there are no independent variability occurrence rate estimates available, we use an non-informative Jeffreys prior. As outlined in Appendix \ref{app:prior}, the suitable Jeffreys prior is given as:
\begin{equation}\label{eq:prior}
J(f) = \sqrt{\sum_j \frac{p_j}{f(1-fp_j)}}
\end{equation}

We calculate the posterior probability distribution, $p(f|{d_j})$, which represents the fraction of variable objects, or variability occurrence rate of brown dwarfs. We calculate credible intervals for the variability occurrence rate, $f$, for a given level of credibility, $\alpha$. 
We follow the method outlined by \citet{Kraft1991} to calculate the confidence intervals $[f_\mathrm{min},f_\mathrm{max}]$, i.e. the upper and lower boundaries for the variability occurrence rate. We calculate the confidence interval by solving for $[f_\mathrm{min},f_\mathrm{max}]$ in the equations:
\begin{equation}
    \alpha = \int_{f_\mathrm{min}}^{f_{\mathrm{max}}} p(f|{d_j})df
\end{equation}
and
\begin{equation}
    p(f_{\mathrm{min}} | d_j) = p(f_{\mathrm{max}} | d_j).
\end{equation}
For the case of a one-sided distribution,  the upper or lower bound of the credible interval is found by solving:
\begin{equation}
    \alpha=\int_{f_{\mathrm{min}}}^{1} p(f|{d_j})df ~\mathrm{or}~   \alpha=\int_{0}^{f_{\mathrm{max}}} p(f|{d_j})df
\end{equation}
We determine the $68\%$ and $95\%$ confidence intervals in this paper.


\subsection{The Variability Occurrence Rates of L and T Spectral Type Objects}

\begin{figure*}[tb]
   \centering
   
   \includegraphics[scale=0.55]{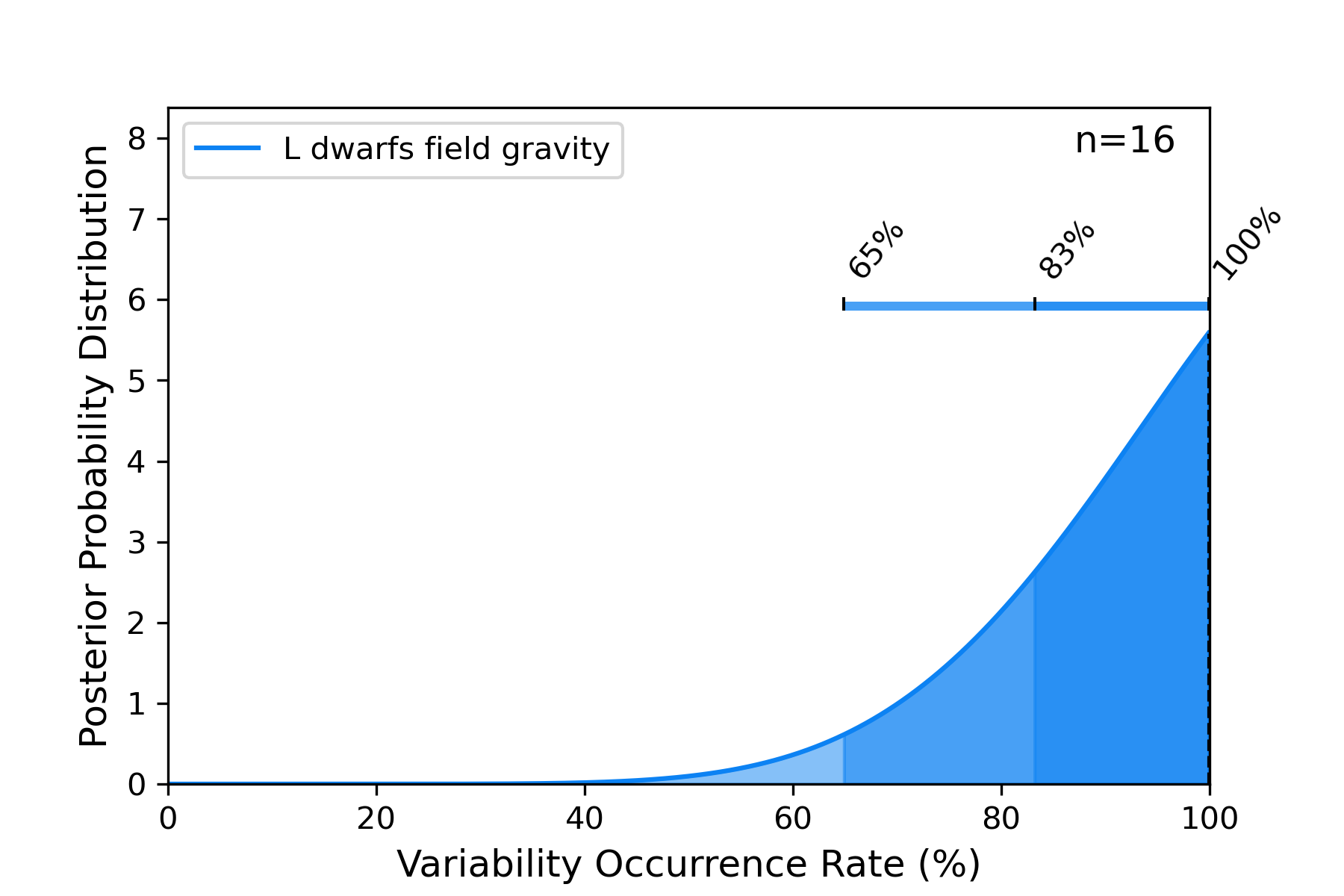}
    \includegraphics[scale=0.55]{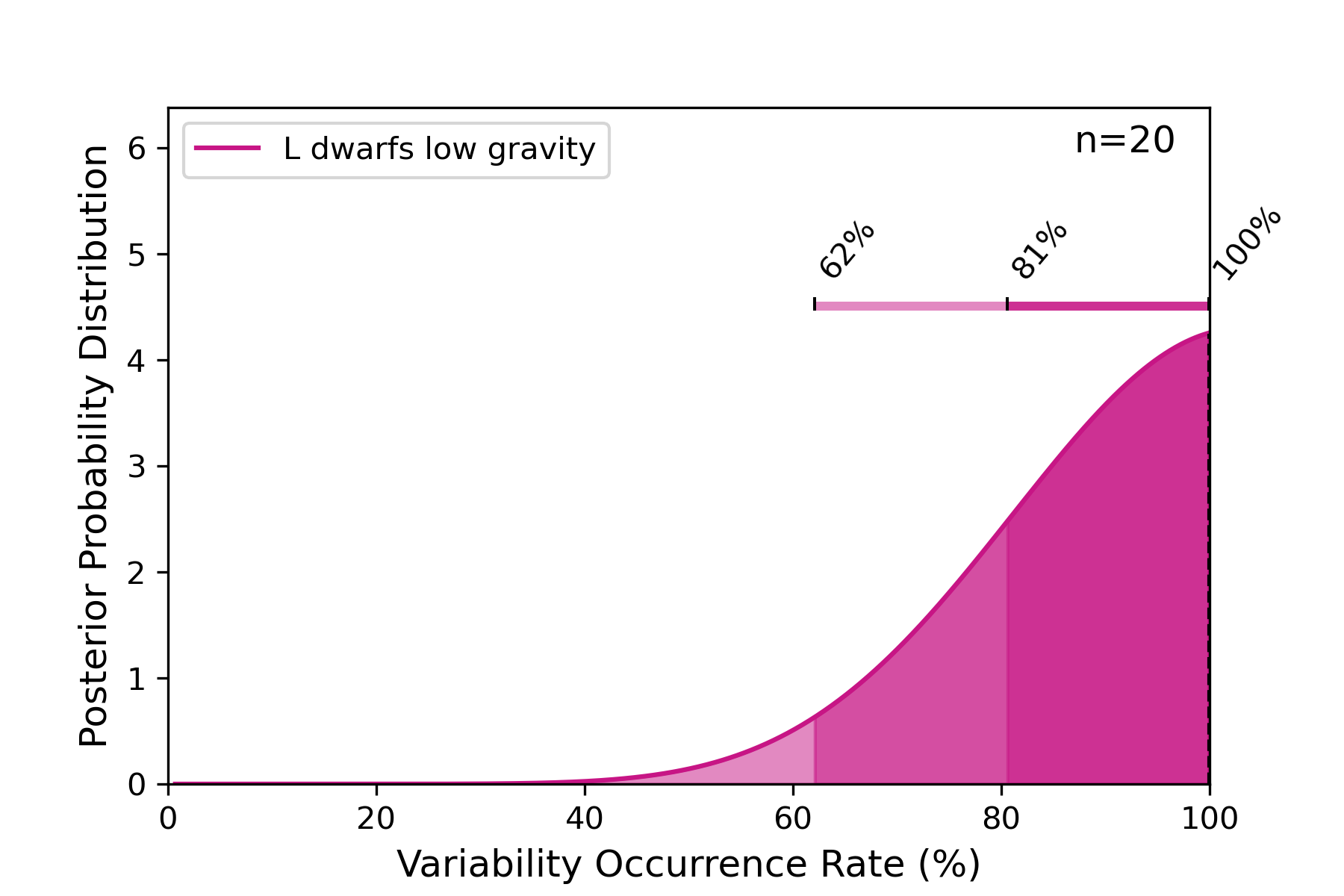}\\
   \includegraphics[scale=0.55]{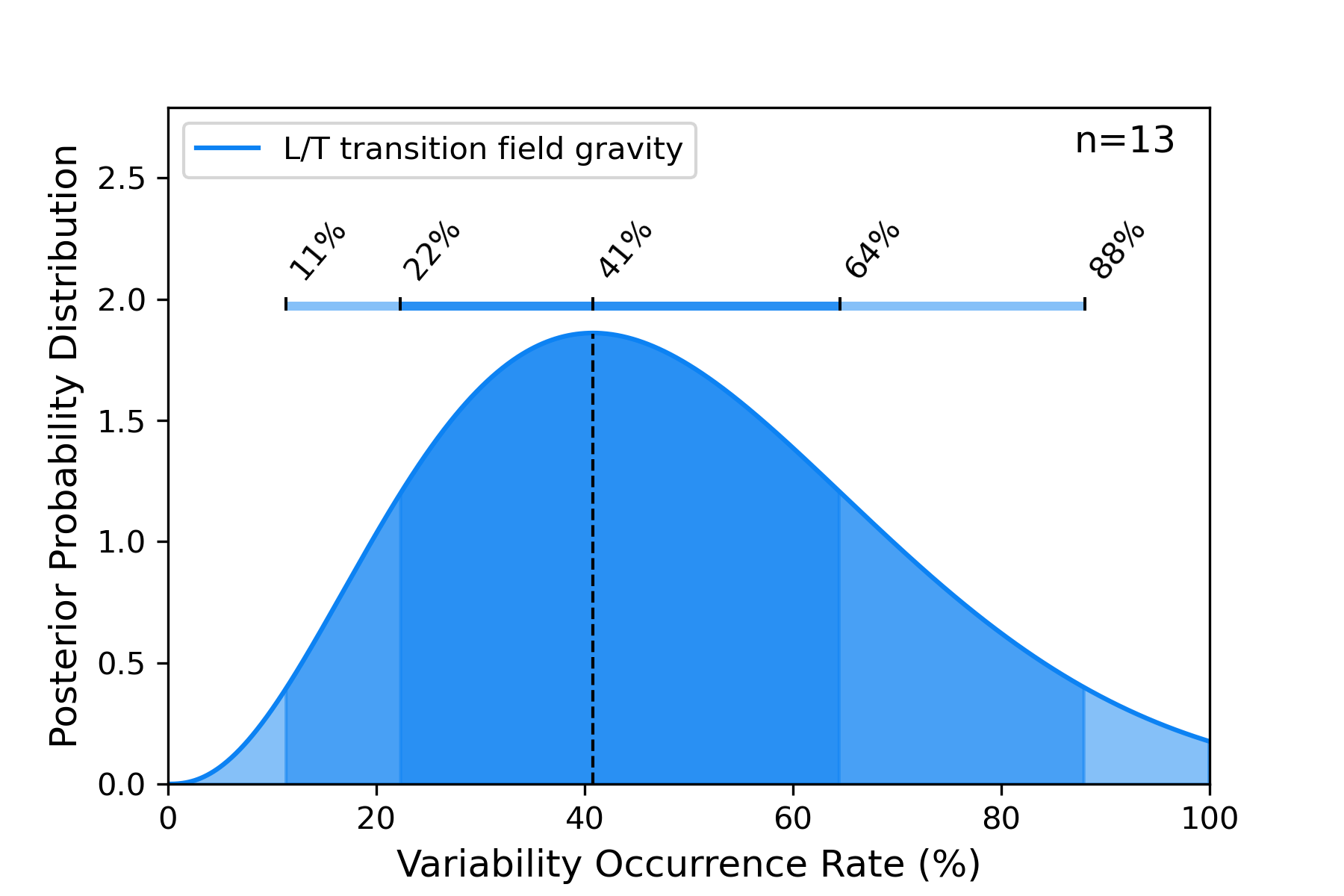} 
    \includegraphics[scale=0.55]{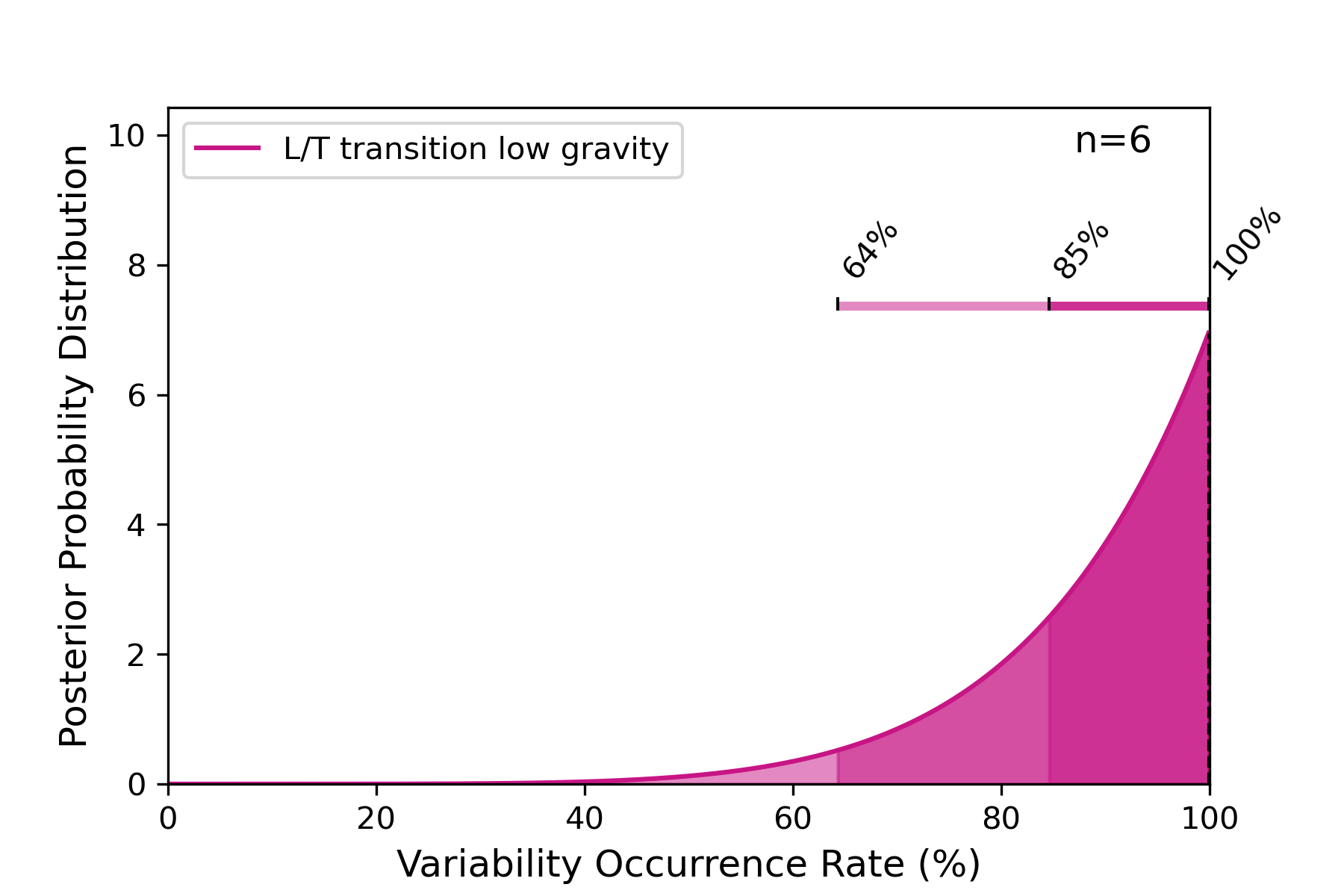}\\
    \includegraphics[scale=0.55]{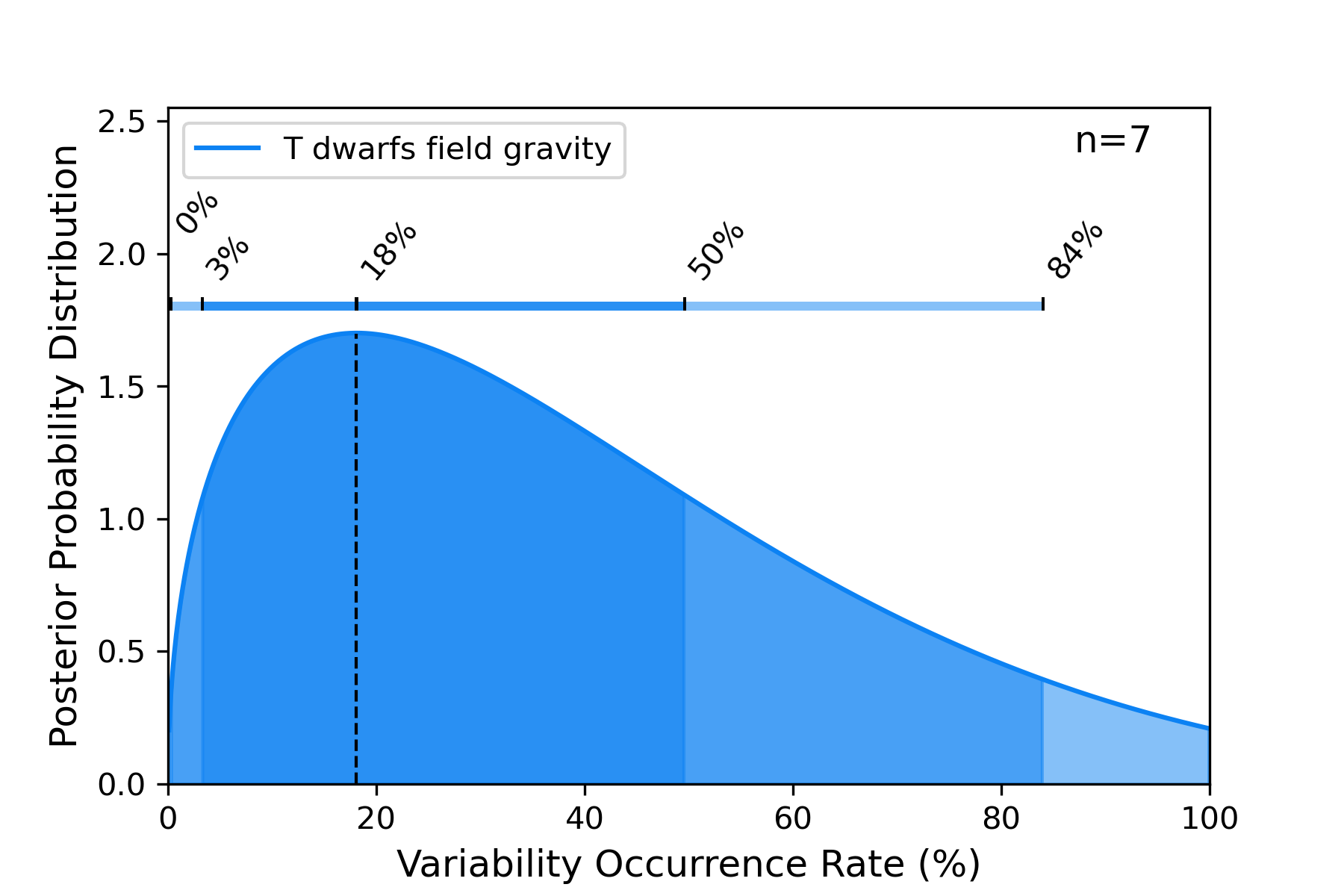}
    \includegraphics[scale=0.55]{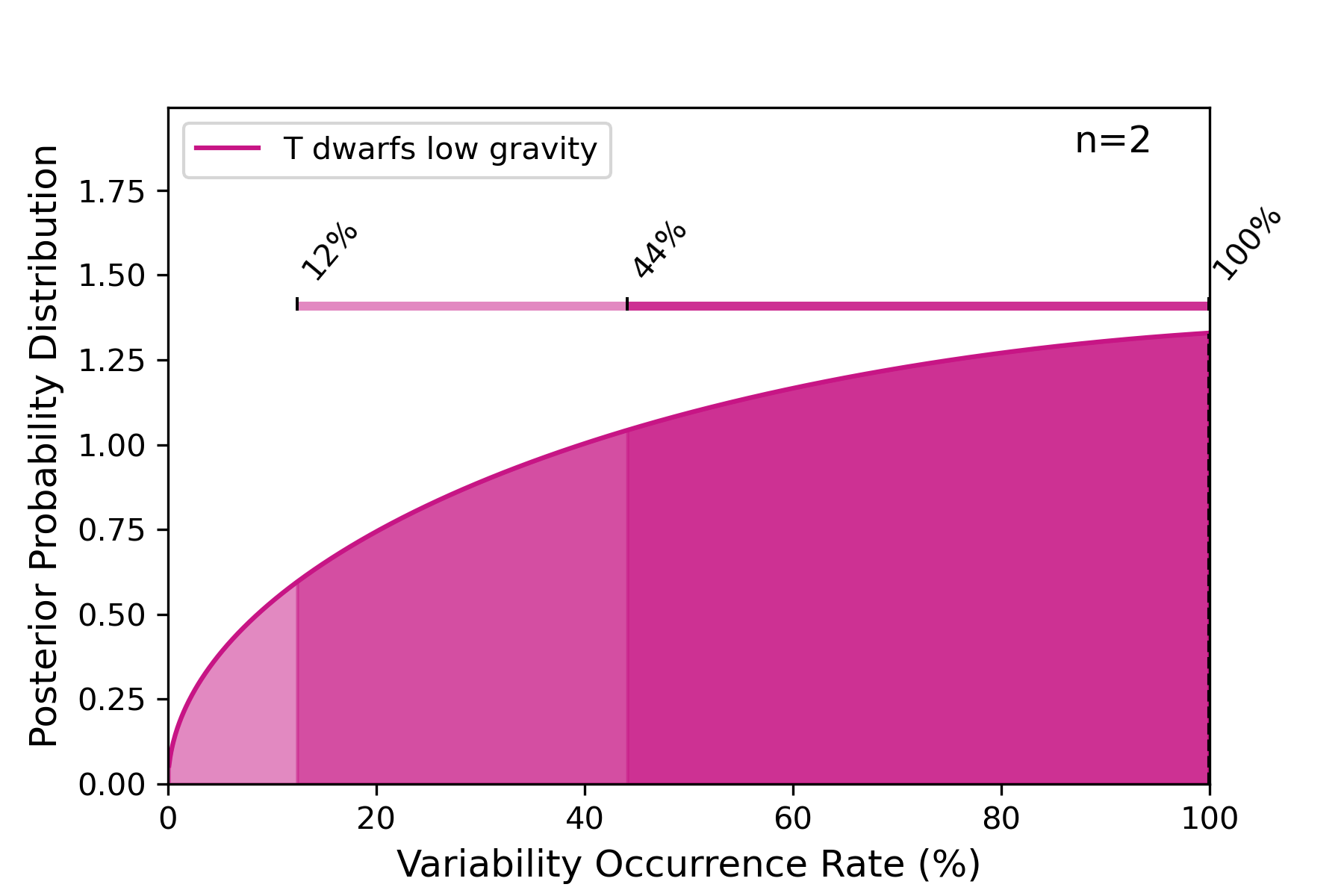}

   \caption{Posterior distribution functions of the variability occurrence rate of field brown dwarfs (blue) from \citet{Metchev2015} and young, low-gravity objects presented in this work. The shaded areas show the variability occurrence rate limits for the $95\%$ and $68\%$ confidence levels. Top panels show the occurrence rates for L dwarfs, the middle panel shows occurrence rates for the L/T transition and the bottom panel shows the variability occurrence rates for T dwarfs.}
   \label{fig:PDF}
\end{figure*}

We calculate the posterior distribution functions of the variability occurrence rates for the low-gravity and field population of brown dwarfs using the low-gravity sample presented here and the field objects from \citet{Metchev2015}. The \citet{Metchev2015} sample included eight low-gravity brown dwarfs, which were added to the low-gravity survey sample presented in this paper and excluded from the field sample. We  discard the magnetically active brown dwarf 2MASSW J0036159+182110 because it was a known variable prior to the \citet{Metchev2015} survey, as well as known binaries. 
We also discard three targets from the low-gravity sample presented in this paper: \obj{2m0447}, \obj{2m0506} and \obj{2m2117}, because their youth  and kinematic membership are uncertain, as discussed in Section \ref{sec:youth+var}.
We calculate the posterior distribution functions of the variability occurrence rate for L dwarfs (SpT $<$L9), L/T transition objects (L9 $\leq$ SpT $\leq$T3.5) and T dwarfs (SpT $>$T3.5). 
We show the variability occurrence rates of the young (pink) and field (blue) samples in Figure \ref{fig:PDF} and discuss each spectral type bin below.  

\subsubsection{L Dwarf Variability Occurrence Rates}
For field L dwarfs, we find a variability occurrence rate of  $83-100\%$ (68\% confidence). 
We find a variability occurrence rate of $81-100\%$ for the low-gravity L dwarfs, which is consistent with that of the field gravity L dwarfs. Thus we find no evidence for an enhancement in variability occurrence rate for the low-gravity L dwarfs. This is in contrast to \citet{Vos2019} who find a higher $J$-band occurrence rate for the low-gravity objects.  This apparent discrepancy could be due to the different wavelength regimes and/or different sensitivities of each survey. \citet{Vos2019} surveyed 30 low-gravity objects using ground-based $J$-band observations. The $J$-band probes deeper pressure levels than the Spitzer [$3.6~\mu$m] band used in this survey \citep{Buenzli2012}. The $J$-band flux arises from pressures of $\sim10$ bar while the Spitzer [$3.6~\mu$m] band probes pressures of $\sim1$ bar. Any difference in the observed variability properties between these wavelength regimes may suggest that they are sensitive to different variability mechanisms that take place at different altitudes. Simultaneous spectroscopic variability monitoring ranging from the near-infrared to the mid-infrared is necessary to test this possibility. 
However, another significant difference is the increased sensitivity achieved by Spitzer surveys \citep[][this work]{Metchev2015} compared to ground-based searches \citep{Radigan2014, Wilson2014, Vos2019, Eriksson2019}. The Spitzer surveys on average probe variability amplitudes that are a factor of $\sim10$ smaller than the ground-based surveys mentioned above, and thus offer a more complete picture of low-level variability. Ground-based surveys thus may probe the variability occurrence rates of high-amplitude variables, while the Spitzer surveys may probe a more complete picture of variability at low to high amplitudes. 

\subsubsection{L/T Transition Variability Occurrence Rates}
At the L/T transition (L9--T3.5), field dwarfs show a lower variability occurrence rate compared to field L dwarf behavior, with a range of $22-64\%$ (68\% confidence interval). 
\citet{Metchev2015} report similar findings, using virtually the same sample \citep[][ combined young and field brown dwarfs for their analysis whereas  we removed low-gravity objects from the field dwarf sample]{Metchev2015}. As discussed by \citet{Metchev2015}, this is in apparent contrast with the ground-based $J$-band survey reported by \citet{Radigan2014}, who find an enhancement in variability occurrence rate at the L/T transition compared to earlier Ls and later Ts. The reasons for this difference may be the same reasons outlined above --- the different wavelength regimes and sensitivities for Spitzer and ground-based near-infrared variability surveys. 
In contrast to the field dwarfs, young L/T transition objects maintain their high variability rate in the range $85-100\%$.  These results suggest that there may be an enhanced variability occurrence rate of low-gravity L/T transition objects compared to their field brown dwarf counterparts. Interestingly, this tentative enhancement may also explain the apparent disagreement between the L/T transition variability occurrence rates between \citet{Radigan2014} and \citet{Metchev2015} discussed above. \citet{Radigan2014} find high-amplitude variability in 4/16 of their L9--T3.5 sample, resulting in a high occurrence rate of high-amplitude variables. However, two of these detections,  \obj{s0136} and \obj{2m2139}, have since been identified as planetary-mass members of the Carina-Near moving group \citep{Gagne2017, ZJZhang2021}. Thus, a reanalysis of the \citet{Radigan2014} variability occurrence rates with updated youth information would be beneficial for evaluating whether field L/T transition objects have an enhanced variability rate. 

\subsubsection{T Dwarf Variability Occurrence Rates}
For field T dwarfs, the maximum likelihood value drops slightly (46\% to 18\%), but the confidence intervals ($3-50\%$) are generally consistent with the L/T transition objects. \citet{Metchev2015} explain the lower rate of variability for field T dwarfs as due to lower sensitivity to this sample, however since our calculations take into account the sensitivity of the sample, we do not believe that the low rate is due to lower sensitivity. For young T dwarfs, our sample is composed of only two T dwarfs -- \obj{2m0718} (variable in this work) and Ross 458C, which is a known variable \citep{Manjavacas2019}, but was observed to be non-variable by \citet{Metchev2015}, so is classed as non-variable for our analysis.  The resulting posterior distribution of the variability occurrence rate is wide, and we find that $>44\%$ of young objects with spectral types greater than T3.5 are likely to be variable at $68\%$ confidence, $>12\%$ at $95\%$ confidence. With the current sample we cannot robustly compare between the variability occurrence rates of high and low-gravity objects at spectral types $>T$4, but as more young T dwarfs are confirmed this will be an interesting question.

\section{The Variability Amplitudes of Young Brown Dwarfs at 3.6 micron} \label{sec:variability_properties}

\begin{figure*}[tb]
   \centering
   \includegraphics[scale=0.8]{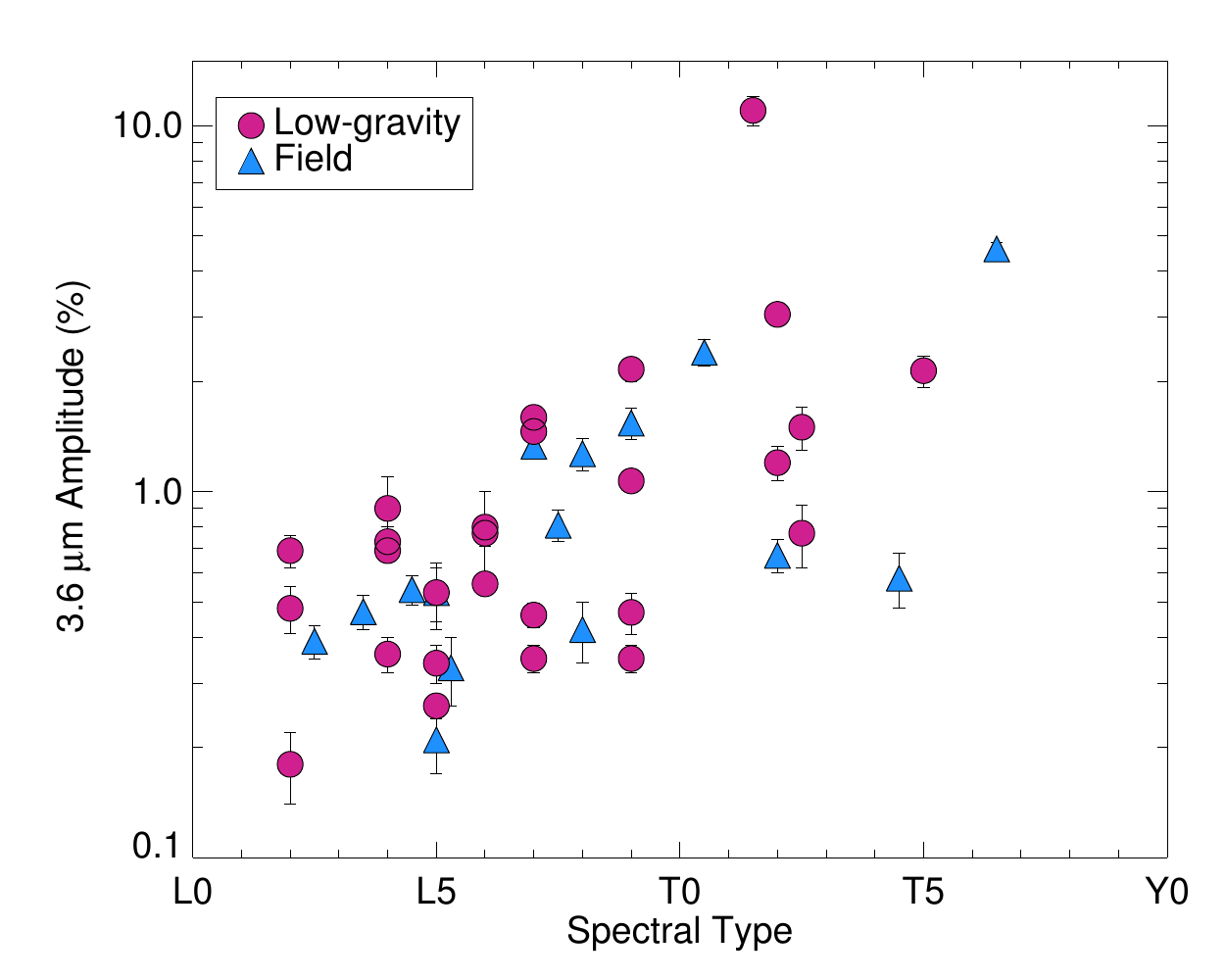}
   \caption{[$3.6~\mu$m] variability amplitudes of young objects (pink circles) and field objects (blue triangles) as a function of spectral type. Sources from this study and those compiled in \citet{Vos2020_cat}.  As is the case for the field brown dwarfs \citep{Metchev2015}, the maximum variability amplitude of the young objects steadily increases with spectral type from early L to the L/T transition.  It is also notable that the highest amplitude for each bin is measured for a young object for spectral types L0--L9. Thus, it seems that the \textit{maximum} variability amplitudes of young brown dwarfs may be enhanced compared to field brown dwarfs.}
   \label{fig:amplitudes}
\end{figure*}

A number of studies have suggested that the variability amplitudes of low-gravity objects may be enhanced compared to the field brown dwarfs. \citep{Biller2015, Metchev2015, Vos2020}.
Clouds in low-gravity atmospheres tend to form at lower pressures/higher altitudes \citep{Marley2012}, and thus higher variability amplitudes might be expected due to a higher photometric contrast between regions of thick and thin clouds. 

We combine the new variable objects detected in this survey with brown dwarfs previously monitored for [$3.6~\mu$m] variability with Spitzer \citep[compiled in][]{Vos2020_cat} to investigate whether  [$3.6~\mu$m] amplitudes are enhanced for low-gravity brown dwarfs. The T2.5 object \obj{2m2139} has been added to the young sample since it is a high probability member of the Carina-Near moving group \citep{ZJZhang2021}. 
Figure \ref{fig:amplitudes} shows the [$3.6~\mu$m] variability amplitudes of the young (pink) and field (blue) L2--T5 brown dwarfs. As is the case for the field brown dwarfs \citep{Metchev2015}, the maximum variability amplitude of the young objects steadily increases with spectral type from early L to the L/T transition.  The range of measured amplitudes in each spectral bin are generally consistent between the low-gravity and field dwarf populations. However, it is notable that the highest amplitude for each bin is measured for a young object for spectral types L0--L9. Thus, it seems that the \textit{maximum} variability amplitudes of young brown dwarfs are enhanced compared to field brown dwarfs. 
These results suggest that variability amplitudes may be enhanced for low-gravity objects, but that secondary effects may affect the observed amplitudes. For example, the viewing inclination \citep[e.g.][]{Vos2017} likely reduce the observed variability amplitudes for objects viewed closer to pole-on, and may be responsible for the observed spread in amplitudes at each spectral type bin. Thus, the maximum variability amplitudes observed in low-gravity brown dwarfs would be higher than those of the field brown dwarfs, while both populations display variability down to the lower detection limit. Measuring the viewing angles of these low-gravity brown dwarfs with maximum amplitudes would allow us to test this possibility, and will be the focus of a future paper. Additionally, using three-dimensional atmospheric circulation models, \citet{Tan2021} find that the rotation rate likely affects the size of atmospheric features in the atmosphere, and thus alters the variability amplitude driven by these features.


\section{Rotation Rates of an Age-Calibrated Sample} \label{sec:rotation_rates}
Identifying variable members of young moving groups is extremely valuable, as it provides an age-calibrated sample of young brown dwarfs with measured rotation rates, allowing us to trace angular momentum evolution {\citep[e.g.][]{Scholz2018, Zhou2019, Vos2020}.} Of particular interest is whether there are rotational evolution trends related to mass, as is seen in our own solar system \citep[e.g.][]{Allers2016}. We estimate masses for objects whose variability properties are presented in this paper (as described in the following section) and gather estimated masses from the literature for the sample presented in \citet{Vos2020}.

\subsection{Estimating Masses with Spectral Energy Distribution Analysis} \label{sec:SED}

We analyze the spectral-energy distributions (SEDs) of our young sample in order to investigate their fundamental parameters. 
We calculate the $L_{\mathrm{bol}}$, $T_{\mathrm{eff}}$, radii, masses and log(g) following the method presented by \citet{Filippazzo2015}. This method performs a numerical integration of the combined empirical visible and near-infrared photometry and spectra as well as WISE and/or Spitzer photometry, accessed through the BDNYC database \citep{bdnyc_db}. We use parallaxes when available, but supplement with kinematic distances when a target is regarded as a high-likelihood member of a moving group. We present the fundamental parameters in Table \ref{tab:SEDs}. Our  sample spans temperatures of $\sim600-1750$ K and model-dependent masses of $2.5-33~M_\mathrm{Jup}$, overlapping with the masses and temperatures of the directly-imaged exoplanets that have been discovered to date \citep[e.g.][]{Marois2008, Marois2010, Lagrange2010, Bohn2020, Bohn2021}.

\input{SED_table_v2.tex}

\subsection{Rotational Evolution as a Function of Time and Mass}

\begin{figure*}[tb]
   \centering
  \includegraphics[scale=0.6]{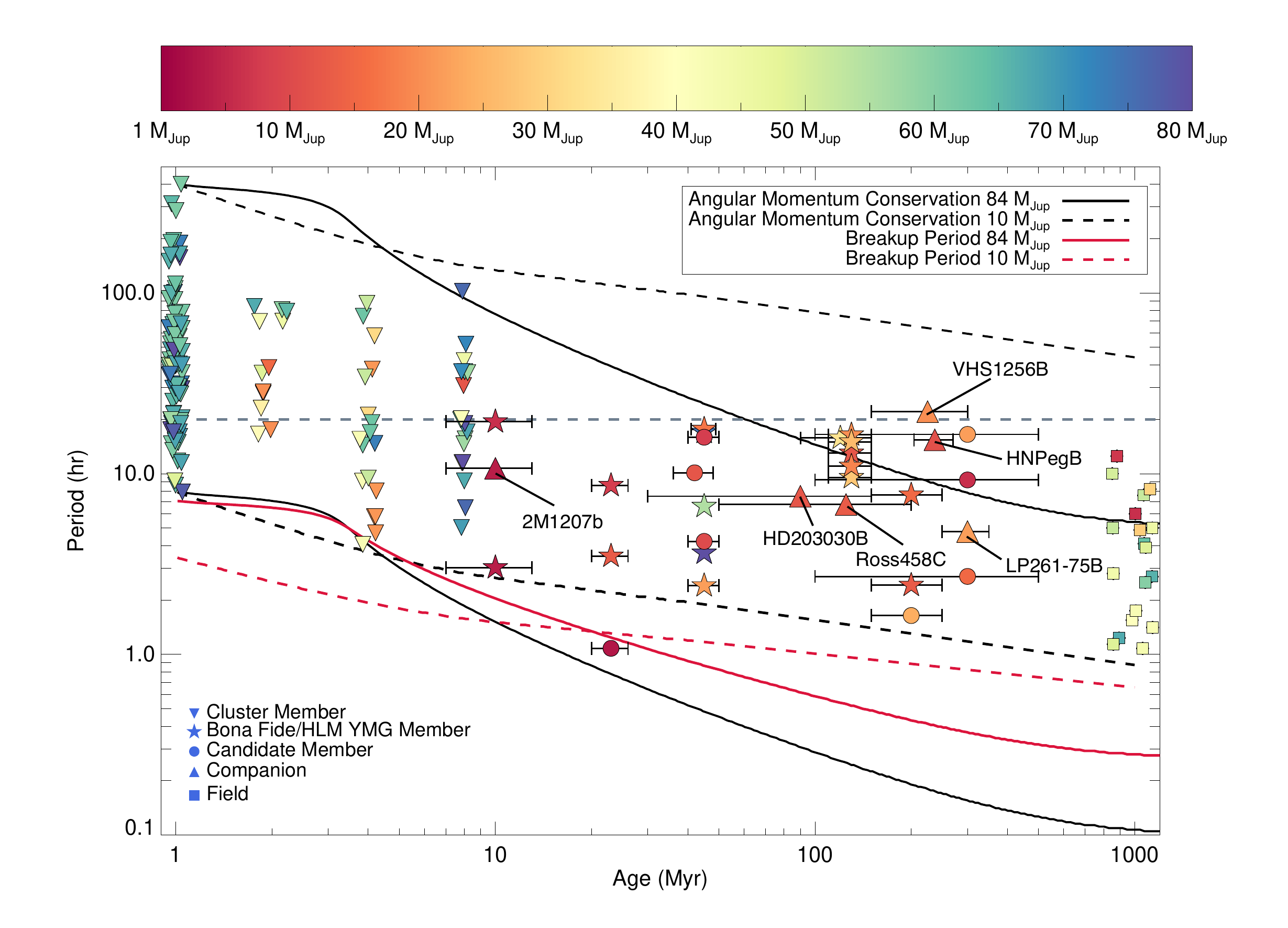}
   \caption{Brown dwarf rotation period as a function of age and color-coded by mass. Stars show bona-fide and high-likelihood members of young moving groups, circles show candidate members of young moving groups, squares show field brown dwarfs, which are plotted at 1 Gyr. Upside-down triangles are brown dwarf cluster members. Triangles show substellar companions with measured rotation rates. The 20 hr observation duration typically used by brown dwarf Spitzer variability studies is shown by the dashed line. While the higher mass brown dwarfs ($>30~M_{\mathrm{Jup}}$) can be seen to spin up with age, this behavior is not observed in the low mass sample ($<30~M_{\mathrm{Jup}}$), most likely because most low-mass objects lie in the intermediate age range. Companion brown dwarfs and planetary-mass objects at $10-500$~Myr have rotation periods consistent with those of isolated objects with similar masses, indicating that they likely have similar formation pathways and subsequent angular momentum evolution histories.} 
   \label{fig:periods}
\end{figure*}

In Figure \ref{fig:periods}, we show the current census of brown dwarf rotation rates as a function of age and color-coded by mass. The majority of objects were originally compiled in \citet{Vos2020}. The $1-10$ Myr sample was vetted and compiled by \citet{Moore2019} using $J$-band cutoffs to identify brown dwarfs. \citet{Scholz2018} calculate masses for the $\sim2$ Myr Taurus sample, and \citet{Scholz2004} provide masses for the $\sim4$ Myr $\sigma$ Ori cluster sample. Although \citet{Rodriguez-Ledesma2009} estimate masses for the 1 Myr sample, they are not provided in their paper. We estimate their masses by comparing their distance and reddening-corrected $I-$band magnitudes with the \citet{Baraffe2015} evolutionary models, as described by \citet{Rodriguez-Ledesma2009}. Similarly, \citet{Moore2019} do not provide their estimated masses for the Upper Scorpius region. We estimate masses for this sample by comparing the targets' $J$-band magnitudes with the 10 Myr \citet{Baraffe2015} models, as described by \citet{Moore2019}. For the Orion Nebula Cluster, the Orion Belt region, Upper Sco and Taurus we have added a random spread in the ages of each cluster member for clarity. 
The rotation periods of brown dwarfs at age $>10$ Myr are compiled in \citet{Vos2020}. We have added the measured rotation rates of young objects measured in this paper and those that have since been published by \citet{Tannock2021}. 
Masses were either gathered from the literature \citep{Faherty2016, Filippazzo2015} or estimated in this paper (Table \ref{tab:SEDs}). As described in \citet{Vos2020}, we also plot angular momentum conservation tracks for objects with masses of $10~M_{\mathrm{Jup}}$ and $84~M_{\mathrm{Jup}}$ using the evolutionary models of \citet{Baraffe2015} as well as the breakup period for objects with these masses.

The majority of brown dwarf rotation rates $>10$ Myr have been measured with Spitzer, and many surveys including \citet{Metchev2015} and this work observe for $\sim$20 hrs. We show this $\sim$20 hr cutoff as the gray dashed line in Figure \ref{fig:periods}. It is notable that we have detected a large number of variables in the $10-500$ Myr age range with periods close to 20 hr, most notably a cluster of AB~Doradus members with periods of 10--20 hrs. This suggests that by limiting our observation length, we are only sensitive to the fastest rotators and may be missing some variable brown dwarfs with rotation rates $>20$ hr.   
However, the high variability occurrence rates achieved by large surveys  suggest that the majority of  rotation rates have been measured. Until we can probe longer rotation rates by observing on timescales of many days, it is important to keep in mind that we may not be measuring the full rotation rate distribution of brown dwarfs and/or isolated and companion planetary-mass objects. 

We can also split the sample into two mass bins: the low-mass sample ($<30~M_{\mathrm{Jup}}$) and the high-mass sample ($>30~M_{\mathrm{Jup}}$). It is evident that the higher mass brown dwarf sample spins faster with age, as expected from angular momentum conservation. However, the spin up of the low-mass sample is not so evident. Based on the angular momentum conservation tracks shown in Figure \ref{fig:periods}, we expect that the high-gravity brown dwarfs should spin up more than their low-gravity counterparts. However, we do still expect to see some spin-up for the low-mass sample if we can probe a large enough age range. In particular, the long rotation periods of the two older planetary-mass Y dwarfs WISE~J085510.83-071442.5 and WISE~J140518.8+553421.3 \citep{Cushing2016, Esplin2016} are somewhat unexpected{, as they are longer than the rotation periods of some planetary-mass objects at younger ages. However, the $\sim10$~hr rotation period of Jupiter suggests that the lowest-mass objects may maintain longer periods at older ages.} A larger survey for variability in Y dwarfs (Cushing, M. C. et al., in preparation) may help to provide more Y dwarf rotation periods, and reveal the rotational behavior of low-mass objects at old ages.

Finally, we have included companion objects with measured rotation rates \citep{Bowler2020, Manjavacas2018, Manjavacas2019, Miles-Paez2019, Zhou2016, Zhou2018, Zhou2020} in Figure \ref{fig:periods} to investigate whether their angular momentum evolution may be different from the isolated objects focused on in this paper. 
Companion brown dwarfs and planetary-mass objects at ages 10--500~Myr have rotation periods consistent with those of isolated objects with similar mass, indicating that they likely have similar formation pathways and subsequent angular momentum evolution histories. A similar conclusion was reached by \citet{Bryan2018} based on rotational velocities of 11 low-mass objects. Based on our current data, it seems reasonable to use the observed distribution of planetary-mass brown dwarfs to estimate the expected rotation rates of directly-imaged exoplanets when planning variability monitoring observations.

\section{Variability Properties of AB Doradus Members}\label{sec:ABDor}

Within our sample, AB Doradus offers a unique opportunity to study the variability properties of a sample of coeval objects. {Four} objects in our survey are either bona fide or high-likelihood members of the AB Doradus moving group: \obj{2m0001}, \obj{2m0355}, 
\obj{2m1741} and \obj{2m2206-42}. {While the high-likelihood member \obj{2m0326} was observed in this survey, its variability results are somewhat inconclusive so we leave it out of the analysis.} {Four} additional AB Doradus BF/HLM members have variability monitoring observations in the literature: \obj*{w0047}, \obj*{2m2244}, \obj*{2m1425} and \obj*{s1110} \citep{Lew2016,Vos2018,Vos2020}. 

\subsection{Variability Occurrence Rate of AB Doradus}
\begin{figure}[tb]
   \centering
  \includegraphics[scale=0.5]{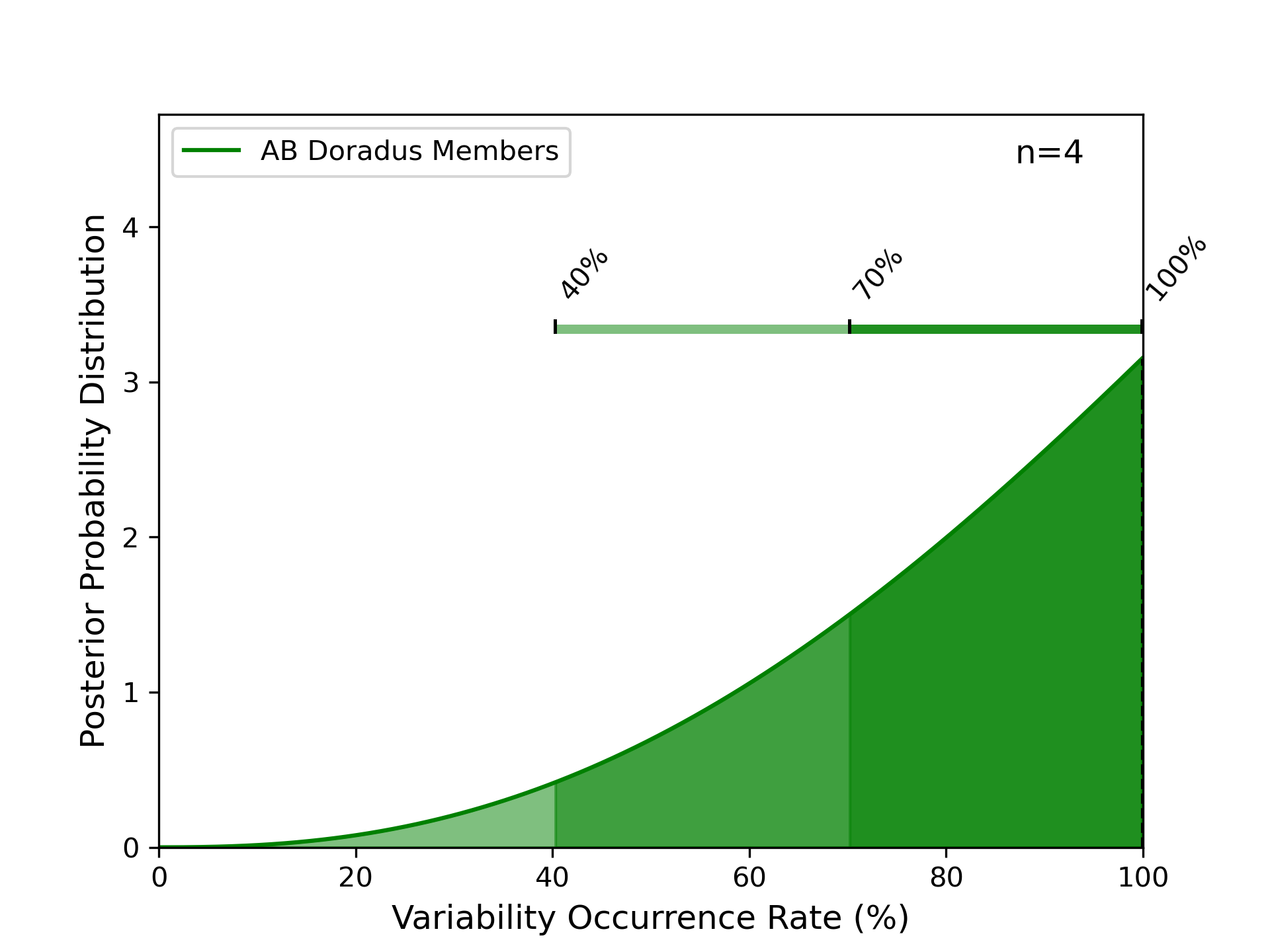}
   \caption{Variability occurrence rate of AB Doradus members in our sample. We determine that the variability occurrence rate for this sample is $>76\%$ at $68\%$ confidence and $>50\%$ at $95\%$ confidence.} 
   \label{fig:ABDMG_occurrence}
\end{figure}

First we use the methods outlined in Section \ref{sec:variability_occurrence} to determine the rate of variability for the {four} AB Doradus members in our sample. We did not include the other AB Doradus members presented in the literature to avoid biasing our sample -- some targets may have been observed with Spitzer after having been previously found to exhibit variability while some non-detections may not have been published. We detect significant variability in 3/4 objects in our sample, {with measured amplitudes from $0.3-0.7\%$}. We show the probability distribution of the variability occurrence rate for AB Doradus member in Figure \ref{fig:ABDMG_occurrence}. We determine that the variability occurrence rate for this sample is $>70\%$ at $68\%$ confidence and $>40\%$ at $95\%$ confidence. As we uncover more L and T type members of young moving groups, we can begin to compare occurrence rates across moving groups, and hence across ages.

\subsection{Variability Amplitudes and Rotation in AB Doradus}
\begin{figure}[tb]
   \centering
   \includegraphics[scale=0.55]{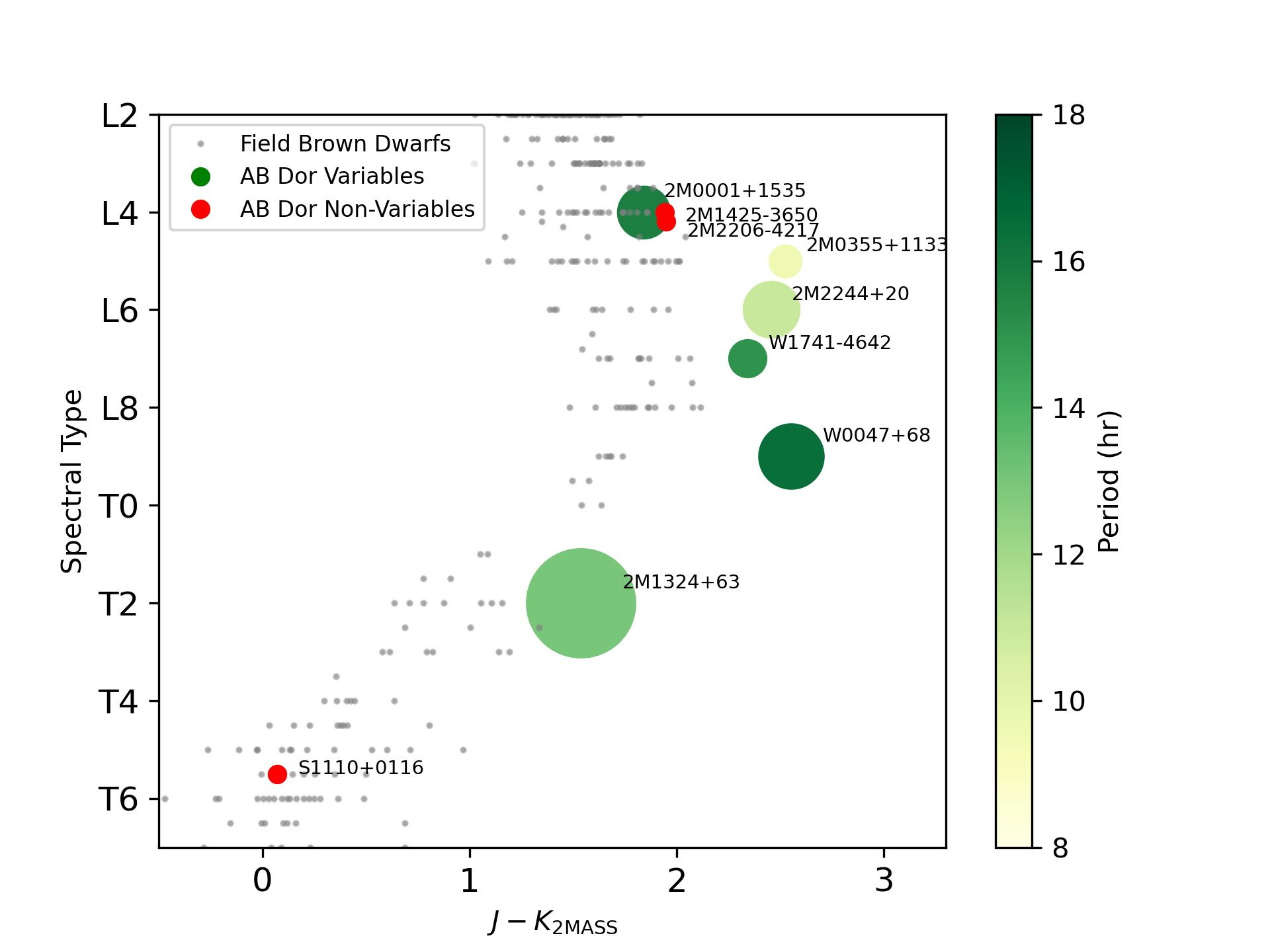}
   \caption{AB Doradus members monitored for variability in the Spitzer $[3.6~\mu$m] band. AB Doradus with detections of variability are shown in green, where the symbol size is proportional to the variability amplitude and the rotation rate is shown by color. Non-detections are shown in red. The field brown sample from \citet{Best2020} is shown in gray for context.} 
   \label{fig:ABDMG_var}
\end{figure}

In Figure \ref{fig:ABDMG_var} we show all {ten} HLM/BF substellar AB Doradus members that have been monitored in the Spitzer $[3.6~\mu$m] band, including this work and literature values from \citet{Vos2018,Vos2020}, plotted on a spectral type color diagram. We see a large range of [$3.6~\mu$m] amplitudes for this sample from $0.4-3.1\%$. It seems that the observed variability amplitude increases as the temperature cools, at least from early-L to early-T spectral types. This behavior is also seen in Figure \ref{fig:amplitudes} for both the low-gravity and field-gravity populations, but it is notable that the amplitudes increase with decreasing temperature for a coeval population of young objects.

{There is a very small range in the measured periods within AB Doradus; all of the variable objects have periods from $9-17$~hr.  }This clustering of rotation periods is also evident in Figure \ref{fig:periods} where there is a pile-up of objects in this region. Since these objects are coeval, with similar masses, the similarity in their periods suggest that they share common angular momentum histories. Comparing these variability and rotation trends across other moving groups will help to reveal the diversity in the atmospheres of coeval objects. Moreover, combining Spitzer variability information for brown dwarfs with observations of high- to low-mass stars with the Transiting Exoplanet Survey Satellite (TESS) will allow us to trace variability properties and rotation from the highest to lowest mass products of star formation in a coeval sample.

\section{Lessons for variability monitoring of directly-imaged exoplanets} \label{sec:lessons}
One of the goals of this survey is to establish expectations for detecting variability in directly-imaged exoplanets. There has been considerable success with variability searches in a handful of young wide-orbit companions with the WFC3 instrument on the Hubble Space Telescope \citep{Bowler2020, Lew2019, Zhou2016,Zhou2019, Zhou2020} as well as some ground-based studies \citep[e.g.][]{Naud2017}. Due to their close proximity to their host star and relative faintness, variability searches on directly-imaged exoplanets such as HR8799~bcde and $\beta$ Pic b present considerable challenges. To date, there have been two ground-based attempts to detect variability in HR8799~bc using the Very Large Telescope (VLT) \citep{Apai2016, Biller2021}, however neither study reached the sensitivity to detect  variability, placing near-infrared upper limits of $5-10\%$ on the variability amplitudes of the planets. Searching for variability in these exoplanets will likely require next-generation telecopes such as the James Webb Space Telescope (JWST) and 30-m class telescopes such as the European-Extremely Large Telescope (E-ELT), the Thirty Metre Telescope (TMT) and the Giant Magellan Telescope (GMT), which will provide unprecedented sensitivity for variability searches in directly-imaged exoplanets. The survey presented here outlines baseline expectations for the behavior of exoplanets with similar temperatures, ages and masses.

Our survey suggests that the directly-imaged exoplanets are likely to have a high variability occurrence rate across the full L and T spectral type regime. Our results also show that we can expect a large range of variability amplitudes for the directly-imaged exoplanets with very large amplitudes for a subset of objects, potentially those with favorable viewing inclinations \citep[e.g.][]{Vos2017}. Comparing the variability behavior of the directly-imaged exoplanets with the low-mass sample presented in this paper will help to put each exoplanet in context with a larger sample of variables with similar temperatures, masses and radii.
Our survey also reveals the expected rotation periods for young, directly-imaged exoplanets. With a median rotation period of $\sim12$~hr for young free-floating  and companion objects, this work highlights the importance of long duration observations in order to measure a full period. For ground-based observations with the E-ELT, TMT and/or GMT multiple nights of observations may be necessary to capture the periodicity, {however the fastest rotators could be prioritized by measuring $v\sin (i)$ values prior to variability searches \citep{Wang2021}.} 
Finally, ongoing variability surveys of Y dwarfs will be critical in revealing the potential for detecting variability in the coldest exoplanets (Cushing, M. C. et al., in preparation).

\section{Discussion and Conclusions}\label{sec:conclusions}

We have presented the most sensitive survey to date for variability in young, low-gravity, giant planet analogs. We surveyed 26 brown dwarfs with  evidence of youth using the Spitzer Space Telescope at $3.6~\mu$m, 23 of which we classify as young after a detailed study of their kinematic and spectroscopic evidence of youth. We report the discovery of three new brown dwarfs in this paper: \obj{2m0349}, \obj{2m0951} and \obj{2m0718}, all of which show signatures of youth. Our variability analysis, which makes use of periodogram analysis and the Bayesian Information Criterion (BIC), identifies 15 variable and 11 non-variable objects in the full sample. 
We compare the variability properties of the low-gravity sample with those of their higher-mass field brown dwarf counterparts reported in the literature, specifically focusing on the variability occurrence rates, the variability amplitudes, and the rotation rates. 

We determine the variability occurrence rates of young and field brown dwarfs in three spectral bins: L2$-$L8, L9$-$T3.5 and $\geq$T4. For L dwarfs, we find that both the young and field objects have variability occurrence rates of $80-100\%$, i.e.  virtually all L dwarfs are likely to be variable at the $0.05-3\%$ level.  
Our variability occurrence rate calculations find a tentative enhancement in the rate of variability in low-gravity objects with spectral types L9-T3.5 compared to the field dwarf populations with similar spectral types. The variability occurrence rate of low-gravity objects remains high from L to T dwarfs, while the occurrence rate appears to drop for field L/T transition and T dwarfs. The small sample size for young objects with spectral types $\geq$T4 prevents us from placing meaningful constraints on their variability occurrence rates. These results are encouraging for future high-contrast searches for variability in directly-imaged exoplanets with JWST. While a number of ground-based searches have been carried out for the HR8799 planets \citep{Apai2016, Biller2021} without detecting variability, JWST will provide unprecedented sensitivity and precision. The results of these searches will be directly informed by the variability studies of isolated low-gravity brown dwarfs such as this one.

We find that the [$3.6~\mu$m] variability amplitudes increase steadily with spectral type from L to early T, as is observed for the field brown dwarfs \citep{Metchev2015}. This similar observed behavior supports our assumptions that the same variability mechanisms are at play in both samples. We also note that the maximum variability amplitudes observed for L dwarfs are higher for the low-gravity objects compared to the field objects. Atmospheric models predict that clouds are likely to form at higher altitudes in low-gravity atmospheres \citep{Marley2012}, and these higher clouds might result in a larger observed contrast between more cloudy and less cloudy regions. If this is the case, one might expect higher variability amplitudes in the low-gravity sample. Indeed, there are a number of young brown dwarfs displaying extremely large amplitudes compared to their field brown dwarf counterparts such as \obj{2m2139} \citep[$\sim26\%$][]{Radigan2012}, VHS~1256~B \citep[$\sim25\%$][]{Bowler2020, Zhou2020} and PSO~J318.5$-$22 \citep[$\sim11\%$][]{Biller2015}. Our work shows that not every young brown dwarf exhibits such high amplitudes. We speculate that while the higher contrast from higher altitude clouds may result in higher intrinsic amplitudes, secondary effects such as viewing inclination \citep{Vos2017} may be responsible for the range in amplitudes that we observe.

We combine our new rotation rates with those compiled by \citet{Vos2020} to investigate the angular momentum evolution of substellar objects from 1 Myr to $>1$ Gyr. Firstly, we find that a large number of objects with ages 10--300~Myr have rotation periods close to 20 hr, which is the observation duration for many variability searches with Spitzer, including this work. This suggests that we may be missing the slowest rotators within this sample. 
Secondly, while the evolutionary spin-up of brown dwarfs with masses $>30~M_{\mathrm{Jup}}$ is evident, this process is not so clear for the lower mass brown dwarfs ($<30~M_{\mathrm{Jup}}$). Measuring rotation rates of a larger number of low-mass brown dwarfs at ages $<10$ Myr and $>1$ Gyr (e.g. Y dwarfs) will help to  reveal their spin-up properties over time and as a function of mass. We also compare the rotation rates of the small number of low-mass companion objects that have been monitored to date, finding that their rotation rates are consistent with those of the isolated low-mass brown dwarfs. These results suggest that we can base future variability monitoring observations of directly-imaged exoplanets on the rotation rates measured to date for isolated low-mass brown dwarfs.

AB Doradus offers a unique opportunity to investigate the variability properties of a sample of coeval objects. Within AB Doradus, we find a variability occurrence rate of $>70\%$ and $68\%$ confidence. We note that the amplitudes of AB Doradus members increase with decreasing temperature, as has been noted for field brown dwarfs \citep{Metchev2015} and low-gravity sources \citep[][this work]{Vos2020}, but shown here for the first time in a coeval sample. We also note an apparent clustering of rotation periods for AB Doradus objects, with all of the measured periods in the range $9-17$~hr. The similarity in their periods may suggest that these objects share common angular momentum histories. Comparing these variability trends across other moving groups will help to reveal how they change with age and mass.

Overall, the results from this survey are encouraging for variability searches in directly-imaged exoplanets with future facilities such as JWST and 30-meter telescopes, and the young brown dwarf sample presented here will serve as an excellent baseline for comparison. Variability monitoring observations with these facilities will enable unprecedented characterization of the weather phenomena and angular momentum properties and directly-imaged exoplanets.


\section{Acknowledgements}

The authors would like to thank Azul Ruiz Diaz, Jai Glazer and Sophia Ameneyro for their help and enthusiasm on this project through the NYC Science Research Mentoring Program at the American Museum of Natural History. {We would like to thank the anonymous referee for thoughtful suggestions that significantly improved the paper.} We would like to thank Dr Seppo Laine and the Spitzer Science Center for their expertise and advice on Spitzer data products and analysis. 
This work is based on observations made with the Spitzer Space Telescope, which is operated by the Jet Propulsion Laboratory, California Institute of Technology under a contract with NASA. The Spitzer data is available at the Spitzer Heritage Archive hosted by IRSA: \dataset[10.26131/IRSA430]{\doi{10.26131/IRSA430}}.
J. M. V. acknowledges support by NSF Award Number 1614527, Spitzer Cycle 14 JPL Research Support Agreement 1627378 and  HST-GO-15924.001-A.   J. V. and J.F. acknowledge support from the Heising-Simons Foundation. 
This work has made use of data from the European Space Agency (ESA) mission {\it Gaia} (\url{https://www.cosmos.esa.int/gaia}), processed by the {\it Gaia} Data Processing and Analysis Consortium (DPAC, \url{https://www.cosmos.esa.int/web/gaia/dpac/consortium}). Funding for the DPAC has been provided by national institutions, in particular the institutions participating in the {\it Gaia} Multilateral Agreement. This work has benefited from The UltracoolSheet, maintained by Will Best, Trent Dupuy, Michael Liu, Rob Siverd, and Zhoujian Zhang, and developed from compilations by \citet{Dupuy2012}, \citet{Dupuy2013a}, \citet{Liu2016}, \citet{Best2018}, and \citet{Best2020}.

\facilities{Spitzer Space Telescope, Gaia DR2 \citep{Gaia2016, Gaia2018}, Magellan I FIRE \citep{Simcoe2013}}
\software{BANYAN~$\Sigma$ \citep{Gagne2018b}, \texttt{SEDkit} \citep{Filippazzo2015}, \texttt{emcee} \citep{fm2013}, \texttt{matplotlib} \citep{Hunter2007}, \texttt{celerite2} \citep{Foreman-Mackey2017}}




\appendix

\section{Jeffreys Prior Derivation}\label{app:prior}

\begin{figure}
   \centering
  \includegraphics[scale=0.6]{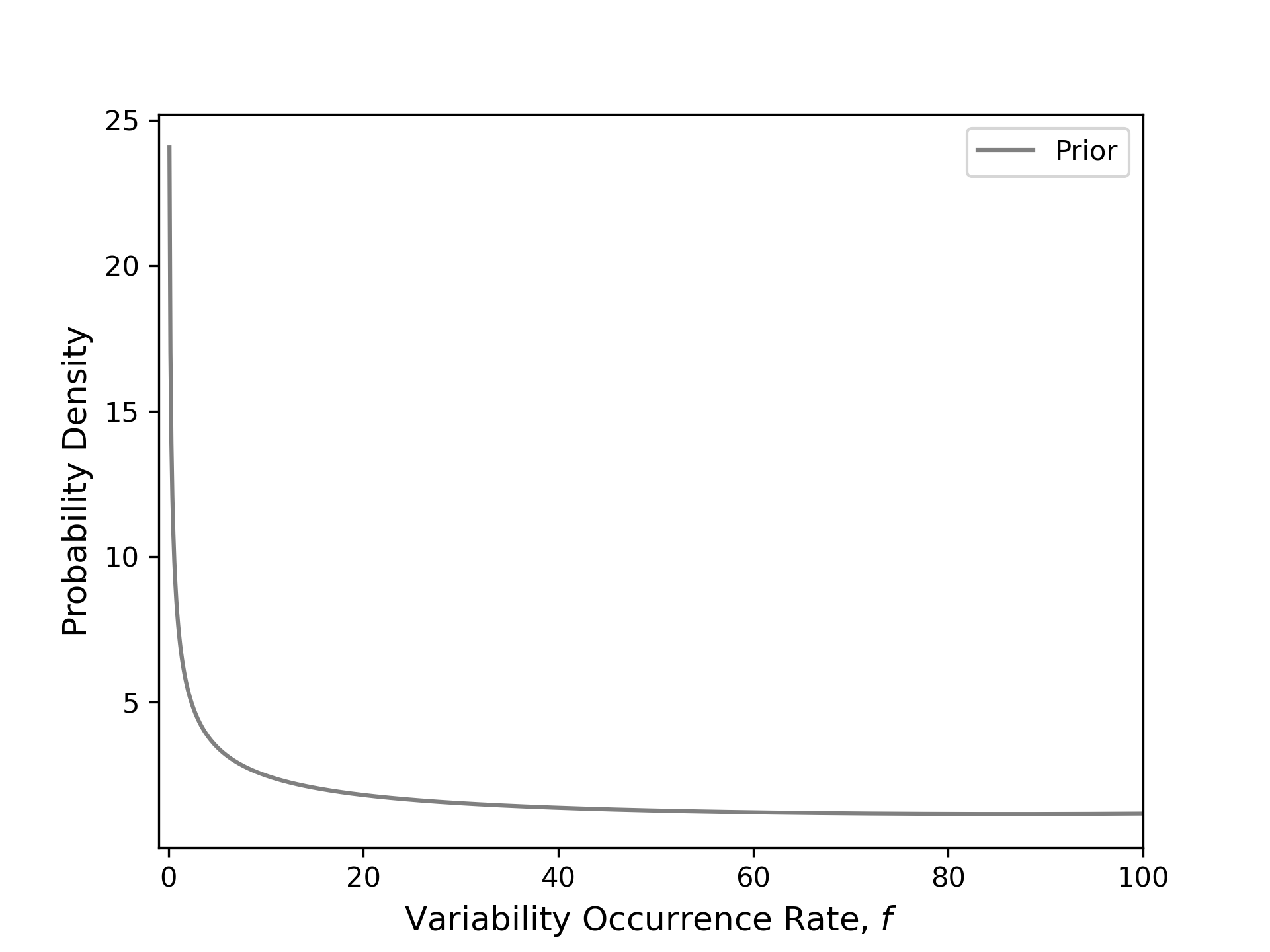}
   \caption{Prior used for estimation of variability occurrence rate in Section \ref{sec:variability_occurrence}.} 
   \label{fig:prior}
\end{figure}

The Jeffreys prior is a non-informative prior which is invariant under parameter transformation. Jeffreys prior is defined in terms of the Fisher information:
\begin{equation}
    J(\theta)\propto \sqrt {I(\theta)}
\end{equation}
where the Fisher information $I(\theta)$ is given by:
\begin{equation}
    I(\theta)=-\mathbb{E}_{\theta}\left[\frac{d^2\ell}{d\theta^2} \right]
\end{equation}
where $\mathbb{E}$ represents the expectation value and $\ell$ is the log likelihood. For the likelihood presented in Section \ref{sec:variability_occurrence}, the Fisher information is given as:
\begin{equation}
    I(f)= -\mathbb{E}_{f} \left[\sum_j \frac{d_j}{f^2 p_j} + \frac{p_j(1-d_j)}{(1-fp_j)^2} \right]
\end{equation}
The expectation value of $d_j$ is simply the $fp_j$, so the Fisher information can be expressed as:
\begin{equation}
   I(f)= \sum_j \frac{p_j}{f(1-fp_j)}
\end{equation}
and the Jeffreys prior is thus given as
\begin{equation}
J(f) = \sqrt{\sum_j \frac{p_j}{f(1-fp_j)}}.
\end{equation}
This represents a noninformative prior for estimation of the posterior distribution of the variability occurrence rate, $f$. We plot the prior in Figure \ref{fig:prior}.

\bibliography{main}{}
\bibliographystyle{aasjournal}

\end{document}

%% file: objects.tex
\makeatletter
\newcommand\DefineObj[3][\@empty]{%
  \expandafter\newcommand\csname pkgwobj@l#2\endcsname{#3}%
  \ifx\@empty#1%
    \expandafter\newcommand\csname pkgwobj@s#2\endcsname{#3}%
  \else%
    \expandafter\newcommand\csname pkgwobj@s#2\endcsname{#1}%
  \fi}%
\newcommand{\pkgw@printobj@long}[1]{%
  \expandafter\ifx\csname pkgwobj@l#1\endcsname\relax%
    \textbf{[unknown object!]}%
  \else%
    \csname pkgwobj@l#1\endcsname%
  \fi}%
\newcommand{\pkgw@printobj@short}[1]{%
  \expandafter\ifx\csname pkgwobj@l#1\endcsname\relax%
    \textbf{[unknown object!]}%
  \else%
    \csname pkgwobj@s#1\endcsname%
  \fi}%
\newcommand{\obj}{\@ifstar{\pkgw@printobj@long}{\pkgw@printobj@short}}%
\makeatother

%% file: kinematics_v2.tex
\startlongtable
\begin{longrotatetable}
\begin{deluxetable}{lcccccccccc}
\tabletypesize{\scriptsize}
\tablecolumns{10}
\tablewidth{0pt}
\setlength{\tabcolsep}{0.05in}
\tablecaption{Kinematic information for sample.}
\tablehead{  
\colhead{Name} &
\colhead{$\mu_{\alpha}$ } &
\colhead{$\mu_{\delta}$} &
\colhead{Ref} &
\colhead{RV} &
\colhead{Ref } &
\colhead{Parallax} &
\colhead{Ref } & 
\colhead{BANYAN $\Sigma$}&
\colhead{Membership}\\
\colhead{} &
\colhead{mas yr$^{-1}$} &
\colhead{mas yr$^{-1}$} &
\colhead{} &
\colhead{km s$^{-1}$} &
\colhead{ } &
\colhead{mas} &
\colhead{} & 
\colhead{}&
\colhead{}}

\startdata
\obj*{2m0001} & $143.5\pm4.5$  & $-174.7\pm2.4$  & 1  &  \nodata   & \nodata       & $31.6\pm1.8$   & 1   &  92\% ABDMG & HLM ABDMG\\
\obj*{2m0030}  & $256.1\pm3.4$ & $-41.7\pm3.3$   & 2$^{\mathrm{b}}$  &\nodata  &\nodata        & $41.1\pm5.3$  & 2     & 93\% ARG  & HLM ARG\\
\obj*{2m0031} & $522.5\pm 3.5$       & $-23.9 \pm4.0$    & 3   & \nodata     & \nodata  & $71\pm5$       & 2      &  96\% CARN & HLM CARN\\
\obj*{2m0153} & $82\pm2.7$       & $-21.3 \pm9.4$& 4   & \nodata       & \nodata   &   \nodata    & \nodata  &  89\% THA  & CAN THA\\
\obj*{2m0326} & $108.0\pm5.0$   & $-156.6\pm4.5$ & 5  &\nodata     & \nodata       &  \nodata       & \nodata   & 92\% ABDMG& HLM ABDMG\\
\obj*{2m0342}  & $73.7\pm2.3$    & $20.6\pm7.8$  & 4  &\nodata      & \nodata       &  \nodata       &  \nodata  &  93\% THA   & HLM THA\\
\obj*{2m0349} & $52.0\pm10.7$   & $-81.9\pm10.3$ & 5  & $13.8\pm3.3$  &    8    &\nodata         &\nodata   & 55\% BPMG, 20\% ABDMG & AM\\
\obj*{2m0355} & $219.8\pm1.6$  & $-631.3\pm0.8$  & 1   & $11.92\pm0.22$    & 9    & $109.6\pm0.7$  & 1     & 100\% ABDMG & BF ABDMG\\
\obj*{2m0447} & $-6.3\pm2.6$     & $-99.1\pm2.3$ & 2   & \nodata &\nodata   & $23.0\pm3.9$     & 2          & 99\% FLD   & FLD\\
\obj*{2m0459} & $53.8\pm7.0$   & $84.9\pm7.2$    & 5  & $23.4\pm3.8$& 8      &  \nodata       & \nodata.        &98\% ARG     & HLM ARG\\
\obj*{2m0506} & $49.8\pm3.3$ & $-205.9\pm3.8$    & 2    & \nodata    & \nodata  &  $59.6\pm5.4$      &\nodata      & 69\% FLD  & FLD\\
\obj*{2m0642} & $-2.0\pm 1.2$   & $-383.1\pm1.2$    & 7    & \nodata   &  \nodata      & $62.6\pm3.1$   & 7   &   86\% ABDMG & CAN ABDMG\\
\obj*{2m0718} & $-175.4\pm13.5$& $254.81\pm13.9$ & 5 & $13\pm2$  & 8    &\nodata         &\nodata             & 85\% BPMG  & CAN BPMG\\
\obj*{2m0809} & $-181.9 \pm5.1$    & $-219\pm5.1$& 5  & \nodata   & \nodata     & $42.4\pm3.6$       & 11         & 89\% FLD    &YNG-FLD\\
\obj*{2m0951} & $-92.3\pm9.9$    & $36.8 \pm9.2$ & 5 & $2.8\pm3.3$     & 8    & \nodata        &\nodata      &  87\% ARG &CAN ARG\\
\obj*{2m1551} & $-62.1\pm0.6$ & $-57.7\pm0.6$    & 6   & $-15.2\pm1.1$  & 8    & $22.1\pm1.5$   & 6    &100\% FLD  & YNG-FLD\\
\obj*{2m1647} & $-84.6\pm1.4$    & $241.0\pm2.6$& 2   & \nodata   & \nodata   & $42.7\pm2.1$   &2  &66\% FLD &YNG-FLD \\
\obj*{2m1741} & $-29.2\pm2.1$   & $-356.5\pm 2.1$& 7& $-0.9\pm3.2$   &  8     & $50.5\pm2.9$   & 7      & 99\% ABDMG  &BF ABDMG\\ 
\obj*{2m2002} &$-111.4\pm3.1$  & $-114.7\pm2.4$  & 1    &  \nodata  & \nodata   & $56.7\pm1.5$   & 1     &100\% FLD  & YNG-FLD\\
\obj*{2m2117} & $139.3\pm4.5$      & $-171\pm4.8$      & 2     & \nodata   &\nodata    & $52.4\pm6.8$   & 2    & 98\% BPMG & AM$^{\mathrm{a}}$ \\
              & $149\pm2.7$      & $-168.1\pm3.2$      & 7     & \nodata   &\nodata    & $76.1\pm3.5$   & 7    & 99\% FIELD & AM$^{\mathrm{a}}$ \\
\obj*{2m2154} & $174.9\pm6.2$    & $-3.3\pm6.1$  & 5     & $-20.7\pm1.0$   & 8       & $32.6\pm1$     & 6    & 80\% CARN & CAN CARN\\
\obj*{2m2206+33} & $176\pm9$      & $16\pm11$    & 3    & \nodata    & \nodata    & \nodata      & \nodata     & 68\% ARG  &CAN ARG\\
\obj*{2m2206-42} & $130.8\pm1.8$ & $-183.2\pm2.3$& 1     &   \nodata   &\nodata    & $33.8\pm1.8$   & 1       &  99\% ABDMG &HLM ABDMG\\
\obj*{2m2216}   & $160.4\pm14.8$ & $-71.53\pm15.0$& 5    & \nodata    & \nodata       & \nodata        &\nodata     & 85\% BPMG &CAN BPMG\\
\obj*{2m2322} & $80.1 \pm1.4$  & $-82.0\pm1.6$    & 1    & $6.8\pm0.8$ & 10       & $23.2\pm1.0$   & 1      &  99\% THA   &BF THA\\
\obj*{2m2343} & $124.5\pm5.5$    & $-105.7\pm4.7$ & 5    & \nodata     &  \nodata  &  \nodata       &\nodata   &  75\% ABDMG & CAN ABDMG\\ [0.5ex]  \hline          \enddata
\tablecomments{BF: Bona fide member, HLM: High-likelihood member, CAN: Candidate member, FLD: Field member\\
$^{\mathrm{a}}$ We classify \obj*{2m2117} as ambiguous because \citet{Best2020} and \citet{Kirkpatrick2020} present discrepant parallax values for this target.\\
\textbf{$^{\mathrm{b}}$ The radial velocities quoted from Gagn\'e et al. in prep are determined following the method presented in \citet{Gagne2017, Gagne2018a}.}}
\tablerefs{
(1)~\citet{Gaia2018};
(2)~\citet{Best2020}; 
(3)~\citet{Best2015};
(4)~\citet{Gagne2015a};
(5)~\citet{Marocco2021};
(6)~\citet{Liu2016};
(7)~\citet{Kirkpatrick2020};
(8)~Gagn\'e et al. in prep;
(9)~\citet{Blake2010};
(10)~\citet{Faherty2016};
(11)~\citet{Faherty2012}}
\label{tab:kinematics}
\end{deluxetable}
\end{longrotatetable}

%% file: SED_table_v2.tex
\startlongtable
\begin{deluxetable*}{lllllllll}
\tabletypesize{\scriptsize}
\tablecolumns{10}
\tablewidth{0pt}
\setlength{\tabcolsep}{0.05in}
\tablecaption{Fundamental parameters for sample from SED analysis}
\tablehead{  
\colhead{Shortname} &
\colhead{Parallax$^{a}$ } &
\colhead{Membership} &
\colhead{Age} &
\colhead{$L_{\mathrm{bol}}$} &
\colhead{$T_\mathrm{eff}$} &
\colhead{Radius } & 
\colhead{Mass} &
\colhead{log(g)}\\
\colhead{} &
\colhead{(mas)} &
\colhead{} &
\colhead{(Myr)} &
\colhead{(dex)} &
\colhead{(K)} &
\colhead{($R_{\mathrm{Jup}}$)} & 
\colhead{($M_{\mathrm{Jup}}$)} &
\colhead{(dex)}}

\startdata
\obj{2m0001} & 31.6 $\pm$ 1.8   & HLM ABDMG      & 110--130     & $-3.90\pm0.05$ & $1725\pm5$  & $1.22\pm0.02$      & $33\pm4$    & $4.73\pm0.06$    \\
\obj{2m0030}  & 41.1$\pm$5.3   & HLM ARG     & 30--50          & $-4.35\pm0.11$ & $1264\pm78$  & $1.36\pm0.04$      & $10\pm2$    & $4.12\pm0.05$   \\
\obj{2m0031}& 71$\pm$5.0     & HLM CARN      & 150--250         & $-4.35\pm0.06$ & $1343\pm58$  & $1.2\pm0.06$       & $24\pm7$   & $4.60\pm0.17$     \\
\obj{2m0153} & (19.9$\pm$1.2)  & CAN THA      & 41--49          & $-3.80\pm0.05$ & $1664\pm53$  & $1.48\pm0.03$      & $17\pm2$    & $4.26\pm0.03$    \\
\obj{2m0326} & (42.7$\pm$2.1)   & HLM ABDMG    & 110--150       & $-4.25\pm0.04$ & $1359\pm46$  & $1.32\pm0.06$      & $18\pm4$   & $4.37\pm0.11$    \\       
\obj{2m0342}  & (19.4$\pm$1.1)   & HLM THA    & 41--49          &$-3.81\pm0.05$   &$1650\pm51$  &$1.48\pm0.03$       & $17\pm2$  & $4.26\pm0.03$   \\
\obj{2m0349} & \nodata    & AM     & 10--150           &\nodata  & \nodata  &\nodata      & \nodata      & \nodata        \\
\obj{2m0355} & 109.6$\pm$0.7   & BF ABDMG      & 110--150       & $-4.12\pm0.01$ & $1527\pm14$  & $1.22\pm0.02$      & $28\pm2$   & $4.66\pm0.05$     \\
\obj{2m0447} & 23$\pm$3.9     & FLD     & 500--1000              & $-4.28\pm0.15$ & $1521\pm139$ &  $1.02\pm0.07$     & $51\pm16$   & $5.05\pm0.22$    \\
\obj{2m0459} & (32.2$\pm$3.1)   & HLM ARG     &30--50          & $-4.54\pm0.15$   & $1152\pm98$ & $1.32\pm0.03$      & $8\pm1$    & $4.06\pm0.06$   \\
\obj{2m0506} & $59.6\pm5.4$  & FLD     & 500-1000               & $-4.87\pm0.08$ & $1065\pm63$  & $1.05\pm0.08$       & $26\pm7$   & $4.74\pm0.20$    \\
\obj{2m0642} & $62.6\pm3.1$  &  CAN ABDMG     & 110--150        & $-4.64\pm0.02$ & $1124\pm11$  & $1.23\pm0.0$       & $12\pm1$   & $4.28\pm0.01$    \\
\obj{2m0718} & ($107\pm11$)     & CAN BPMG      & 16--28        & $-5.70\pm0.09$ & $596\pm32$   & $1.29\pm0.02$      & $3\pm1$    & $3.57\pm0.07$    \\
\obj{2m0809} & 42.4$\pm$3.6     & YNG-FLD      & 10--150        & $-4.37\pm0.07$  & $1253\pm74$ & $1.35\pm0.11$      & $13\pm8$   & $4.12\pm0.36$   \\
\obj{2m0951} & ($17.2\pm1.4$)       & CAN ARG      & 30--50    & $-4.04\pm0.07$ &  $1463\pm65$ & $1.45\pm0.05$       & $ 15\pm2$  & $4.22\pm0.07$     \\
\obj{2m1551} & 22.1$\pm$1.5     & YNG-FLD       & 10--150       & $-3.86\pm0.06$ & $1653\pm126$ & $1.39\pm0.19$      & $24\pm14$  & $4.41\pm0.40$    \\
\obj{2m1647} & 42.7$\pm$2.1  & YNG-FLD     & 150--250           & $-4.47\pm0.04$ & $1244\pm47$  & $1.22\pm0.07$      & $18\pm6$    & $4.45\pm0.19$    \\
\obj{2m1741} & 50.5$\pm$2.9    & BF ABDMG      & 110--150       & $-4.17\pm0.05$ & $1471\pm44$  & $1.23\pm0.02$      & $26\pm3$   & $4.63\pm0.07$    \\
\obj{2m2002} & 56.7$\pm$1.5     & YNG-FLD       & 10--150       & $-4.30\pm0.02$ & $1301\pm56$  & $1.36\pm0.11$      & $13\pm8$   & $4.14\pm0.34$    \\
\obj{2m2117}$^{b}$ & 52.4$\pm$ 6.8    & AM/HLM BPMG  & 16--28  & $-4.44\pm0.11$  & $1192\pm80$  & $1.38\pm0.05$      & $7\pm2$    & $3.95\pm0.12$ \\
             & $76.1\pm3.5$    & AM/FLD    & 500--1000          & $-4.77\pm0.04$  & $1130\pm50$  & $1.05\pm0.08$      & $27\pm7$    & $4.75\pm0.21$    \\
\obj{2m2154} & 32.6$\pm$1.0     & CAN CARN       & 150--250     & $-4.14\pm0.03$ & $1539\pm31$  & $1.17\pm0.03$      & $33\pm4$    & $4.76\pm0.08$    \\
\obj{2m2206+33}& (35.8$\pm$2.4) & CAN ARG      & 30--50         & $-4.67\pm0.06$  & $1076\pm37$  & $1.3\pm0.02$       & $8\pm1$     & $4.03\pm0.05$    \\
\obj{2m2206-42}  & 33.8$\pm$1.8 & HLM ABDMG    & 110--150       & $-3.95\pm0.05$ & $1684\pm46$  & $1.22\pm0.02$      & $32\pm3$   & $4.72\pm0.06$     \\
\obj{2m2216}   & (44.2$\pm$3.9) & CAN BPMG       &   16--28     & $-4.95\pm0.08$ & $909\pm41$   & $1.32\pm0.02$      & $5\pm1$   & $3.79\pm0.12$   \\
\obj{2m2322} & 23.2$\pm$1.0     & BF THA    & 41--49            & $-3.71\pm0.04$ & $1740\pm43$  & $1.5\pm0.03$       & $18\pm2$   & $4.27\pm0.04$    \\ 
\obj{2m2343} &  (24.7$\pm$1.4)      & CAN ABDMG     & 110-150   &$-3.87\pm0.05$  &  $1755\pm53$ &  $1.22\pm0.02$     &$34\pm4$     &   $4.74\pm0.06$        \\   \enddata
\label{tab:SEDs}
\tablecomments{The fundamental parameters listed in this table assume membership in the groups listed in the 'Membership' column. {Fundamental parameters for candidate and ambiguous members should be treated with caution since their membership is uncertain and may change with updated kinematics.}\\  $^{a}$ Parallaxes in parentheses are kinematic distances that assume moving group membership indicated in the membership column. \\ 
$^{b}$ \obj{2m2117} has two discrepant parallax values which result in either high-likelihood $\beta$ Pictoris membership or field membership. We show fundamental parameters for both cases in this table for completeness.}
\end{deluxetable*}